\documentclass[12pt,preprint]{aastex}

\def\lsim{\mathrel{\rlap{\lower4pt\hbox{\hskip1pt$\sim$}}
    \raise1pt\hbox{$<$}}}                % less than or approx. symbol
\def\gsim{\mathrel{\rlap{\lower4pt\hbox{\hskip1pt$\sim$}}
    \raise1pt\hbox{$>$}}}                % greater than or approx. symbol

\shorttitle{A HST/Spitzer Survey for Gravitationally-Lensed Galaxies beyond Redshift Seven}

\begin{document}

\title{A Hubble \& Spitzer Space Telescope Survey for Gravitationally-Lensed Galaxies: Further Evidence for a Significant Population of Low Luminosity Galaxies beyond Redshift Seven}


\author {
Johan Richard \altaffilmark{1},
Daniel P. Stark\altaffilmark{1}, 
Richard S. Ellis\altaffilmark{1}, 
Matthew R. George\altaffilmark{1,3},
Eiichi Egami\altaffilmark{4},
Jean-Paul Kneib \altaffilmark{1,2},
Graham P. Smith \altaffilmark{1,5}
} 

\altaffiltext{1}{Department of Astrophysics, California Institute of Technology, 
MS 105-24, Pasadena, CA 91125; johan@astro.caltech.edu}
\altaffiltext{2}{Observatoire Astronomique de Marseille-Provence, 
Traverse du Siphon - BP 8, 13376 Marseille Cedex 12, France}
%\altaffiltext{3}{Harvard-Smithsonian Center for
%Astrophysics, 60 Garden Street, Cambridge, MA02138}
\altaffiltext{3}{Current address: Institute of Astronomy, Madingley Road, Cambridge, CB3 0HA, UK}
\altaffiltext{4}{Steward Observatory, University
of Arizona, 933 N. Cherry Avenue, Tucson, AZ 85721}
\altaffiltext{5}{School of Physics and Astronomy, University of Birmingham, Edgbaston, Birmingham B15 2TT, UK}

\begin{abstract}

We present the results of a systematic search for gravitationally-lensed continuum Lyman break `drop-outs'
beyond a redshift 7 conducted via very deep imaging through six foreground clusters undertaken with the 
Hubble and Spitzer Space Telescopes. The survey has yielded 10 $z$-band and 2 $J$-band drop-out 
candidates to photometric limits of $J_{110}\simeq$26.2 AB (5$\sigma$). Taking into account the 
magnifications afforded by our clusters (1-4 magnitudes), we probe the presence of $z>$7 
sources to unlensed limits of $J_{110}\simeq$30 AB, fainter than those charted in the Hubble 
Ultradeep Field. To verify the fidelity of our candidates we conduct a number of tests for 
instrumental effects which would lead to spurious detections, and carefully evaluate the likelihood of 
foreground contamination by considering photometric uncertainties in the drop-out signature, the 
upper limits from stacked IRAC data and the statistics of multiply-imaged sources. Overall, we conclude 
that we can expect about half  of our sample of $z$-band drop-outs to be at high redshift. 
An ambitious infrared spectroscopic campaign undertaken with the NIRSPEC spectrograph at the
WM Keck Observatory for seven of the most promising candidates failed to detect any Lyman $\alpha$ 
emission highlighting the challenge of making further progress in this field. While the volume density of 
high redshift sources will likely remain uncertain until more powerful facilities are available, 
our data provides the first potentially interesting constraints on the UV luminosity function at $z\simeq$7.5 
at intrinsically faint limits. We discuss the implications of our results in the context of the hypothesis 
that  the bulk of the reionizing photons in the era $7<z<12$ arise in low luminosity galaxies undetected 
by conventional surveys. 

\end{abstract}


\section{Introduction}

Very little is currently known about the abundance and luminosity distribution of star-forming sources 
beyond $z\gtrsim7$. The two principal techniques used to locate distant star forming sources at lower 
redshifts, the Lyman-break `drop-out' technique \citep{Bouwens06} and the location of Lyman $\alpha$ 
emitters \citep{Kashikawa,Shimasaku}, become challenged by the lower performance of infrared instruments.  
In addition, the likely sources are much fainter, particularly if an increasing fraction are sub-luminous as 
might be expected given mass assembly is at an early stage \citep{Loeb,Choudhury}. Despite these hurdles, 
it seems reasonable to expect that there is an abundance of star-forming galaxies at these epochs. The 
improved measurement of the optical depth to electron scattering derived from temperature-polarization correlations in the microwave background \citep{WMAP} suggests reionization occurred around 
$z_{reion}=10.8\pm1.4$ assuming it happened instantaneously; more probably it proceeded over 
an extended redshift window $7<z<12$ \citep{WMAP4}. Moreover, the detection of galaxies at $z\sim6$ 
with significant stellar masses and mature stellar populations \citep{Stark07a,Eyles07} and the ubiquity of 
ionized carbon in the intergalactic medium probed by the highest redshift QSOs \citep{Songaila,Ryan}, 
together demand a significant amount of star-formation at earlier times, possibly enough to cause 
reionization. Although uncertainties remain, these independent arguments strongly motivate the search for 
 $z>$7 star-forming sources.

Most of the early progress in this quest has been made through the publicly-available deep Hubble
Space Telescope (HST) images. The Great Observatories Origins Deep Survey (GOODS, \citet{goods}), the 
Hubble Ultra Deep Field \citep[UDF]{UDF} and its associated parallel fields have been used to search 
for `drop-outs' in the $i_{775}$ \citep{Bouwens06,Bunker}, $z_{850}$ \citep{BouwensNature,Bouwens08} and 
$J_{110}$ \citep{Bouwens05,Bouwens08} bands, corresponding to effective source redshifts of z$\simeq$6, 
7.5 and 10. These studies found a highly uncertain number density of candidates, none of which has been confirmed spectroscopically at $z>7$. However, taken at face value, the overall conclusion from these 
ultradeep images is that the declining abundance of luminous star forming sources beyond $z\simeq$7 
is insufficient to account for reionization. Although there is no guarantee that star-forming sources did 
reionize the Universe at $z\simeq$10, a possible solution is that the bulk of the early star formation resides 
in an undetected population of intrinsically sub-luminous sources \citep{Stark07a}.

Prior to the availability of the next generation of telescopes, gravitational lensing is an effective means 
to evaluate this hypothesis. Depending on the method, foreground massive clusters can provide a 
magnification boost of $\times5-30$ in flux (for unresolved sources) or in size (for resolved sources).  
An analysis by \citet{Stark07c} concluded that lensing surveys should
be able to find ample candidates at $z\gsim 7$, permitting follow-up
spectroscopy and detailed studies at sensitivity limits that would be
unachievable otherwise. As pointed out by \citet{Broadhurst}, this gain is offset by a reduction 
in the sky area surveyed, producing an overall increase or decrease in the number of 
lensed sources, depending on the slope of the luminosity function.

A first attempt at constraining the abundance of lensed drop-out galaxies at $6\lesssim z \lesssim 10$ 
was made by \citet{Richard06} using deep ISAAC near-infrared images obtained at the ESO VLT. A 
number of faint (intrinsic $H (AB) \sim 26$) candidates were identified in two clusters, where the 
{\it magnification factor} $\mu$ ranged from 0.4 to 2.5 magnitudes. This analysis suggested a star formation 
rate density at  $z\simeq$7.5  $\sim\times 10$ higher than that derived by \citet{Bouwens04}. Deeper 
NICMOS images have failed to confirm some of these sources.  The number densities of faint candidates are currently
being revised using follow-up imaging and spectroscopy for a small fraction of the ISAAC field of view 
(Richard et al., in preparation). The deeper NICMOS imaging undertaken in the present study illustrates 
the difficulties in making progress beyond $z\simeq$7 using ground-based facilities.

In a parallel effort, \citet{Stark07b} concentrated on the much smaller regions of very high magnification 
($\mu>3$ mags) termed the `critical lines' of the lensing clusters. Using NIRSPEC on the Keck telescope 
they undertook a `blind' spectroscopic search for lensed Lyman-$\alpha$ emitters in the redshift range
$8.5<z<10.4$.  Despite the very small volumes probed in this unique  survey, 6 faint candidates emerged 
across 9 clusters. 
Exhaustive follow-up imaging and spectroscopy has, so far, been unable to provide 
unambiguous confirmation of the nature of these sources. 
Taking into account the uncertainties, \citealt{Stark07b} concluded that the 
abundance of low-luminosity emitters in this 
redshift window may exceed 0.2 Mpc$^{-3}$, suggestive of a major contribution of low luminosity star-forming 
galaxies to cosmic reionization. 

The caveats concerning this conclusion were discussed in detail by Stark et al. Although representing 
a unique search for early star-forming sources at limits well beyond those probed otherwise, the volumes 
addressed are modest and significantly affected by cosmic variance. The 6 candidates were found in only 
3 of the 9 lensing clusters; 6 clusters had no convincing candidates. Although Stark et al.\ were unable 
to prove, unambiguously, that the detected emission is due to Lyman $\alpha$, the null detection of 
associated lines was used as an argument for rejecting lower redshift emission for most of their
candidates.. Their conclusion that the bulk of the reionizing photons arose from low luminosity 
($\simeq0.1\,M_{\odot}$ yr$^{-1}$) star-forming sources can be verified by this independent search 
for lensed {\it continuum} drop-outs at $z>$7.

In searching for faint lensed drop-outs, the advantage of HST over a ground-based survey such as that
undertaken by \citet{Richard06} is considerable. The ACS and NICMOS cameras are much more sensitive, 
not only because of the reduced background level relative to that produced by the atmosphere and by ambient temperature optics, but also because the typical sources have 
angular sizes of 0.2\arcsec\ or less \citep{Ellis}. With similar exposure times, NICMOS can readily attain a 
depth of 26.5 AB mag, $\sim1$ magnitude deeper than the earlier VLT/ISAAC project. Viewed through
a typical $z\sim 0.2$  cluster, the NICMOS field (0.8 $\times$ 0.8 arcmin) closely matches the area of 
moderate to high magnification factors ($\mu\sim1-4$ mag). 
The effective increase in sensitivity to faint sources provided by the lensing magnification along the 
line-of-sight to each NICMOS field probes limits fainter than those in the UDF \citep{Bouwens08}, 
albeit over a considerably reduced area. 

Early studies of lensed drop-outs with HST have served to illustrate the potential.  \citet{Kneib} located a
source at $z\simeq$6.8 behind the cluster Abell 2218. This source forms a triply-imaged system with 
two bright elongated arcs, easily recognized as morphologically-similar in ACS and NICMOS images. 
Follow-up observations with Spitzer \citep{Egami} provided improved constraints on the photometric
redshift, stellar mass ($M\sim 10^9$ M$_\odot$) and past star formation history.  Very recently, a
similar $z\simeq$7.6 lensed source was found by \citet{Bradley} but this was not multiply-imaged. 
A semi-analytic analysis, empirically calibrated using the luminosity functions of Lyman-$\alpha$ emitters 
and drop-out galaxies at $z\sim5-6$ \citep{Stark07c} predicts that such a NICMOS/ACS lensed imaging 
survey should typically detect 0.5-1 sources per cluster in the redshift range 7.0$<z<$8.5.

The present program represents the logical next step: a concerted effort to verify the hypothesis 
advocated by Stark et al.\  as well as its associated predictions via deep imaging of 6 lensing clusters 
with HST (GO 10504: PI: Ellis) and Spitzer (GO-20439, PI: E.Egami). The primary goal is to determine 
the abundance of intrinsically sub-luminous $z$ and $J_{110}$ drop-outs, and to derive 
constraints on the possible contribution of  low luminosity sources to cosmic reionization, 
independently of Lyman-$\alpha$ searches in blind or narrow-band surveys. 

The paper is organized as follows. In $\S2$ we present the new HST/Spitzer and associated ground-based 
imaging observations and their data reduction. A discussion of the means of selecting the drop-out candidates
is presented alongside a catalog in $\S3$. Issues of completeness and contamination by spurious and
foreground sources are discussed. We conclude that a significant fraction of our candidates are possibly 
at high redshift. We then describe a Keck spectroscopic follow-up campaign in $\S4$ which attempts
to detect Lyman $\alpha$ emission from some of the most promising candidates. In $\S5$ we discuss
the UV luminosity function at $z\simeq$7.5 and review the implications in the context of the possible
contribution of low luminosity to cosmic reionization. Our conclusions are summarized in $\S6$. 

All magnitudes given in this paper are standardized within the AB magnitude system \citep{AB}. We assume 
a  flat universe and ($\Omega_m$, $\Omega_\Lambda$, $h$)=(0.3, 0.7, 0.7)  whenever necessary. 

\section{Survey Strategy, Observations and Data Reduction}

We begin by discussing the selection of lensing clusters, the imaging datasets we have secured to select
the various drop-out candidates and validate their high redshift nature, and the image processing
steps taken to produce photometric catalogs. In addition to the HST and Spitzer datasets which form
the fulcrum of this study, we have added ground-based imaging in both the optical and $K$ band.
In general terms, the HST data acts as the primary source of drop-out candidates and the associated
data is used to constrain the likely redshift distribution.

\subsection{Lensing Cluster Sample}

Our primary criterion in selecting foreground clusters for this lensing survey is the value of the
magnification factor expected for sources beyond $z\simeq$7 and the uncertainty implied in this
estimate based on an understanding of the mass model.  We considered a number of massive 
clusters at $z\simeq 0.1-0.5$ with well-understood mass models capable of producing regions of 
strong magnification which match the imaging area of the HST cameras.  Even though the area
enclosed by the critical line for a $z\simeq$7 source is sufficient for the design of the program, 
a precise mass model is a clear advantage in determining the magnification of a particular source, 
as well as for predicting accurately the location of any multiple images.

Our strategy parallels that discussed in some detail by Stark et al. Indeed that Lyman $\alpha$
critical line survey placed greater demands on the reliability of the cluster mass models as
the magnifications are more extreme and positional uncertainties in the image plane are
critical. Notwithstanding this challenge, Stark et al.\ found that the typical magnification uncertainties
for their candidates were only $\simeq$30\% and that errors in the critical line location were 
usually only $\pm$1 arcsec or so. 

In the present survey, six clusters was considered the minimum number necessary to mitigate 
the effects of cosmic variance \citep{Stark07c} while being consistent with the observing time available. 
In considering the final tally of clusters, we included clusters which we have modeled with the latest 
version of the Lenstool\footnote{Publically available, see {\tt http://www.oamp.fr/cosmology/lenstool} 
to download the latest version} software \citep{Jullo}. This improved code provides a new Bayesian 
optimization method to derive error estimates for each optimized parameter. Following the discussion in 
\citet{a68} and \citet{Limousin}, this optimization can be used to compute error estimates on 
the individual magnification factors for each of our sources.

Wherever possible, we included clusters with usefully deep archival HST and Spitzer/IRAC data. Deep 
optical (AB$>$27.0) ACS and/or WFPC2 images, previously used for the identification of 
multiple images during the development of the mass models, allow low redshift 
contaminants to be identified reliably. IRAC images at 3.6 and 4.5 $\mu m$, to the same 1 $\mu$Jy sensitivity as our 
previous work in Abell 2218 \citet{Egami}, are available  as part of a Spitzer Lensing Cluster Survey 
(GTO 83, PI: G. Rieke) or publicly from the archive (GTO 64, PI: G.Fazio).  Finally, we required that 
each of the selected clusters be visible from the northern hemisphere, in order to facilitate 
spectroscopic follow-up and further $K$-band imaging with the Keck and Subaru telescopes.

The six clusters satisfying the above criteria are presented in Table \ref{cluster}. Although the references
cite the most recently-published mass models, as described above, in each case we have utilized
the available multiple images and their redshifts in improving these mass models using Lenstool.
Four out of the six clusters are in common with the sample adopted by Stark et al.

\subsection{Hubble Space Telescope Data and Reduction}

\begin{figure*}
\centerline{\mbox{\includegraphics[width=12cm]{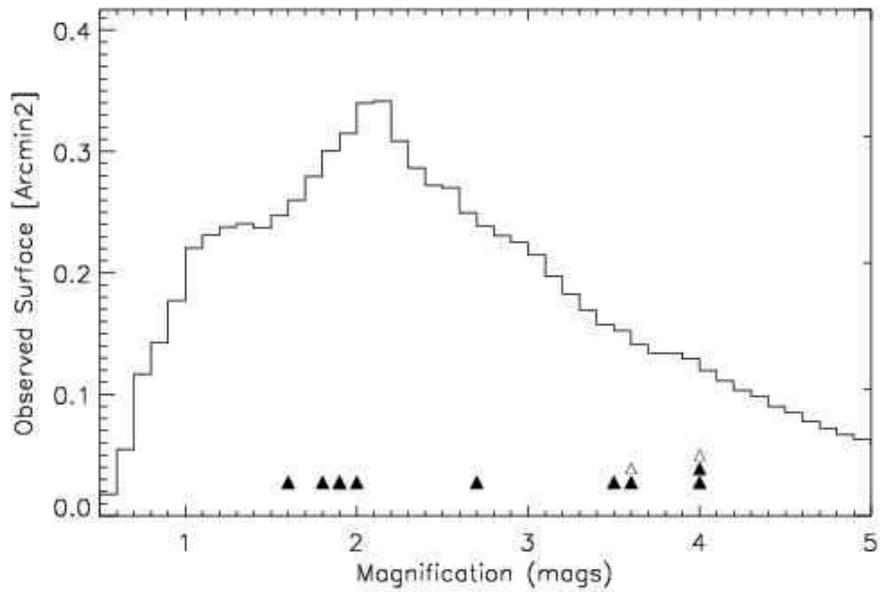}}}
\caption{\label{hmag}
Distribution of magnification factors $\mu$ (in magnitudes) for the survey, 
as predicted from the mass models assuming a point source at very high redshift ($z>7$). 
The vertical axis represents the observed surface area in each 0.1 mag (0.04 dex) magnification bin.
The peak of this distribution indicates our typical magnification factor is $\sim 2.0$ mags. 
Filled triangles mark the individual magnification factors for the 10 $z$ drop-out and 
open triangles that of the $J$ drop-out candidates. 
}
\end{figure*}

Our large program with HST (GO 10504, PI: Ellis) comprised deep observations in the $z_{850}$, 
$J_{100}$ and $H_{160}$ bands, using the Wide Field Camera of ACS and the NIC3 configuration
of NICMOS.  The region enclosed by the critical line for the putative $z>$7 sources was typically covered 
by two NICMOS pointings per cluster, usually adjacent. This ensured a magnification gain of
$\mu\sim 1$ to 4 magnitudes, with a typical value $\mu\sim2$ mags, throughout the NICMOS imaged 
field (Fig. \ref{hmag}). The total sky area covered by the NICMOS observations is 8.9 arcmin$^2$ for the six
clusters.

The ACS images were reduced with the {\tt multidrizzle} software \citep{multidrizzle}. This removes 
cosmic rays and bad pixels and combines the dithered frames  to correct for camera distortion. The output 
pixel scale was fixed at 0.04\arcsec and we used a {\tt pixfrac} value of 0.8 for reducing the area of the 
input pixels. We made small corrections to the absolute astrometry to allow for ACS frames taken at different 
epochs (e.g. Abell 2218 in the F850LP band, see Table \ref{clusterimages}). These corrections were 
computed by correlating the catalog positions of bright objects detected in the overlapping regions.

Each single set of NICMOS observations consists of 8 (in F110W) and 10 (F160W) frames of $\sim$ 
1000 secs, taken with the NIC3 camera using the SPARS64 or SPARS128 sampling sequences. A basic 
reduction was performed by adopting the procedures given in the NICMOS data reduction 
handbook\footnote{\tt http://www.stsci.edu/hst/nicmos/documents/handbooks/handbooks/DataHandbookv7/}. 
Starting with the post-calibrated frames, bad pixels are flagged and rejected based on individual histograms, 
cosmic ray are rejected using the {\tt LACOSMIC} \citep{lacosmic} IRAF procedures, frame-to-frame shifts 
are measured using a cross correlation technique, and all frames are drizzled onto a NICMOS pixel scale 
(0.2\arcsec) to produce an initial reduced image. This then serves as a comparison for each individual
frame so that deviant pixels can be flagged to improve image quality in a second drizzling operation, 
this time undertaken with an image pixel of 0.1\arcsec, to obtain a better sampling. 

These initial images reveal a number of cosmetic effects (bias and flat residuals, bad columns and 
bad pixels, quadrant-to-quadrant variations, background variations) that led to a second stage of image 
reduction. We used an improved pixel mask to flag several bad columns and bad pixels close to the 
frame edges, and we examined each frame individually in order to remove bias and flat residuals and 
quadrant-to-quadrant variations. Finally, we subtracted a smoothed background obtained by averaging 
the frames of all the observed clusters, masking every pixel lying on a physical object. In order to combine 
all NICMOS observations of a given cluster, usually taken at different epochs and with slight variations in 
the sky orientation, the individually-reduced  images were registered onto the wider field ACS data prior
to the final drizzling procedure and combination into a NICMOS mosaiced image.

The NICMOS data acts as the primary basis for selecting our drop-out candidates in association with
non-detections in very deep optical data. In addition to our own ACS data, we reduced deep ACS and 
WFPC2 images from the archive for each cluster in our sample (Table \ref{clusterimages}). As with the 
ACS F850LP data undertaken in our own program, we reduced these data using the IRAF procedures 
{\tt multidrizzle} and {\tt drizzle} as discussed above.

\subsection{Ground-based Data and Reduction}

As mentioned, optical non-detection to deep limits is a key necessity in considering the validity of our
drop-out candidates. Most clusters have complete optical coverage with HST (from $\sim400$ to 
$\sim 850$nm ) useful for this purpose. However,  Abell 2219 has F702W only, and Abell 2390 has
F555W and F814W only. For these two clusters, additional deep ($R\sim26.7$ and $I\sim25.6$ at 5$\sigma$ 
in 1.2\arcsec diameter apertures) ground-based images taken with the CFH12k camera on CFHT \citep{Bardeau} 
were also examined to check for non-detections at the locations of the drop-outs.

Ground-based photometry in the $K$ band can likewise provide additional information for $z$ drop-outs, 
improving the photometric redshift estimates,  and for the reliability of $J$ drop-outs where otherwise 
only a single color would be available (see Section \ref{dropouts}). Although a challenging undertaking
given the depth of the ACS and NICMOS data, we conducted various $K$-band imaging observations 
of clusters using the Keck and Subaru telescopes.

The Near Infrared Camera (NIRC) on Keck-I was used on July 2006 to observe the central regions of 
the clusters Abell 2218 and Abell 2219.  The square field of NIRC (38\arcsec a side) is slightly smaller 
than that of NICMOS therefore we concentrated these observations on our best $z$ and $J$-band 
drop-outs in each cluster (see section \ref{dropouts}). The seeing was stable and in the range 
0.5-0.6\arcsec. We used dithered exposures of 6 coadds $\times$ 10 secs exposure time each. 
The NIRC pointing in Abell 2218 was partially covered, in the region of one of the drop-outs, by previous 
NIRC observations obtained in 22/23 July 2002 (A. Blain \& N. Reddy 2003, private communication).\\

The Multi-Object InfraRed Camera and Spectrograph (MOIRCS) \citep{moircs} at Subaru was also used 
during two observing campaigns, in August 2006 and May 2007. The larger field of view (7' $\times$ 4') 
of this camera ensures a complete coverage of the NICMOS, ACS and IRAC data in each cluster. The four clusters in common with the Stark et al.\ sample (Abell 2218, 2219, 2390 and CL0024) were imaged for 
$\sim 5$ hours each under very good seeing conditions (0.3-0.4\arcsec) using dithered exposures of 
50 secs duration.  We used the MCSRED software package\footnote{available from 
http://www.naoj.org/staff/ichi/MCSRED/mcsred.html} to perform the flat-fielding, sky-subtraction, 
distortion correction and mosaicing of individual images. The MCSRED package also includes 
MOIRCS-specific tasks which correct for quadrant shifts and to  fit sky residuals. The depth reached by 
the MOIRCS imaging data ($K_{AB}\sim26.1$) in Abell 2218 and ABell 2219 is similar or deeper 
than the NIRC observations of the same fields. Therefore, we used the NIRC images as independent 
check for the MOIRCS photometric measurements, performed under better seeing conditions.

\subsection{Spitzer Data and Reduction}

%Each cluster has been observed in all four bands of Spitzer/IRAC, but for practical reasons we concentrate 
%here on the reduction and analysis of only the first two channels (3.6 and 4.5 $\mu$m) where the depth
%achieved is potentially useful in comparison with our  ACS and NICMOS data. A single set of IRAC 
%observations consists of individual exposures of 100 to 200 seconds each, usually observed during two 
%different epochs, delivering a total integration time of 40 minutes in the majority of the cases 
%(GTO 83 and 64, PI: Rieke). Abell 2218 was the subject of a deeper campaign (GO 20439, PI: Egami) 
%and was observed for 10 hours per band, split into 4 epochs (see Table \ref{clusterimages}), and  
%the corresponding depth (25.5 AB) in this particular cluster is much closer to the magnitude of our 
%high redshift candidates.

%In reducing the IRAC data, we used the basic calibrated data (BCD) of each individual frame and 
%combined them using a custom IDL mosaicing routine, following the procedures described in 
%\citet{Egami}. The final pixel size of the IRAC mosaics is 0.6 \arcsec/pixel, half that of the original 
%data.

Each cluster has been observed at 3.6, 4.5, 5.8, and 8.0 $\mu$m using
the Infrared Array Camera (IRAC; \citet{Fazio}) on board the {\em
Spitzer} Space Telescope \citep{Werner04}.  In this paper, we only
discuss the first two channels (3.6 and 4.5 $\mu$m) where the depth
achieved is potentially useful in comparison with our ACS and NICMOS
data.  Each of the IRAC channels uses a separate detector array, and
the 3.6 $\mu$m channel ($\lambda_{c}=3.56 \mu$m; $\Delta\lambda = 0.75
\mu$m) and 4.5 $\mu$m channel ($\lambda_{c}=4.52 \mu$m; $\Delta\lambda
= 1.01 \mu$m) use $256 \times 256$ InSb arrays with a pixel scale of
1\farcs2 pixel$^{-1}$, producing a field of view of
5\farcm2$\times$5\farcm2.  A frame time of 200 seconds was used with
the small-step cycling dither pattern initially, but this was later
changed to 100 seconds with the medium-step cycling pattern for a
better removal of cosmic rays and other artifacts.

Most of the IRAC data come from the Spitzer GTO program PID:83 (PI:
G.~Rieke).  The total integration time is 2400~s per channel, usually
obtained at two different epochs.  CL0024 was observed as part of
another GTO program PID:64 (PI: G.~Fazio) with the integration times
of 2400~s at 3.6 $\mu$m and 3600~s at 4.5 $\mu$m.  Abell 2218 was the
subject of a deeper GO campaign (PID:20439, PI: Egami), and was
observed for $\sim$10 hours per channel, split into 6 separate
observations (AORs).  When these GO data are combined with those of
the GTO program (PID:83), the resultant total integration time is
37700 seconds per channel.  The corresponding depth (25.5 AB) in this
particular cluster is much closer to the magnitude of our high
redshift candidates.  (see Table \ref{clusterimages} for a summary of
the data).

In producing the IRAC images, we started with the basic calibrated
data (BCD) of each individual frame produced by the SSC pipeline, and
combined them using a custom IDL mosaicing routine as presented in
\citet{Egami}.  The final pixel size of the IRAC mosaics is 0.6
\arcsec/pixel, half that of the original data.  A conservative
estimate for the absolute calibration uncertainty is 10\%.

\subsection{Foreground Subtraction}

\begin{figure}[ht]
\centerline{\mbox{\includegraphics[width=4cm]{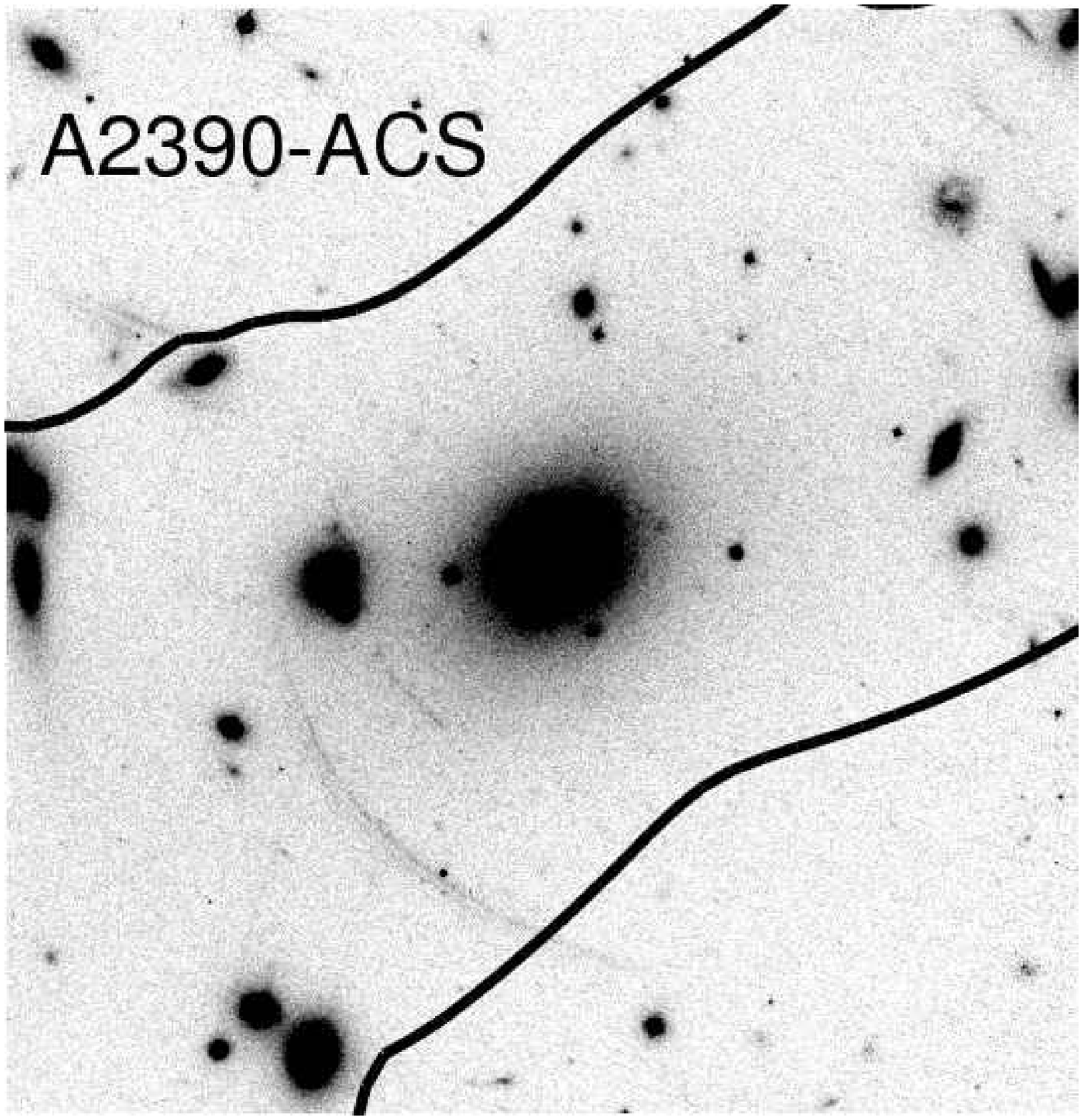}}\hspace{0.2cm}\mbox{\includegraphics[width=4cm]{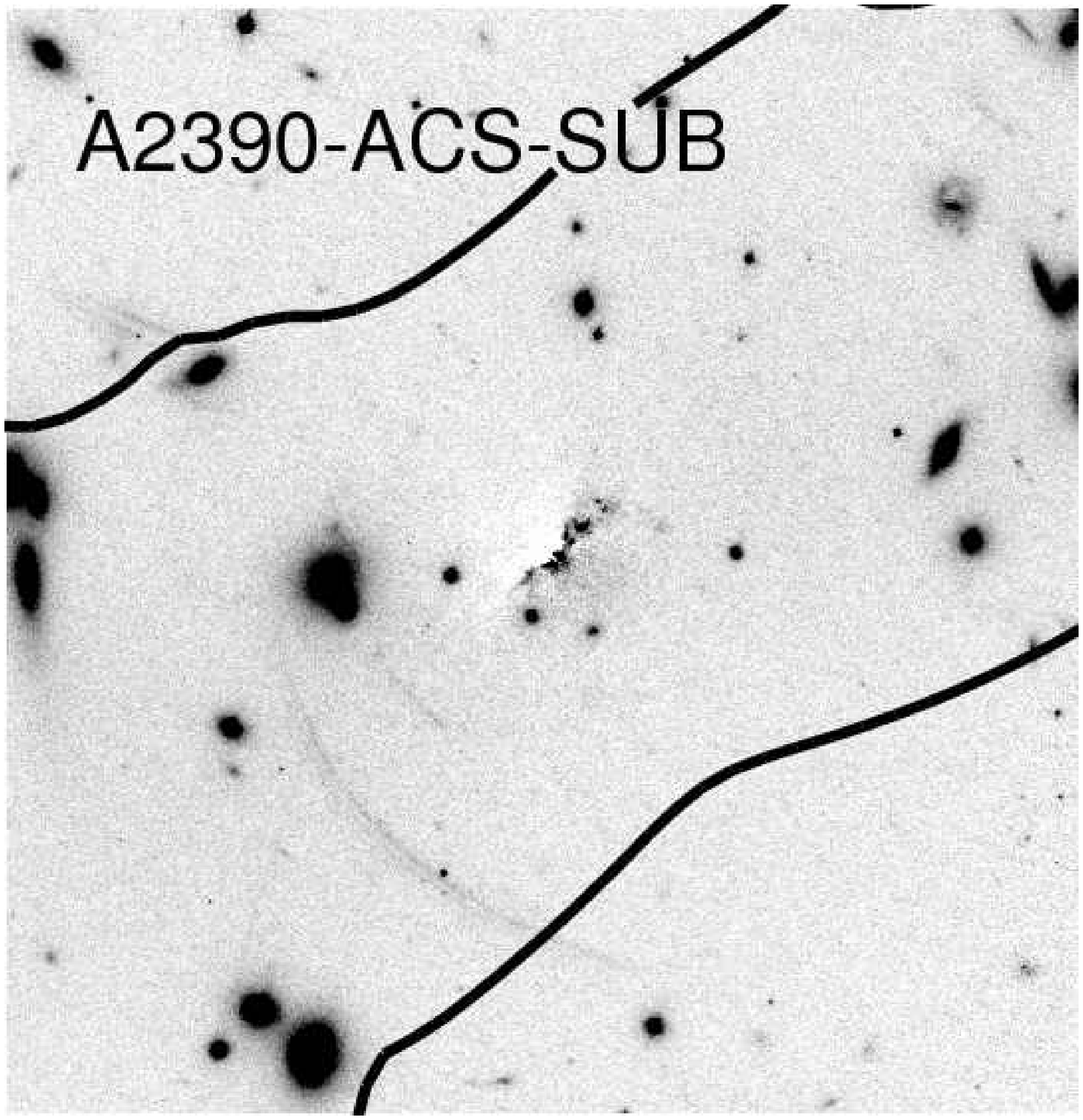}}\hspace{0.2cm}\mbox{\includegraphics[width=4cm]{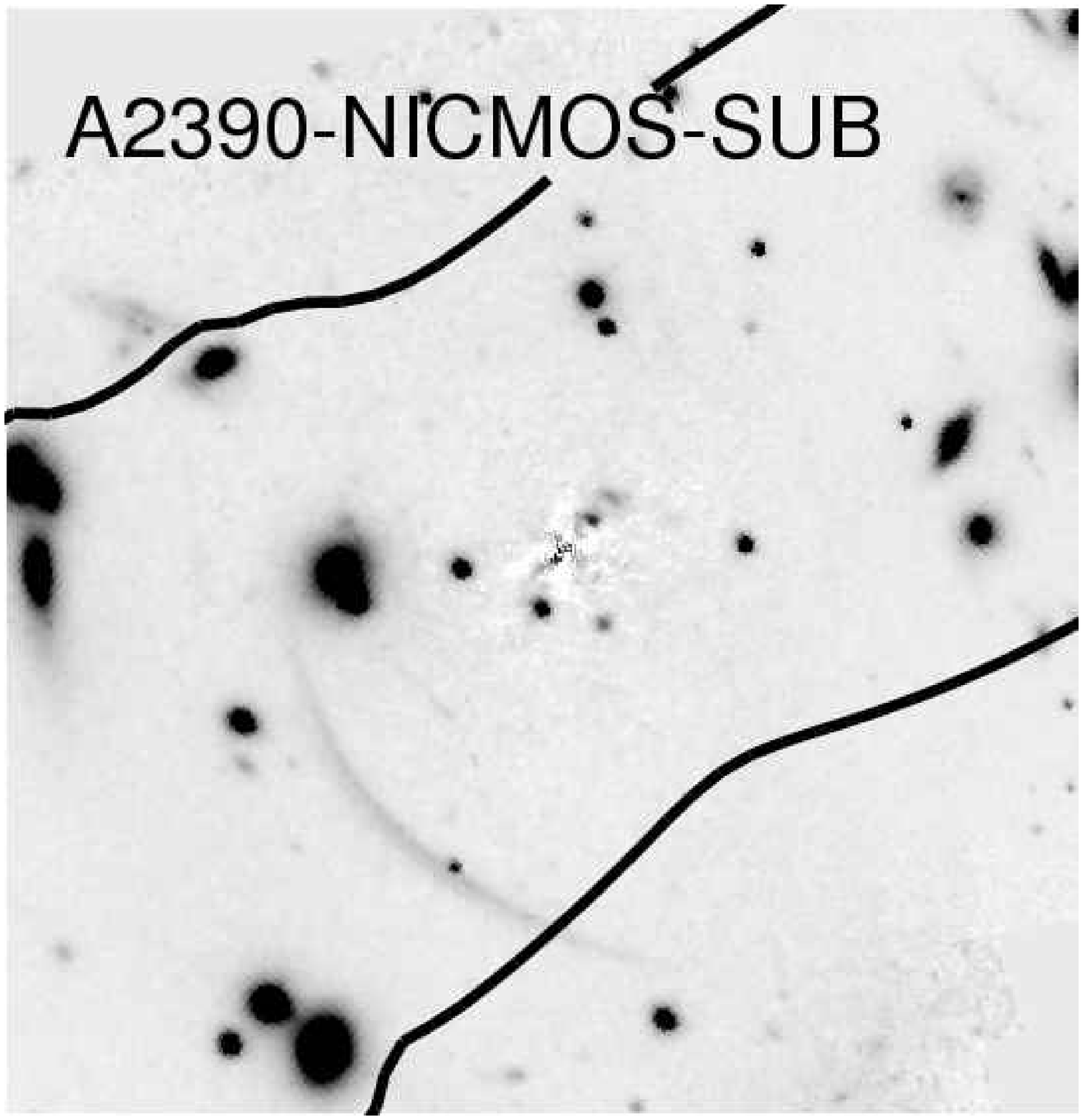}}\hspace{0.2cm}\mbox{\includegraphics[width=4cm]{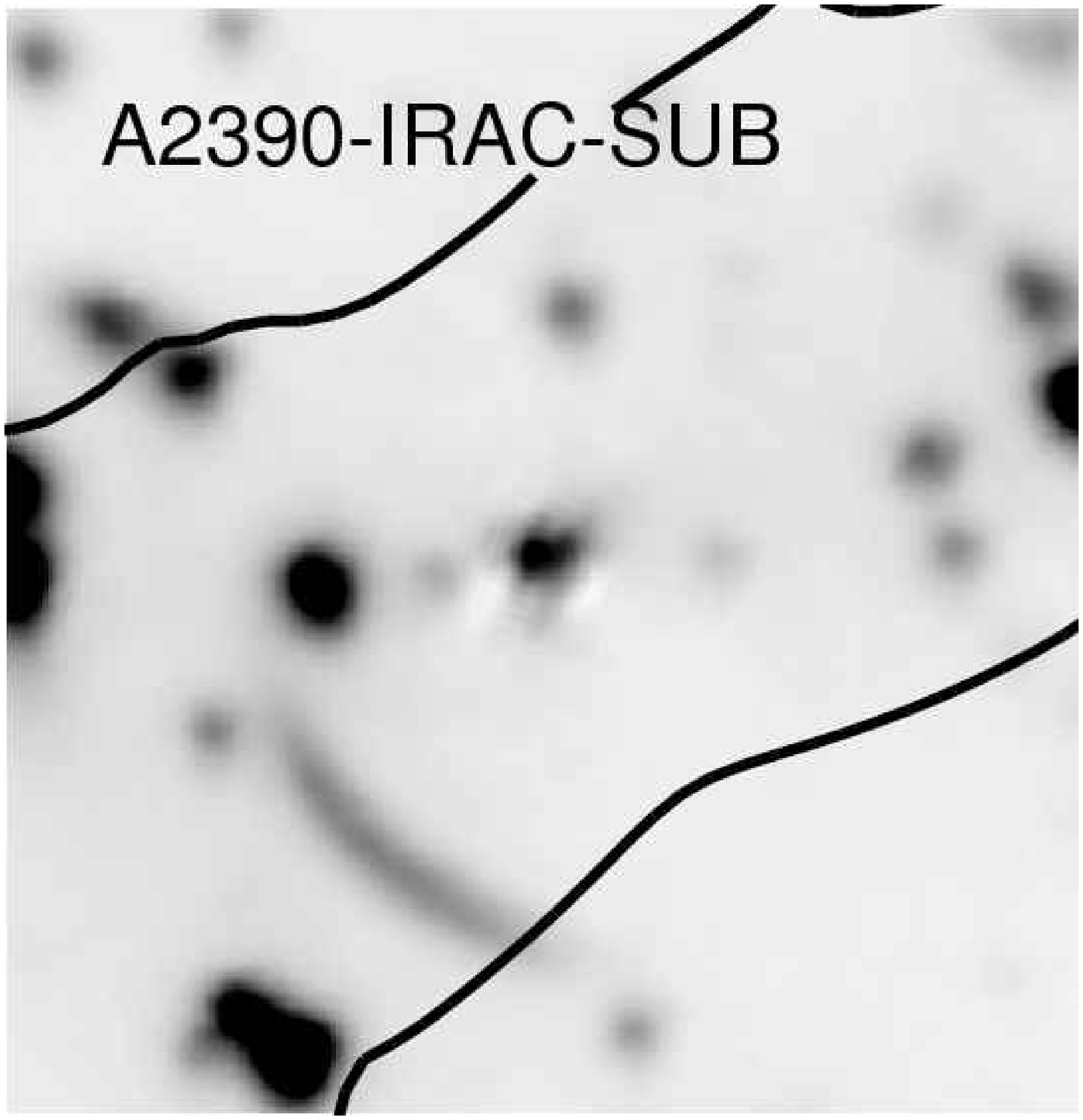}}}
\centerline{\mbox{\includegraphics[width=4cm]{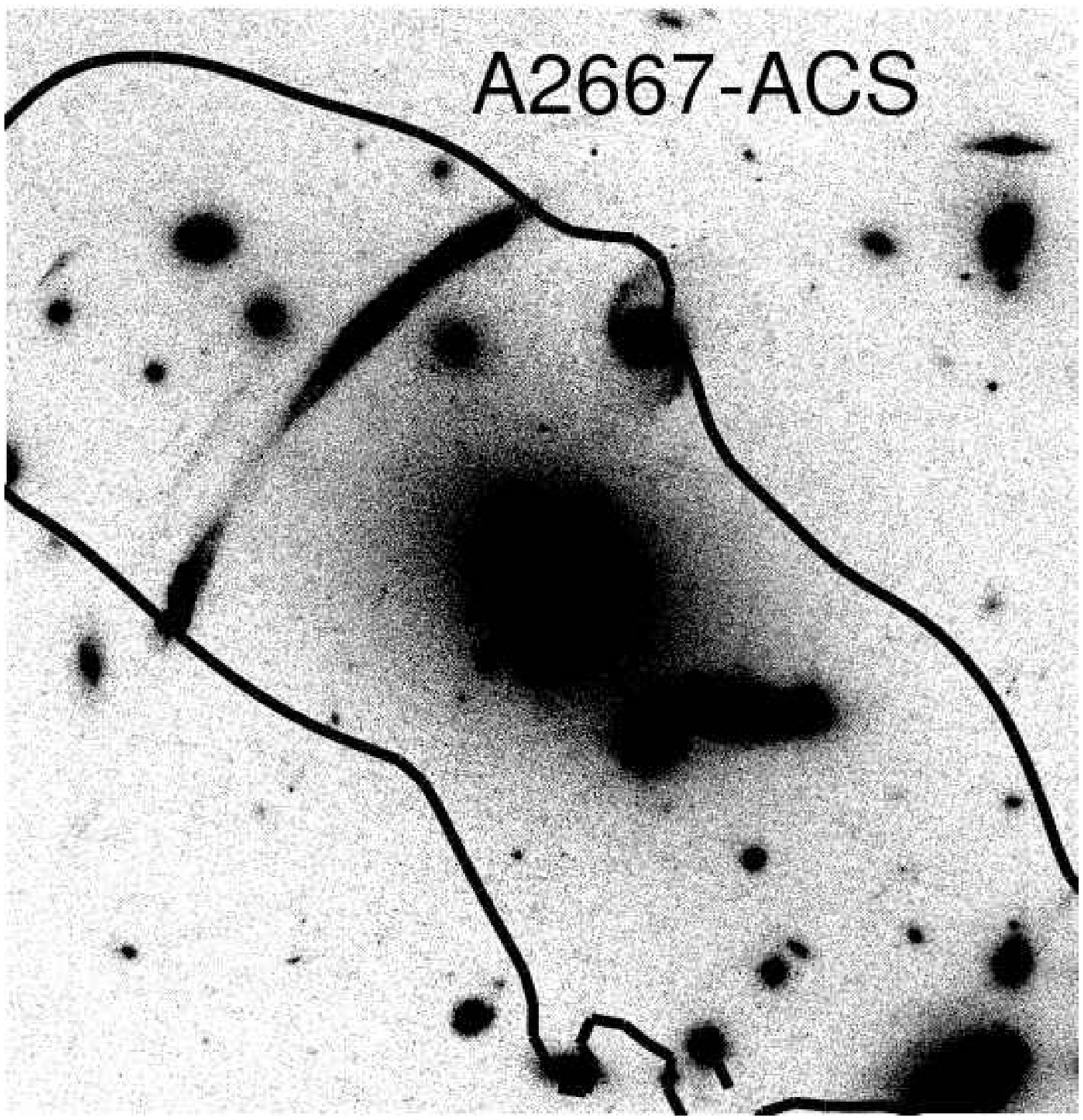}}\hspace{0.2cm}\mbox{\includegraphics[width=4cm]{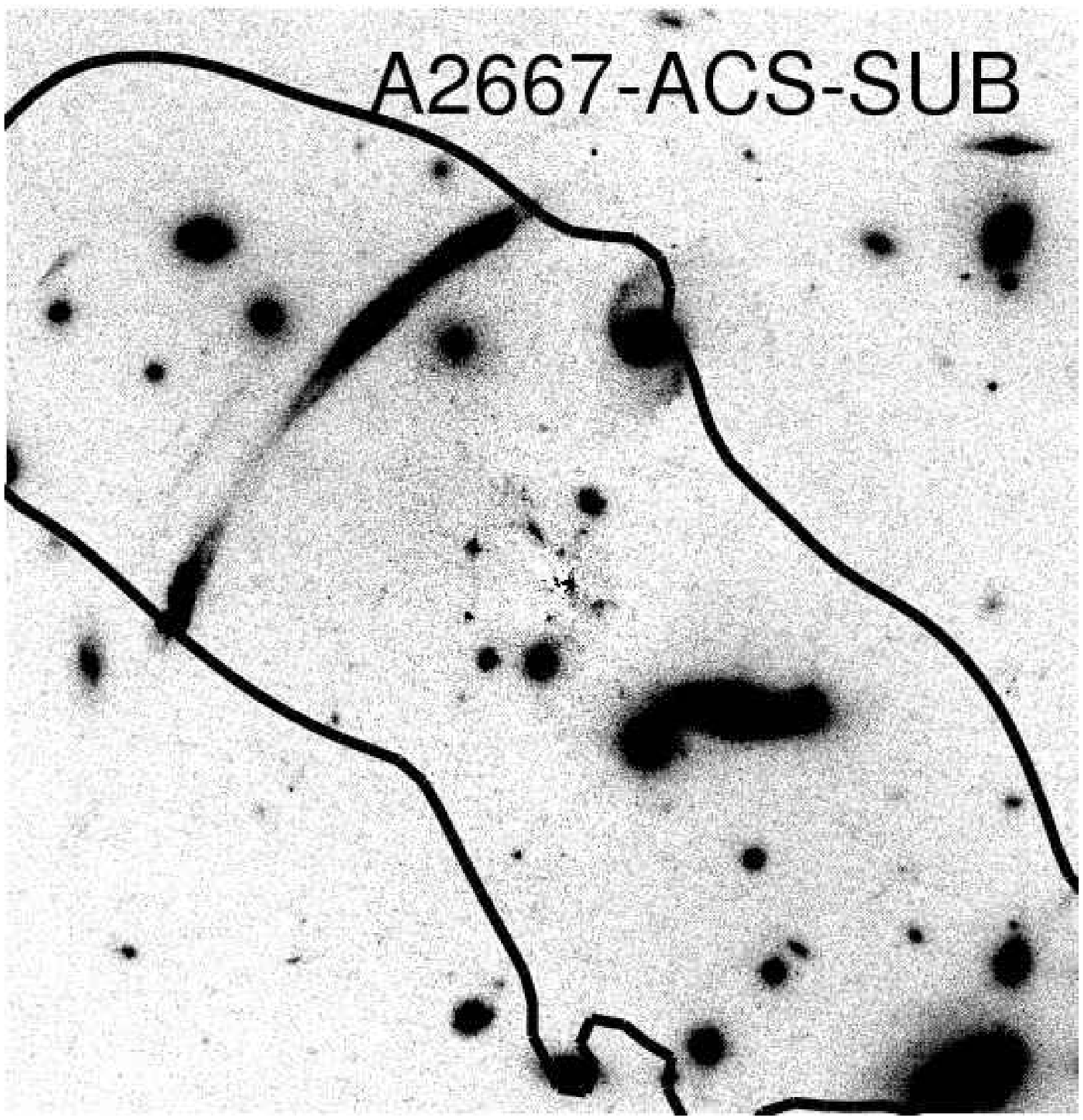}}\hspace{0.2cm}\mbox{\includegraphics[width=4cm]{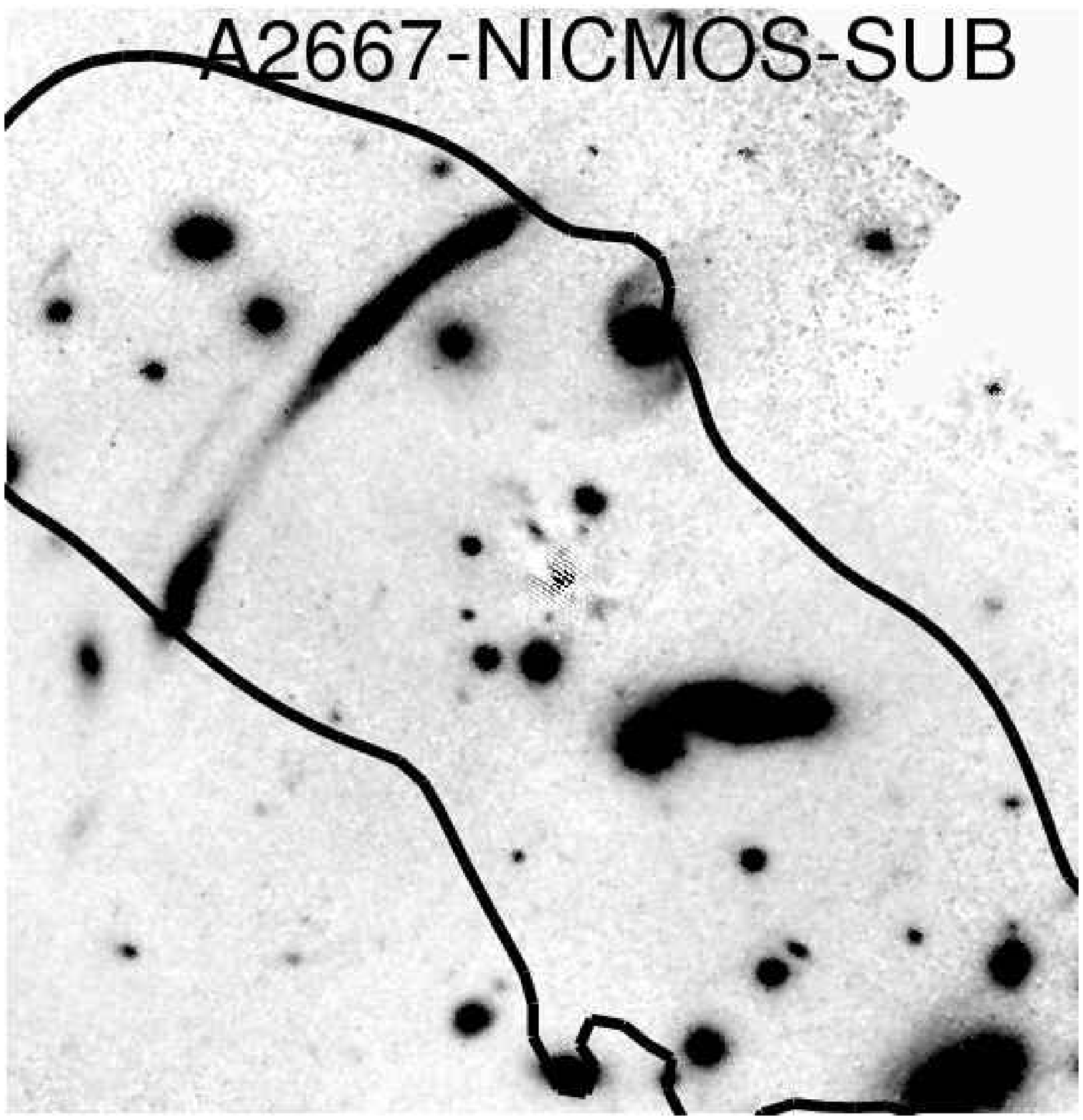}}\hspace{0.2cm}\mbox{\includegraphics[width=4cm]{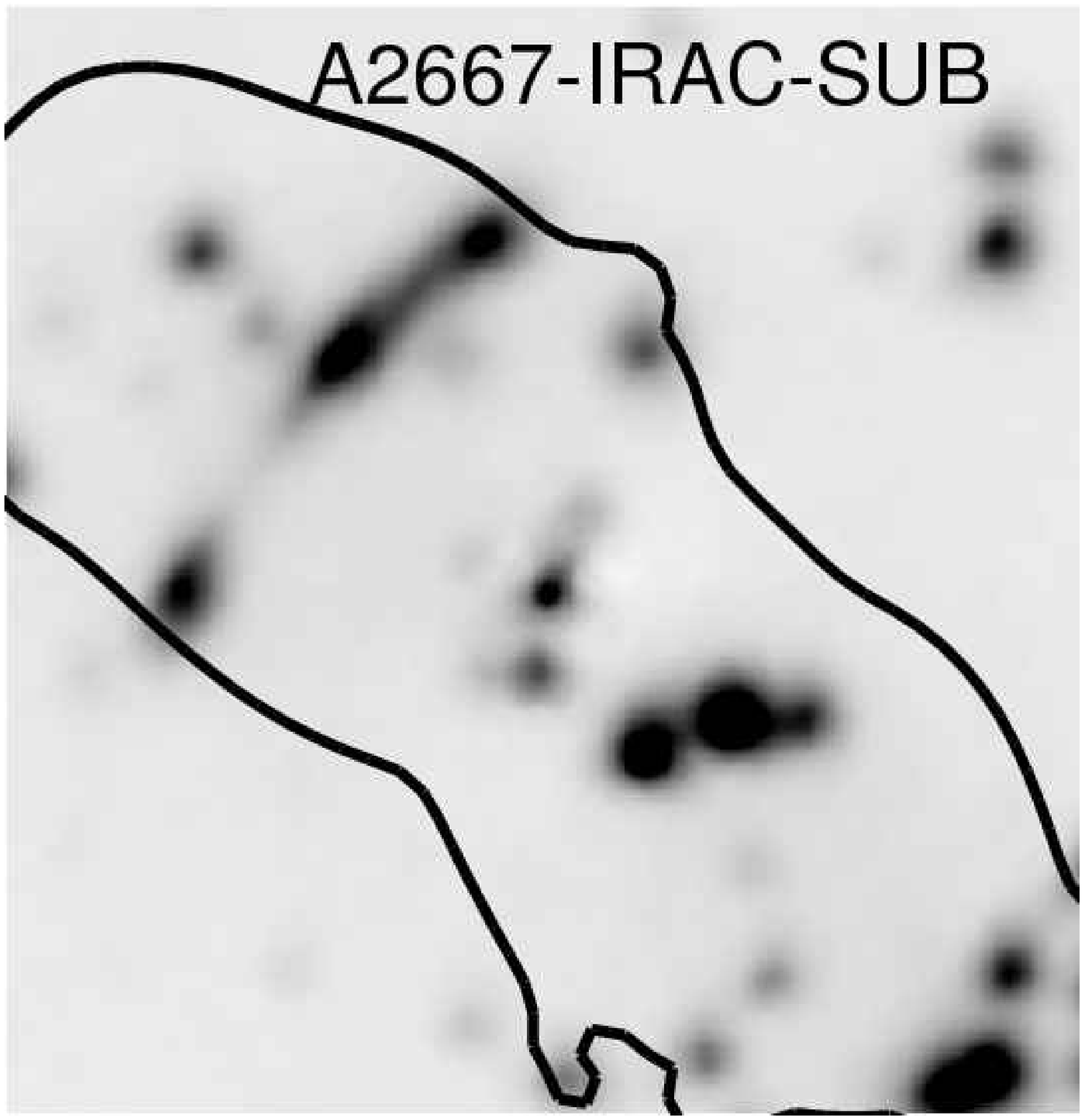}}}
\caption{\label{bcgsubtract} Example of BCG subtraction in the clusters A2390 and A2667. Each panel is 
45\arcsec on a side. (Left to right): ACS image without subtraction, BCG-subtracted images in ACS, NICMOS 
and IRAC respectively. Subtraction residuals appear within $\sim1$ arcsec of the center in the ACS and 
NICMOS case, $\sim2$ arcsec for IRAC. The critical line for a high redshift source is show by the black curve.
}
\end{figure}

Although the magnification afforded by lensing clusters offers a unique gain in probing the distant
universe, the central regions are dominated by bright, extended spheroidal galaxies which obstruct
and whose light increases the background level. More importantly, it also affects the photometry 
and color measurements of any underlying fainter source. Fortunately, the majority of these galaxies 
are also good light deflectors, their morphology is usually regular and their light distribution can be 
accurately modeled by a sequence of elliptical isophotes. Since the 
major contributor to this deficiency is the brightest cluster galaxy (BCG), we have modeled and 
subtracted its light for each cluster in the ACS ($I_{850}$ band) and NICMOS images using the 
IRAF task {\tt ellipse}. Apart from a small region ($\sim 1.0\arcsec$ radius) around the core 
of the removed BCG, this frees the image from the majority of the contamination in highly magnified 
regions close to the critical line, and aids in the detection of underlying sources (see Figure 
\ref{bcgsubtract}).\\

Foreground subtraction is more challenging for the IRAC images because of the much coarser 
PSF. The smaller number of available pixels per galaxy prevents the use of the {\tt ellipse} modeling 
technique discussed above.  Instead we adopted the {\tt ellipse} model of the BCG derived from the
NICMOS data and, on the assumption that the spectral energy distribution is identical across
the galaxy at these long wavelengths, convolved this model with the IRAC PSF derived by 
stacking $\sim 50$ bright isolated point sources. This model of the BCG was subtracted after an 
appropriate scaling factor. We found this procedure to be very effective, with residuals from the 
subtraction confined within a 2.0\arcsec radius region (Figure \ref{bcgsubtract}).

\subsection{Final Photometry}

\label{quality}

The primary filters we will use for identifying the near-infrared drop-outs are the ACS $z_{850}$
and NICMOS $J_{110}$, $H_{160}$ bands available for each cluster. The bulk of the rest-frame 
UV flux is contained in the two reddest filters at high redshift, so we require all sources to be detected 
in both. The signal to noise ratio is improved by combining the $J$ and $H$ images, once normalized 
to a similar noise level, to form a single $<J+H>$ detection image. This provides a more accurate 
measurement of the centroid and geometrical parameters of these objects,

We use the ``double-image'' mode of the SExtractor package \citet{sex} to detect objects and compute 
magnitudes within a 0.6\arcsec (NICMOS images) or 0.3\arcsec (ACS images) diameter aperture. Corrections 
to total magnitudes, assuming a point source, were estimated using bright isolated unsaturated stars. The 
corresponding values are 0.3 and 0.6 magnitudes for ACS and NICMOS, respectively.

The drizzling procedure used in the HST reductions, while conserving flux, does introduce correlations 
between neighboring pixels and hence unreliable error estimates \citep{casertano}. We applied their 
equation $[$A20$]$ to the SExtractor photometric errors.  Dithered exposures also introduce a varying 
effective exposure time across the mosaic, this effect is clearly apparent close to the boundaries of 
the field. We constructed a corresponding weight map, proportional to the effective integration time 
at each pixel position in the detection image, and used it as an input parameter of SExtractor. 
This ensures source detection at a uniform noise level across each image.

The background noise level $\sigma$ was measured for the ACS and NICMOS images in order to 
estimate the achievable photometric limits in each band. The 5$\sigma$ values 
are reported in Table \ref{clusterimages} as total magnitudes. 
We computed the completeness limits in each NICMOS filter used for source detection by adding
artificial unresolved sources in the magnitude range 23-28 AB. Such sources were added 1000 times 
at 30 different random locations on the image and then extracted using the same photometric 
techniques as applied bf to the science image. In this manner we compute the point source completeness 
as a function of source magnitude. Only blank regions were chosen for this exercise, defined 
after application of a 5$\sigma$ threshold to mask pixels associated with bright objects. The 50\% 
completeness values derived in this way are listed for each cluster and band in Table \ref{clusterimages}.
Typically, our NICMOS data is 50\% complete to $J_{110}\simeq$25.9 AB and to 
$H_{160}\simeq$26.05 AB. For each eventual drop-out candidate ($\S$3.3), we will assign a 
completeness factor, $S_{comp}$, based on its magnitude, that will be taken into account in
estimating the high redshift luminosity function. 

Our NICMOS survey reaches limits of typically 26.2 and 26.5 AB in $J$ and $H$, respectively, over 
8.9 arcmin$^2$. For comparison, the UDF limits are 27.8 AB over 7 arcmin$^2$ \citep{Bouwens04}. 
However the magnification provided by the foreground clusters (Fig. \ref{hmag}) enables us to reach unlensed 
source magnitudes (assuming $z\sim7$) of 28-30 AB over a total area of 0.1-1 arcmin$^2$.

\section{High Redshift Candidates}

We now turn to the selection of our high redshift candidates. The primary selection will be based upon 
the $z_{850}-J_{110}$ color for the $z$ drop-outs at redshifts $z\simeq$7-8, and the $J_{110}-H_{160}$ 
color for the $J$-band drop-outs at redshifts $z\simeq$8-10. This section has two components. First
we discuss the optimum color criteria for drop-out selection, the degree of completeness and issues of 
possible contamination from spurious artefacts  at these faint limits. In this way we establish a robust 
set of candidates whose likelihood of being at high redshift  we then assess in the second part
of this section using additional criteria including their photometry at other wavelengths and location
with respect to the cluster mass model.  The catalog of candidates is summarized in $\S$3.3.

\subsection{Photometric Selection and Completeness}
\label{dropouts}

\begin{figure}[ht]
\centerline{\mbox{\includegraphics[height=0.52\textwidth,angle=270]{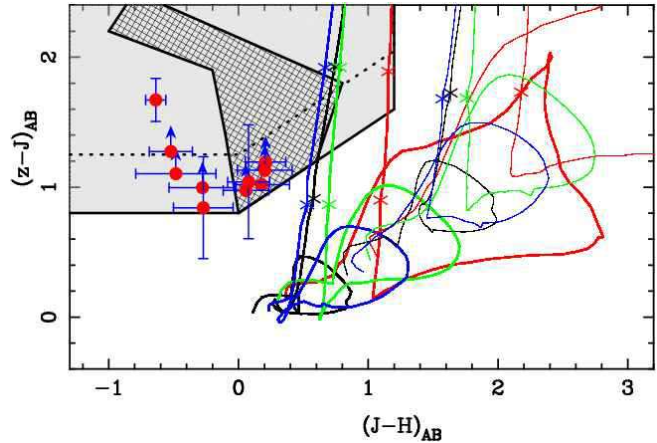}}}
\caption{\label{cz} 
Color-color diagram used for selecting high redshift $z$-band drop-out candidates. The final set of candidates is presented in
red (see $\S$3.3). Color tracks represent predicted colors of Hubble
sequence galaxies (colored tracks) \citep{Coleman,Kinney}. Thick tracks assume no extinction, thin lines
show the effect of including a selective extinction of $A_V=1.0$ magnitudes. 
The observed location of L and T dwarfs is shown as a cross-hatched region \citep{Knapp}. Two possible
$z-J$ color selections ($>0.8$ and $>1.25$) are shown (see text for details).
}
\end{figure}

\begin{figure}[ht]
\centerline{\mbox{\includegraphics[height=0.48\textwidth,angle=270]{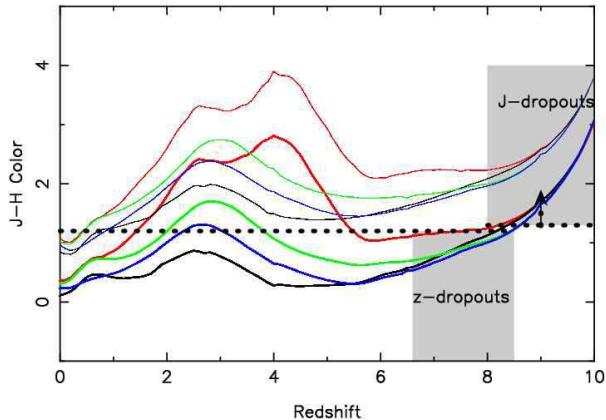}}}
\caption{\label{cz2} Optimizing the selection of high redshift $z$-band and $J$-band drop-out 
candidates using near-infrared colors (dotted lines) as compared to the expected colors of Hubble 
sequence galaxies (as in Fig.\ref{cz}).
}
\end{figure}

The primary concern in selecting high redshift drop-outs from photometric data alone is the issue of
contamination from lower redshift objects, including $z\simeq$2 early-type galaxies, dust-reddened 
objects over a wide redshift range or low mass Galactic stars with deep molecular absorption bands.
Figure \ref{cz} illustrates the problem. A single color-cut fails to isolate $z$-band drop-outs from
a variety of $z\simeq$2-4 sources and the confusion is worse for $J$-band drop-outs (Fig. \ref{cz2}). 

This problem has formed the basis of much discussion in the literature. For the $z$-band drop-outs, 
contamination can be reduced by considering a second color \citep{Stanway04}, of which the most useful with NICMOS data is $J_{110}-H_{160}$
\citep{Bouwens04}. Star-forming galaxies at high $z$ should display a prominent discontinuity in 
$z-J$ while remaining blue in $J-H$, as illustrated in Fig. \ref{cz}. The first issue we address 
is the optimum cut in both colors, on which depend both the redshift range explored and the amount 
of contamination by lower redshift objects.

Figure \ref{cz} shows color-redshift tracks for various galaxy classes (and also includes, for
convenience, the colors of our eventual candidates discussed in $\S$3.3). The location of these 
tracks suggests the following prescription for selecting sources with redshifts 
$6.8 \lesssim z \lesssim 8.0$:

$$(z_{850}-J_{110})>0.8; \  (z_{850}-J_{110})>0.66\ (J_{110}-H_{160})+0.8; \ (J_{110}-H_{160}) <1.2$$ 

The baseline $(z-J)>0.8$ color selection we adopt above is similar to that used by \citet{Bouwens04} and 
\citet{Bouwens08}. However, as the photometric errors for our candidates are typically 0.2-0.3 
magnitudes, the probability of low-redshift contamination remains significant in the range $0.8<(z-J)<1.1$. 
We also explore the use of a more restrictive color cut $(z-J)>1.25$, close to the criterion used by 
\citet{BouwensNature} ($(z-J)>1.3$), also shown on Figure \ref{cz}.

Application of this color selection reveals 10 possible
candidates of which only two satisfy the more rigorous
$(z-J)_{AB} >$1.25 color cut. We discuss the merits of
each of these candidates in more detail in the following
sections. A striking feature is that all have J-H colors
bluer than predicted by the redshifting of local spectral
energy distributions. Similar claims for blue restframe
colors have been made for $i$-band drop-outs (Stanway et al 2004).

Concerning the $J$ drop-outs, corresponding to the redshift range
$8.0\lesssim z \lesssim 10.5$, cuts of $(J-H)>1.8$ and $(J-H)>1.3$ were
adopted respectively by \citet{Bouwens05} and \citet{Bouwens08}.
In their shallower, larger area survey aimed at locating luminous
J drop-out candidates, \citet{Henry} adopted a more restrictive
$(J-H) >2.5$ cut. The large variation in these color cuts reflects the
differing depths of optical exclusion in the various samples.
The sample selected by Henry et al.\ is limited to NICMOS-detected
sources in 2 bands only, with no deep supporting optical observations.
Therefore, despite the apparent stringent color cut in $(J-H)$, it is still
more likely to suffer from contamination by lower redshift sources.
Noting our deep optical data, we adopted a $(J-H)>1.3$ cut, revealing
two candidates (Fig. \ref{cz2}). Neither would satisfy the $(J-H)>$1.8 cut.

We now turn to the important question of how complete our likely drop-out sample will be given 
our adopted magnitude limits with ACS and NICMOS. We can easily imagine that genuine drop-outs
will be missed because photometric errors will scatter points outside our selection region; likewise, 
lower redshift sources will be scattered into our color selection box.

Using the procedures adopted to determine the 50\% magnitude completeness limits in Section \ref{quality},
we can estimate the \textit{color selection completeness} and \textit{color selection contamination} by 
introducing artificial objects with a range of $J$ magnitudes (AB=24 through 27) whose $z-J$ breaks 
have a range (0.6 through 1.2) and determining what fraction of objects which would lie outside of the box
with perfect photometry but are scattered in,  and what fraction of objects lying truly inside are scattered out. 
We assume a flat $f_\nu$ spectrum for the UV continuum between the $J$ and $H$ bands so that, with 
reference to Figure 3, the problem becomes effectively one dimensional. 

\begin{figure}
\begin{minipage}{0.5\textwidth}
\centerline{\mbox{\includegraphics[width=\textwidth]{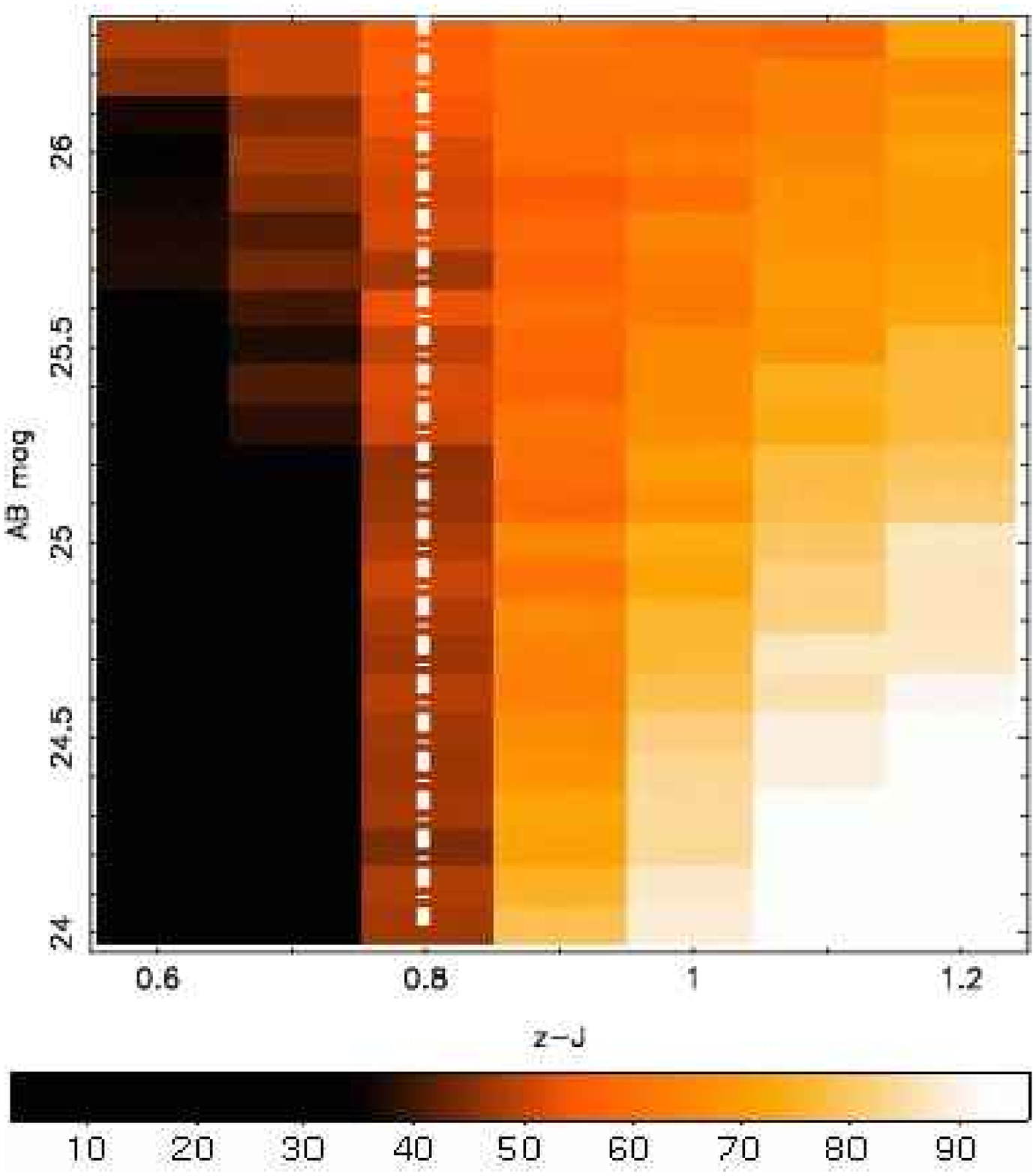}}}
\end{minipage}
\begin{minipage}{0.55\textwidth}
\centerline{\mbox{\includegraphics[height=\textwidth,angle=0]{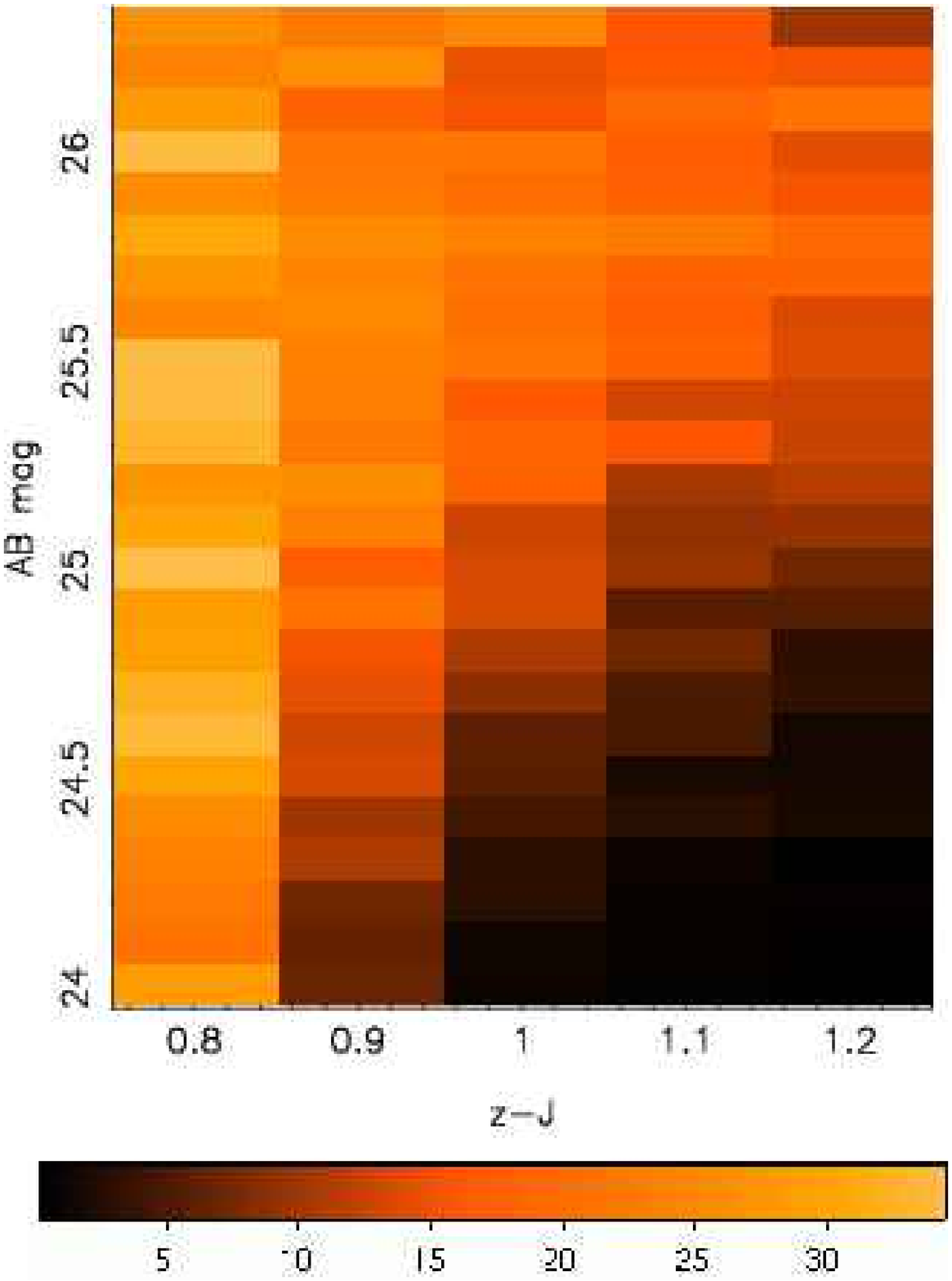}}}
\end{minipage}
\caption{\label{select} Testing the color-selection of $z$-band drop-outs. (Left) Completeness in the 
color-selection ($f_{comp}$), as a function of the $J$ or $H$ magnitude (assuming a flat spectrum in 
AB) and the $(z-J)$ color-break. Values represent the fraction (\%) of simulated objects whose 
photometry satisfies our color-selection criteria as indicated in the color bar beneath. (Right) As for the left panel but referring to the contamination fraction $f_{cont}$.
}
\end{figure}

The two panels in Fig. \ref{select} show the results of this test. The {\it selection completeness fraction} 
$f_{comp}$ (left) represents the fraction of objects of a given magnitude and break color that we are able to 
recover in our selection. The {\it contamination fraction} $f_{cont}$ (right) accounts for the fraction of 
objects with a lower break ($z-J<0.8$) that are photometrically scattered in the color-color diagram 
so that their {\it observed} magnitude and color would allow them to enter our selection window.
As with our magnitude completeness function, $S_{comp}$ ($\S$2.6), both of these correction factors 
will be used, for each candidate, to correctly estimate the true number density of objects having colors 
corresponding to $z>6.8$ galaxies (i.e. a break $z-J$ above 0.8). Individual correction factors are 
reported in Table \ref{magtable}.

Examining Fig. \ref{select}, as expected we find negligible difficulties for the brightest sources, but 
contamination and incompleteness become more troublesome at fainter magnitudes, depending on the $(z-J)$ color.  
We find that the selection completeness ranges from 50 to 95\% and the contamination fraction is typically 15\%.

\subsection{Verifying the Candidates}

The selection techniques discussed above yields a list of candidates for more careful examination.
Here we discuss further tests to determine the possibility that some might be spurious prior to
establishing a catalog of genuine sources whose redshift distribution we will explore using our additional
photometric data.

\subsubsection{Spurious Detections}
\label{reality}

As the signature of both our $z$ and $J$-band drop-outs consists of a non-detection in the optical
band, we must seriously consider the possibility of spurious detections in the NICMOS data. This is
particularly the case for the $J$ drop-outs where only a single band is involved.
An optical non-detection is defined as a implied flux lower than the photometric limit (5 $\sigma$ 
in a 0.3\arcsec-diameter aperture, for ACS and WFPC2 data, see Table \ref{clusterimages}). These
measurements were made using SExtractor in its ``double-image'' mode after the data was resampled 
and aligned to the NICMOS images. Further measures were made using the original multidrizzled images 
(see Sect. \ref{mult}).

A visual inspection was performed for each candidate in order to reject obviously spurious detections in the NICMOS images or false non-detections in the optical bands. The astrometric position of the candidates 
was used to perform this examination on the original images, to prevent biases arising from resampling and geometrical transformations. During this process, we rejected a number of candidates due to their proximity 
to the center of the removed BCG, or due to obvious contamination from very bright galaxy haloes, both 
leading to noisier or biased photometric measurements.

Our photometric detection is based on the combination of 10 individual NICMOS images per pointing 
(4 and 6 exposures in the $J_{110}$ and $H_{160}$ band, respectively). Because of a significant number 
of remaining hot and dead pixels in these individual frames, we investigated the fraction of spurious 
sources that would contaminate our photometric catalogs. To quantify this problem, we constructed a 
\textit{noise image} for each cluster and near-infrared band whose purpose is to remove signals from
all genuine sources while maintaining the same noise properties as the real data. This was done by  
subtracting in pairs the individual frames prior to eventual coaddition. We then applied our usual
photometric detection software, using the same parameters as in the original images. 

\begin{figure}
\centerline{\mbox{\includegraphics[width=15cm]{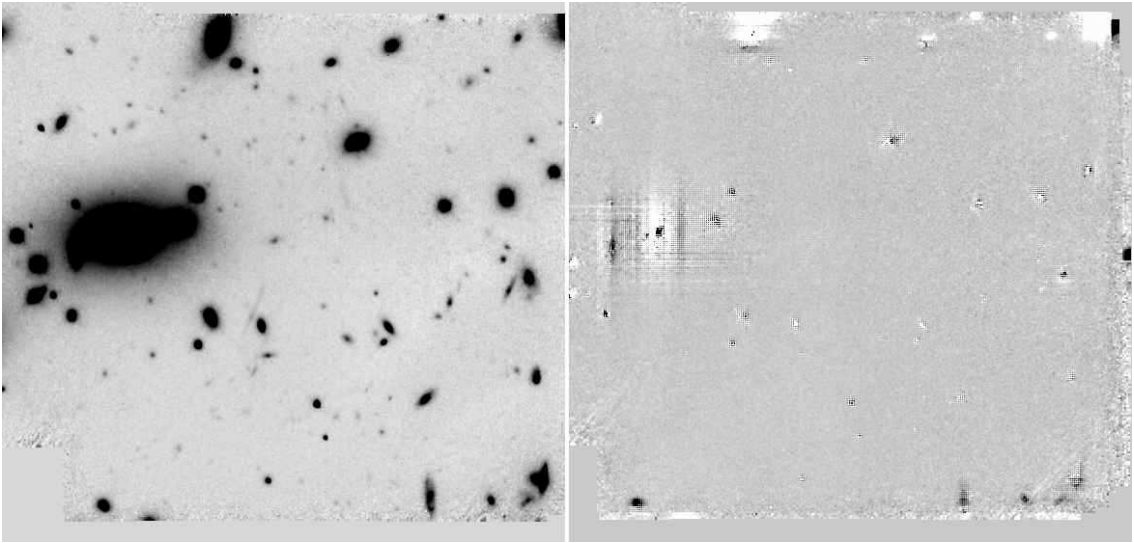}}}
\caption{\label{spurim} NICMOS image of the cluster CL1358 (left) compared with a noise image
(right) in order to estimate the fraction of spurious sources in our sample (see text for details). Except 
in the vicinity of the edges and central cores of bright objects (which are masked out by applying a 
simple threshold), the noise properties of the two images are very similar.}
\end{figure}

This noise image reveals residuals near the frame edges (due to the dithering process undertaken during 
the observations) and in the centers of the brightest objects, which were masked out in the detection process 
(Figure \ref{spurim}). The affected area accounts, in total, for $\sim 20\%$ of the NICMOS field. By comparing 
the number of spurious sources detected in the noise image with the number of objects present in the 
original catalogs, we estimate the spurious fraction in the magnitude range of our candidates ($24.7<J<26.3$ 
and $25.3<H<26.7$) to vary from 4\% to 18\% from cluster to cluster, with an average value of 10\%. 
Typically, therefore, we can expect around 90\% of our candidates to be robust astronomical detections.

\subsubsection{Detector Remnance}

One specific worry, not addressed in the tests mentioned above, concerns the possibility of image persistence
or ghosts appearing in the NICMOS frame during, or shortly after, an overexposure by a very bright 
source \citep{persistence}. The source producing the largest number of counts ($\sim 25000$ ADU) in our 
dataset is a $J_{AB}=17.9$ star in Abell 2390. We do not detect any measurable electronic ghost for this 
particular source, but persistence is seen at the level of a $J_{AB}=24.9$ spurious source in an exposure 
taken 20 minutes later. Once averaged over the entire sequence (6 exposures), this persistence corresponds to 
a $J_{AB}=26.8$ source, i.e. fainter than our detection limit. We also verified, for each pair of successive
exposures, that none of our candidates is coincidently located at the relevant position of similarly bright 
($J<19$) stars in all clusters. 

Although NICMOS exposures from independent orbits are usually separated by a $>30$ minute delay, 
persistence might arise as a result of much brighter sources observed immediately prior to execution
of our program. To eliminate this possibility, we located all preceding exposures and found no 
measurable persistence in our data, even in the case of a calibration program (GO 10726, PI: de Jong) 
aiming at measuring non-linearity effects by repeatedly saturating the NIC3 detector.

\subsection{Catalog of Drop-out Candidates}

\begin{figure}[ht]
\begin{minipage}{11.cm}
\includegraphics[width=\textwidth]{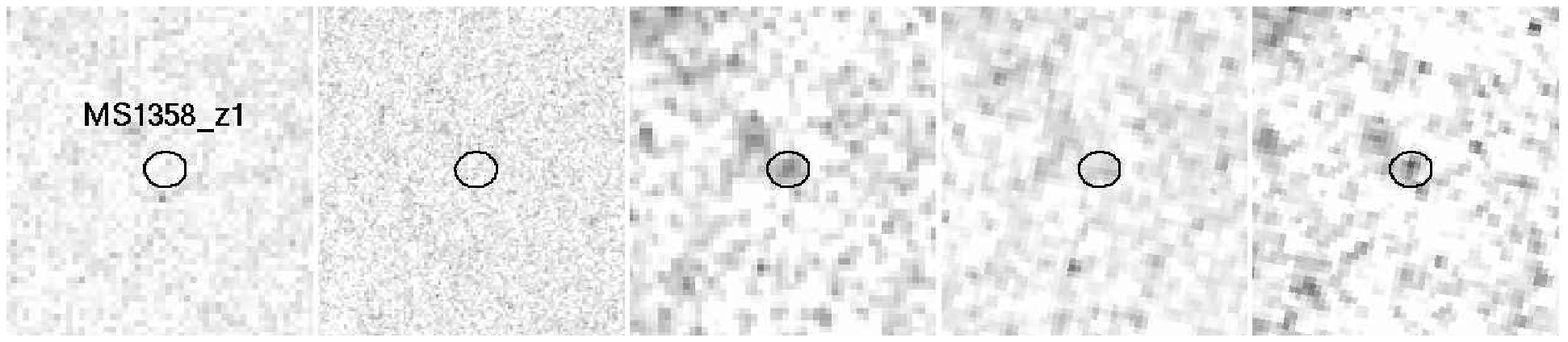}
\end{minipage}
\begin{minipage}{5.cm}
\includegraphics[width=\textwidth]{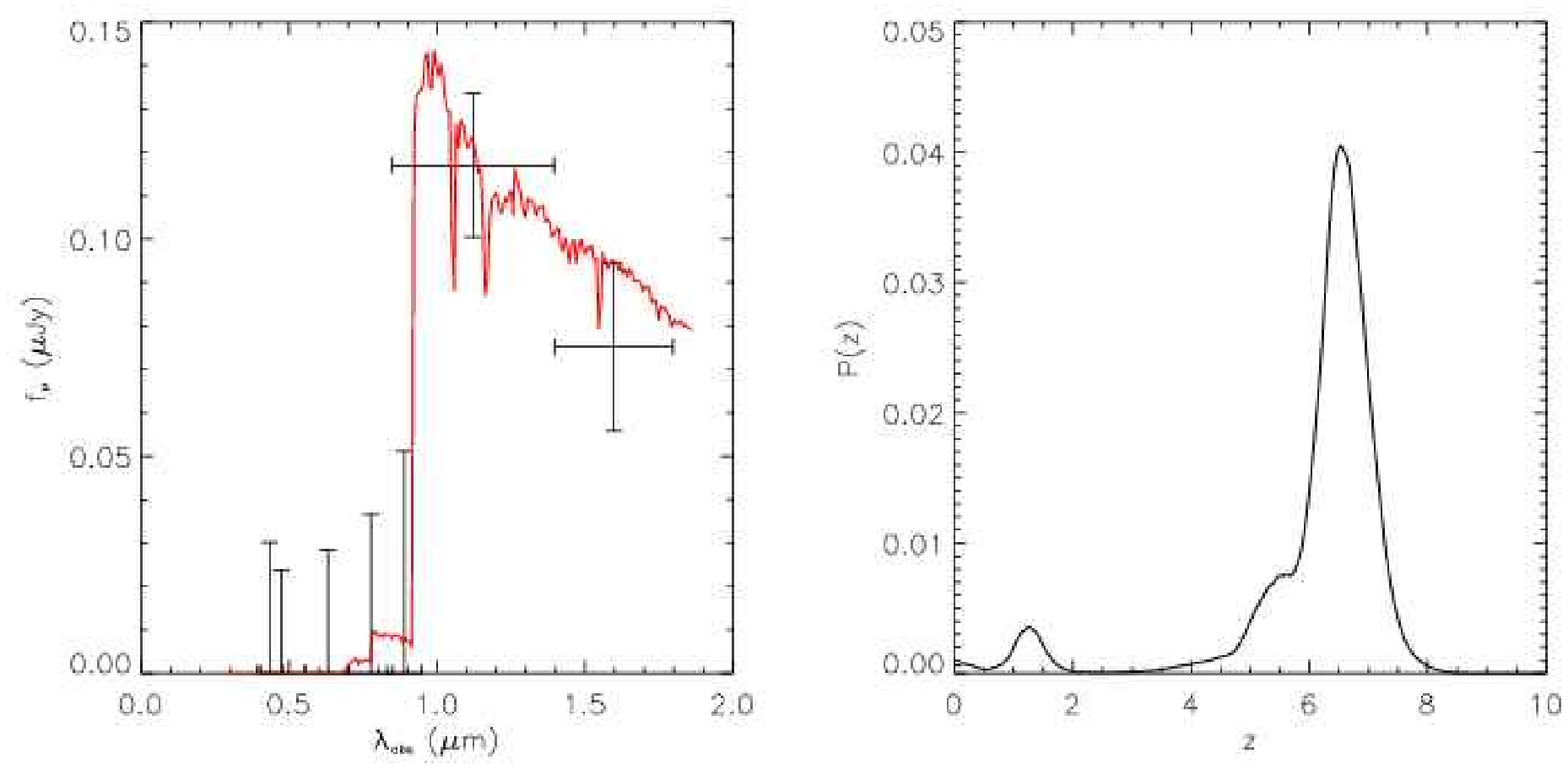}
\end{minipage}
\begin{minipage}{11.cm}
\includegraphics[width=\textwidth]{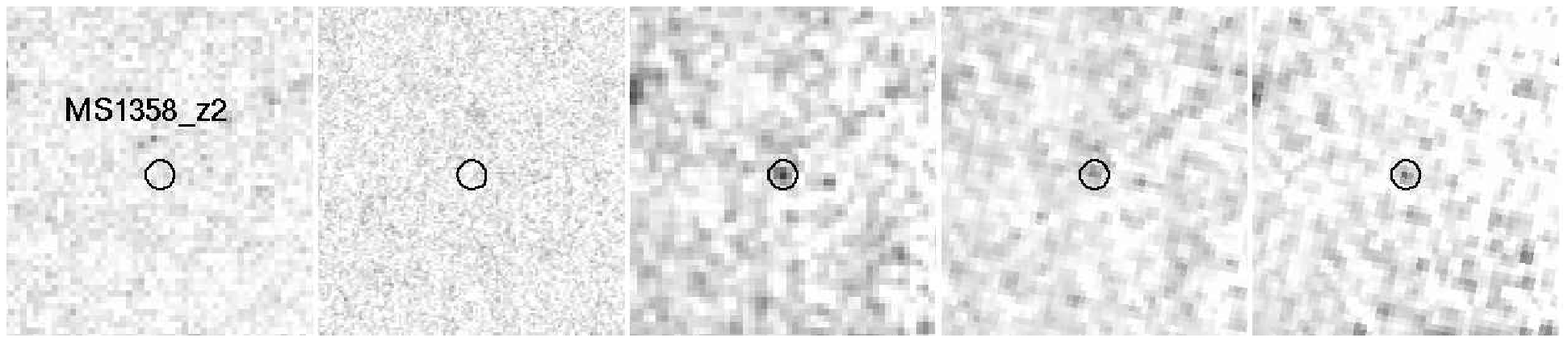}
\end{minipage}
\begin{minipage}{5.cm}
\includegraphics[width=\textwidth]{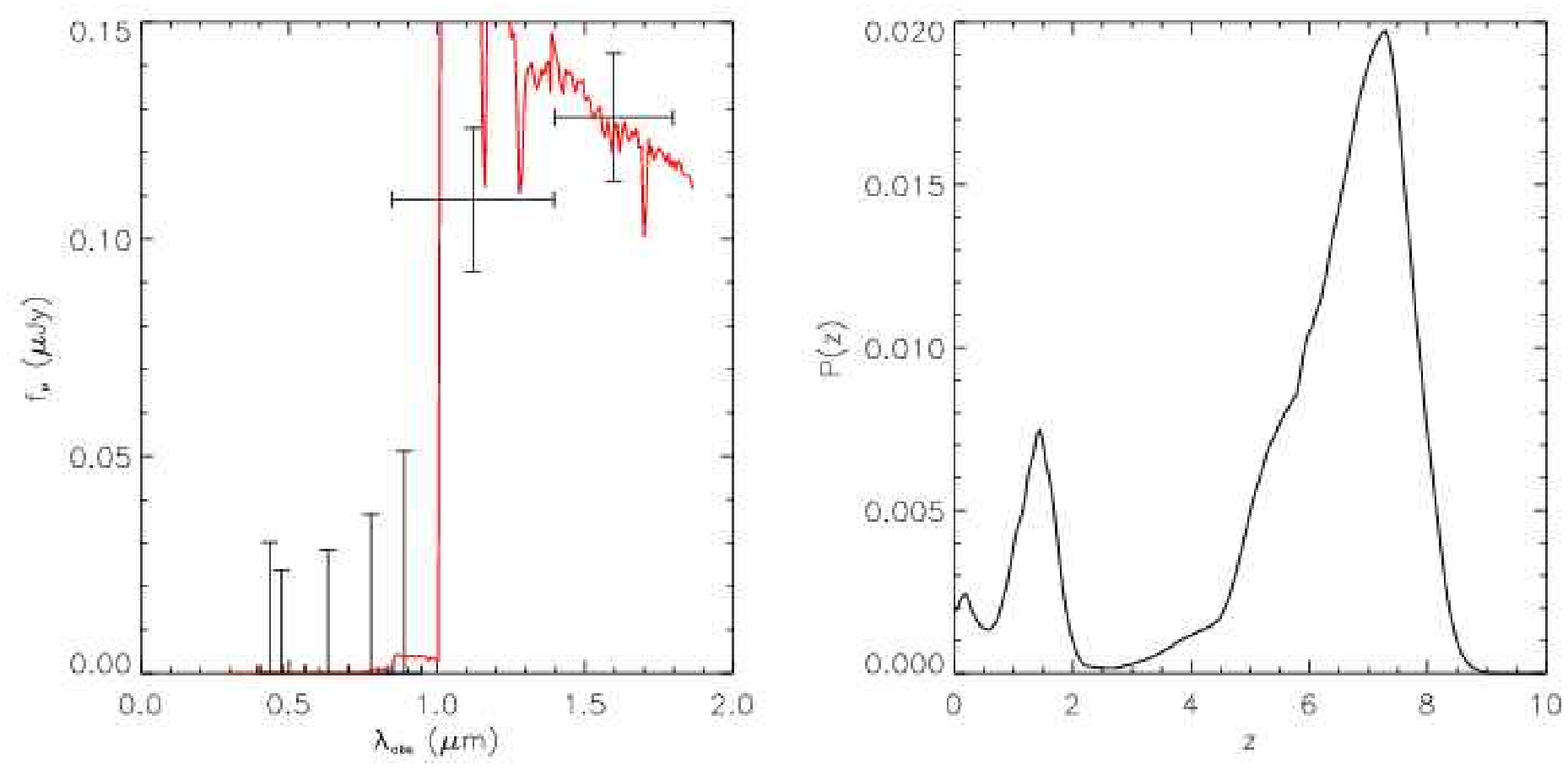}
\end{minipage}
\begin{minipage}{11.cm}
\includegraphics[width=\textwidth]{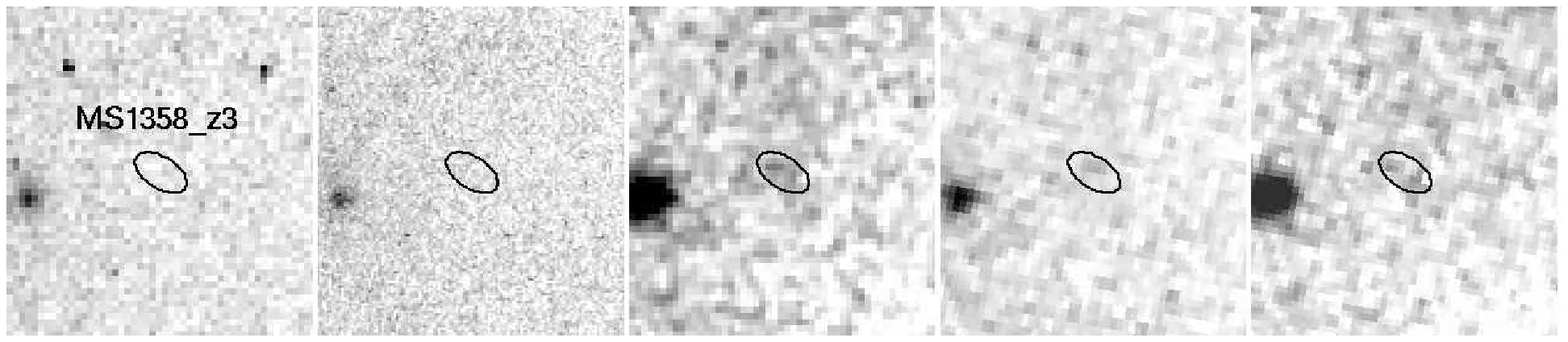}
\end{minipage}
\begin{minipage}{5.cm}
\includegraphics[width=\textwidth]{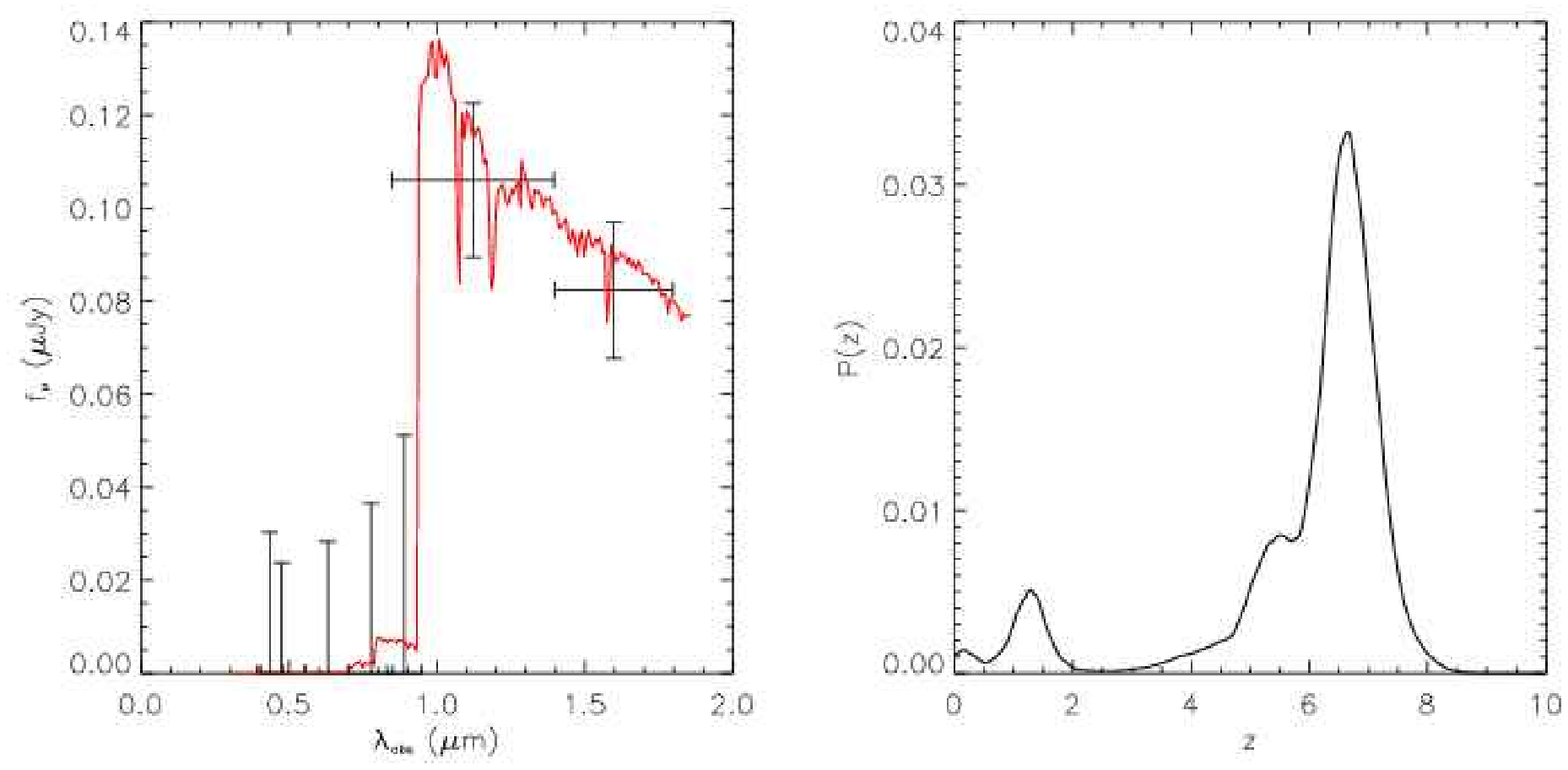}
\end{minipage}
\begin{minipage}{11.cm}
\includegraphics[width=\textwidth]{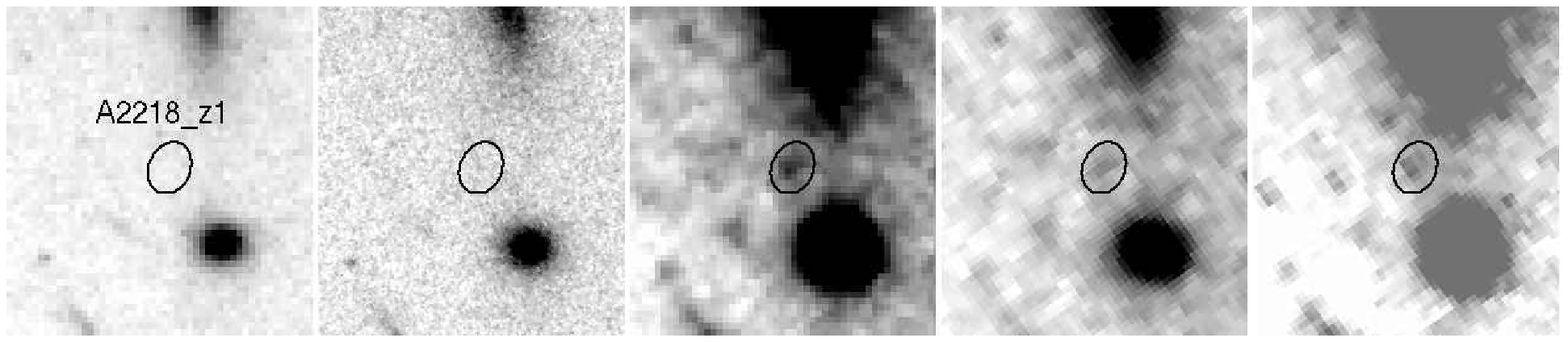}
\end{minipage}
\begin{minipage}{5.cm}
\includegraphics[width=\textwidth]{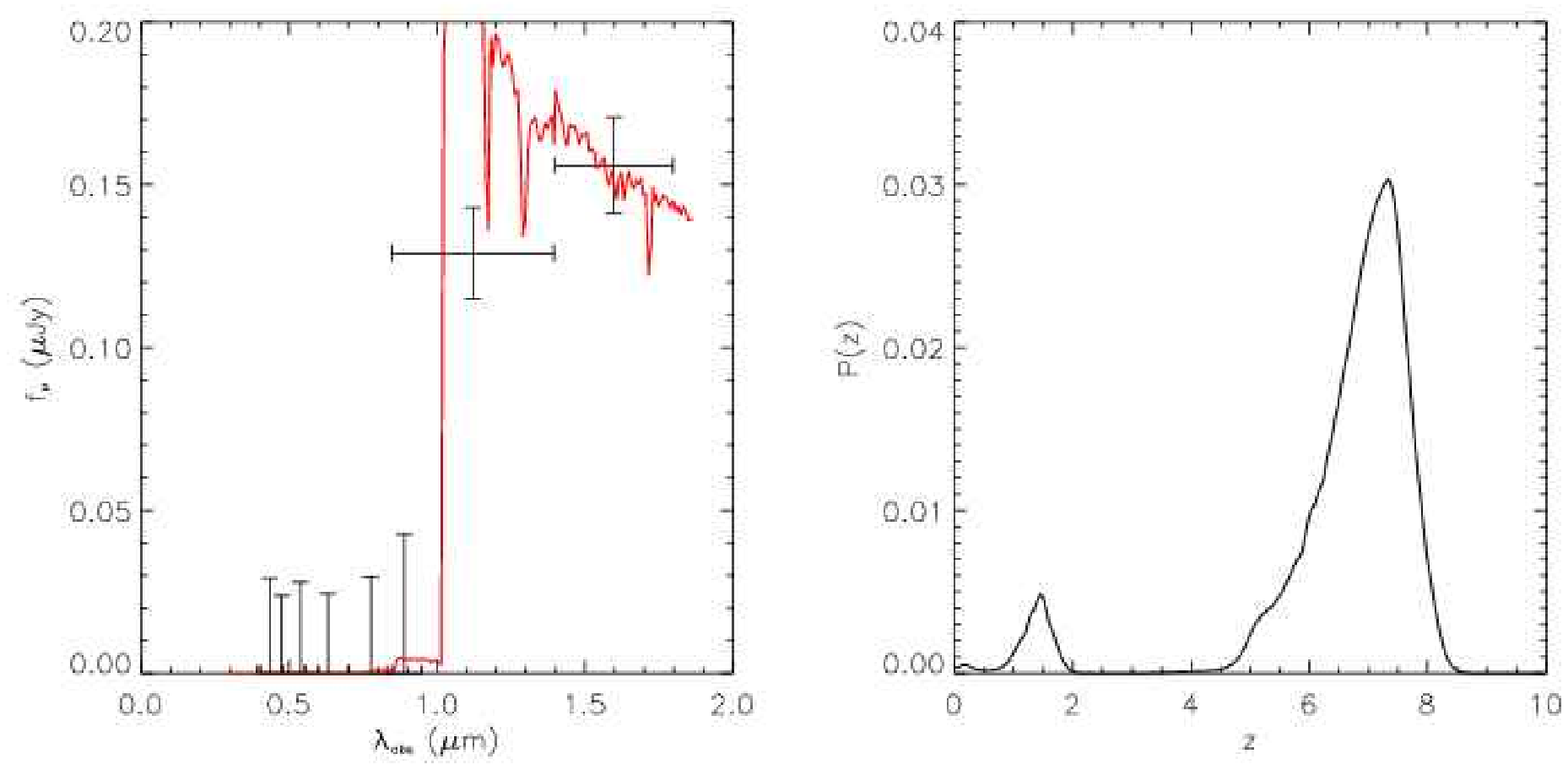}
\end{minipage}
\begin{minipage}{11.cm}
\includegraphics[width=\textwidth]{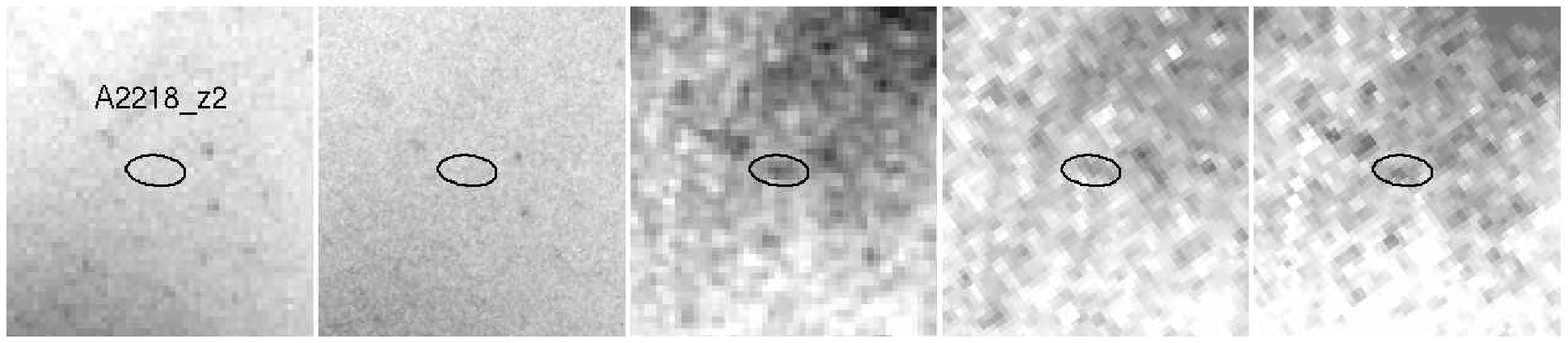}
\end{minipage}
\begin{minipage}{5.cm}
\includegraphics[width=\textwidth]{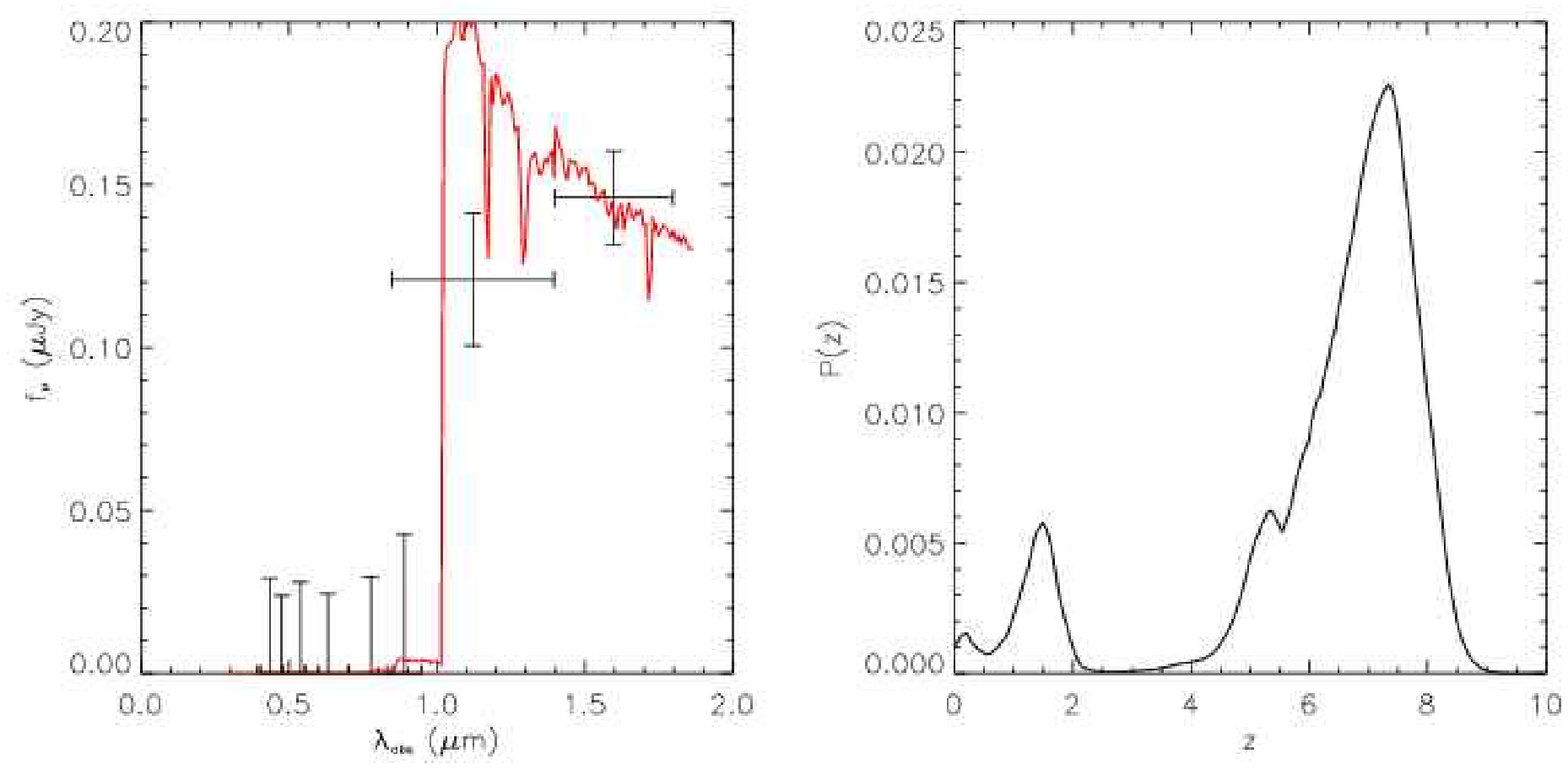}
\end{minipage}
\begin{minipage}{11.4cm}
\includegraphics[width=11.cm]{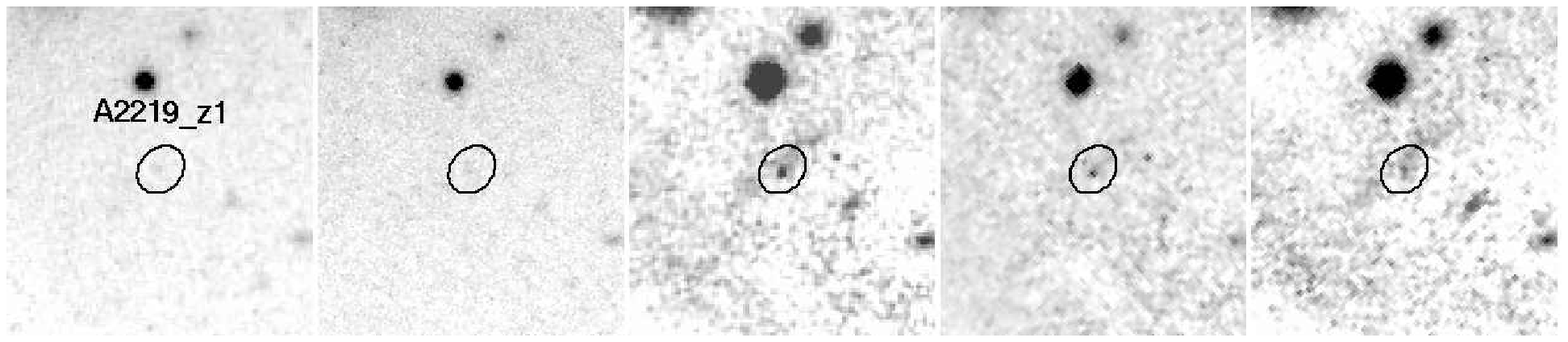}
\end{minipage}
\begin{minipage}{5.cm}
\includegraphics[width=\textwidth]{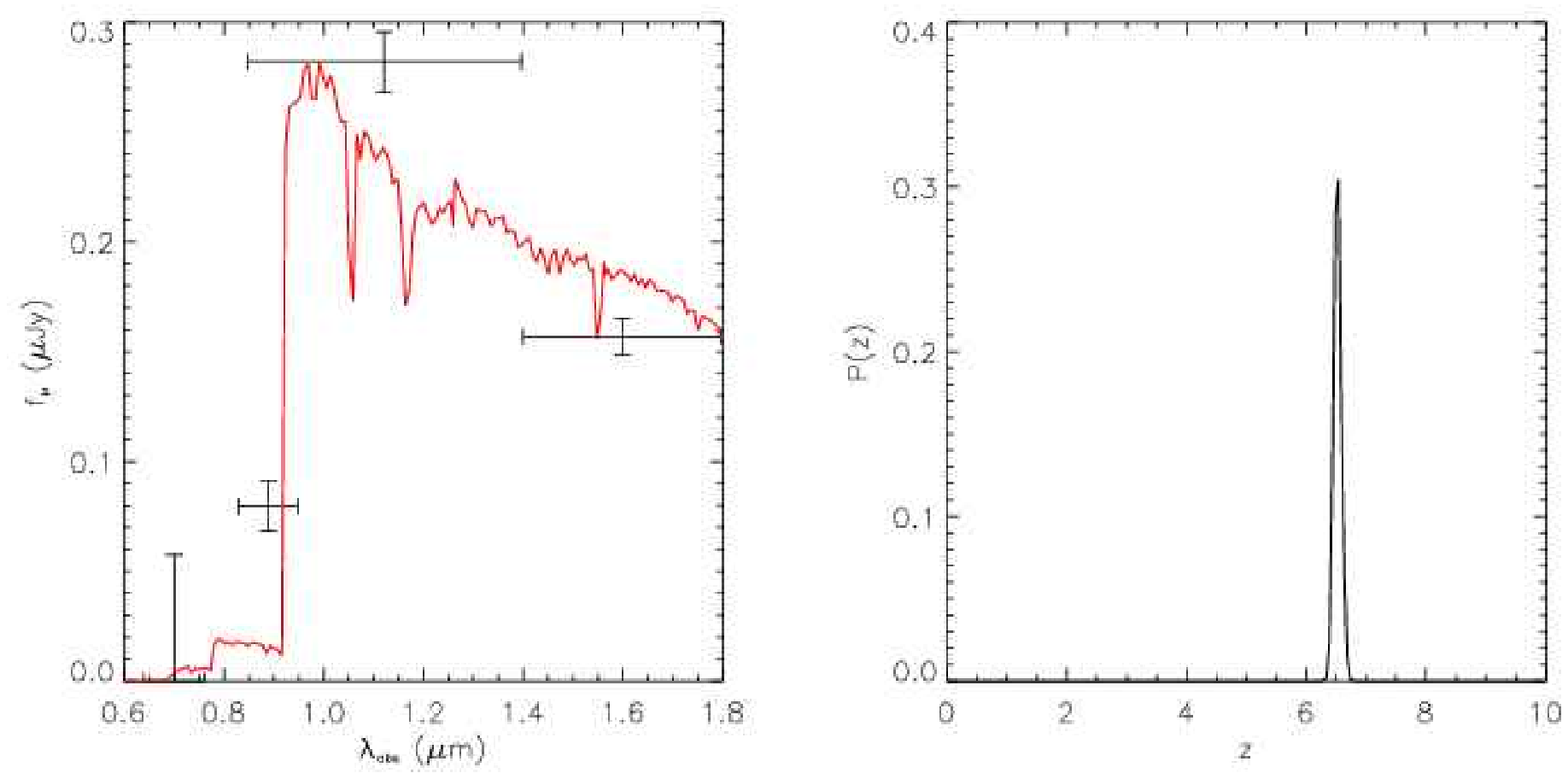}
\end{minipage}
\caption{\label{candfigs}Snapshot images our final catalog of candidate drop-outs. (Left to right): HST 
images in the optical (ACS or WFPC2), in the ACS/F850LP (z) band, in the detection image (sum of 
F110W and F160W bands), in the NIC3/F110W and the NIC3/F160W images. To the right are the observed 
SED with the overplotted best fit template for HyperZ, and the redshift probability distributions (see 
Sect. \ref{hyperz}). Shown separately are the MOIRCS $K$ band images.).
}
\end{figure}

As a result of our visual inspection of the initial candidate list selected according to the precepts of
Section 3.1, we emerge with 10 candidate $z$ drop-outs and 2 $J$ drop-outs. As seen in Fig. \ref{cz}, 2 of 
the 10 $z$-dropouts satisfy the most restrictive color-selection $(z-J)>1.25$. The photometric 
measurements for these are summarized in Table \ref{magtable} and the relevant detection images 
are presented in Fig. \ref{candfigs}. Following the tests described above we can expect over 90\% of these
to be genuine astronomical sources. 

\subsection{Redshift Estimation}

We now turn to the important problem of contamination by lower redshift sources. We first use the 
spectral energy distribution (SED) of each candidate to estimate the individual photometric redshift. We then
consider statistical arguments that can be applied to our entire candidate population.  

\subsubsection{Photometric Redshifts}
\label{hyperz}

As we have seen, the two color selection presented in Sect. \ref{dropouts} enables us to select high 
redshift galaxies with some confidence. However, we can use the multiwavelength data available for 
each source, including all upper limits arising from non-detections, in order to derive a 
{\it photometric redshift probability distribution}. To accomplish this, we used an updated version of 
the photometric redshift software HyperZ \citep{hyperz}. Best-fit redshift distributions were computed 
using a standard SED procedure with a variety of templates, 
including empirical data \citep{Kinney, Coleman} and evolutionary synthesis models \citep{BC}. 
We searched the redshift range $0<z<10$, while reddening was kept as a free parameter ranging 
between $A_V=0$ and 2 mag, assuming the \citet{Calzetti00} law. The effect of Lyman forest blanketting is included following the prescriptions of \citet{Madau}.

This approach is only practical for the $z$ drop-out candidates and the main results are presented
in Fig. \ref{candfigs} (rightmost panels) where we overplot the best fit templates on each SED and 
present the redshift probability distribution P(z), marginalized over the entire parameter space (templates and reddening). As has been found by many workers \citep{Lanzetta,Richard06}, the
likelihood function reveals two peaks with different relative intensities, the relative height of the lower 
redshift peak acting as a valuable measure of foreground contamination, as it is linked to the color 
degeneracy between high-redshift star-forming galaxies and lower redshift early-type or reddened objects.

Encouragingly, each candidate is more likely to be at high redshift and the probability of a foreground
source is negligible for 4 out of our 10 sources. Integrated over all our candidates, we use
$P(z)$ to compute the probability that each object lies beyond a redshift of 6, $\alpha_6=P(z>6.0)$, 
after normalizing $P(z)$ to unity over the redshift range $0<z<10$. We find $\alpha_6$ values ranging 
from 0.46 to 0.91, with an average value of $0.65$. 

\subsubsection{Stacked Imaging}

As is evident from Table \ref{magtable}, the individual candidate SEDs are mostly restricted
to detections in only one or two bands, with upper limits at other wavelengths. Although this precludes
precise photometric redshift measures for each candidate, we can make further progress by combining 
data over several adjacent wavebands, and also by examining the average SED of the population to 
see if it is statistically consistent with that expected for a high redshift source. 

Several of our clusters have ACS data in multiple bands (Abell 2218, CL1358, Cl0024) which we
combined after aligning the images with integer ACS pixel shifts and normalization each to a constant 
signal to noise ratio. The depth of this combined image, covering the wavelength range 4500 - 8000 \AA\ , 
is typically 0.6 to 1.0 magnitudes deeper than the individual bands. Yet in each of the 5 $z$ drop-outs,
no optical detection is seen, increasing our confidence level in the corresponding drop-outs.

We likewise generated stacked $z$, $J$ and $H$ images for all 10 z drop-outs, as well as a stacked $K$ 
image for the 4 sources observed with MOIRCS. In this case, We selected a 10\arcsec$\times$ 
10\arcsec\ region around each candidate, and averaged the data rejecting 20\% of the outlier pixels. 
For 6 objects free from contamination from nearby objects, the same stacking procedure was performed 
in the 3.6 and 4.5 $\mu$m IRAC bands. The stacked images are presented in Fig. \ref{stackim}. 
No significant flux was detected in either the stacked MOIRCS or IRAC images.

Using the SExtractor parameters adopted for processing the individual images, an average SED was 
constructed for 3 populations depending on the availability of multi-band ACS, $K$ band and IRAC
imaging. The photometric properties of each are listed in Table \ref{stackphot} and the SEDs are shown 
in Fig. \ref{stacksed}. These various stacked datasets offers a new opportunity to address the
question of foreground contamination.
 
Using HyperZ, we fit these three SEDs exactly as described in Sect. \ref{hyperz}. The optimal 
photometric redshift is consistent in each case (Fig. \ref{allpz}), with $z_{ph}=7.35\pm0.07$, and a reduced 
$\chi^2$ lower than 1. We also fitted the same SEDs, but now restricting the photometric redshift 
to the range $0<z_{ph}<3$. The best fit at low redshift is found at $z_{ph}=1.75\pm0.03$, but with 
a much higher $\chi^2$, between 5 and 10. To the extent that the low redshift solutions are credible,
they imply galaxies with typically ages of 500-700 Myrs  and extinction values of $A_V\sim1.0-1.4$.

One question that arises is whether the near-infrared $J-H$ color for our $z$ drop-outs is consistent 
with expectations, and also with that of the other limited detections at high redshift \citep{Bouwens08}.  
Using the stacked images, we find this color is typically very blue: $J_{110}-H_{160} \sim 0.0$. We define 
the rest-frame ultraviolet slope $\beta$ as $f_\lambda\propto\lambda^{-\beta}$ between the restframe
wavelengths 1500 and 2000 \AA\ \citep{Calzetti94}, and estimate the uncertainty in $\beta$ for our 
sources either from the dispersion within the range of plausible redshifts (at $\pm1$ $\sigma$), or by 
using the adopted photometric error bars in $J$ and $H$. In both cases, we find a consistent value: 
$\beta\sim2.8^{+0.05}_{-0.2}$. The mean slope is somewhat bluer than the average value of 
2.0$\pm$0.5 for a sample of $i$ band drop-outs measured at $z\sim6$ by \citet{Bouwens06} but 
within the range of $\beta\sim3.0$ found by \citet{Bouwens04} in their sample of $z$ drop-outs, and 
marginally consistent with $\beta\sim2.5$ given for several objects by \citet{BouwensNature}. In order 
to produce such a blue ultraviolet slope, the SED can only be fit by very young models 
(typically $<100$ Myrs) with little or no reddening ($A_V<0.1$).  

\subsubsection{Galactic Stars}
\label{stars}

Our next test for foreground contamination concerns the question of low mass Galactic stars.
Various authors \citep{Stanway04,Bouwens04} have pointed out the difficulty of using optical-near 
infrared colors to distinguish between cool stars and breaks arising from Lyman absorption at high 
redshift. L \citep{Cruz} and T \citep{Burgasser04} dwarfs exhibit metal and H$_2$O absorption 
features which produce features similar to the drop-out signature. Indeed, L and T dwarfs from the 
SDSS and 2MASS surveys \citep{Knapp} lie well within our color-color selection region (Fig. \ref{cz}).
Likewise, we find $\chi^2$ values similar to the best template spectrum from HyperZ when fitting the 
SEDs of the candidates with a library of L0 and T dwarf templates spectra.

In such circumstances, HST data has occasionally been used to evaluate whether the sources have 
half-light radii $R_h$ consistent with being extragalactic objects \citep{Stanway04}. However, this
is not a definitive criterion as we already know that many spectroscopically-confirmed low luminosity
high redshift sources are unresolved \citep{Ellis}.  We measured $R_h$, defined as the radius enclosing 
half of the flux in the detection (NICMOS) images, and compared the values with that derived for 
bright not-saturated stars (0.2\arcsec). Figure \ref{sizemag} represents the location of our candidates 
in a $R_h$ vs $J_{AB}$ diagram, together with all other objects in our photometric catalog. Unfortunately,
surface brightness dimming generates a cut-off at large values of $R_h$, which does not allow to 
distinguish resolved and resolved sources at the faintest magnitudes. This is the case for 4 of the 
$z$-dropouts. At most we can say that 2 dropouts are unresolved, whereas 4 are resolved. 

Noting the difficulty of separating stars from galaxies at these faint limits, a more practical approach
is to examine the likely contamination statistically.  Using simulations by \citet{Burgasser04}, we 
computed the number of expected low-mass stars in the total area surveyed with NICMOS (7.7 arcmin$^2$), 
for all spectral types between L0 and T8, up to the extent of the thick disk (1 kpc). Assuming 
a slope $\alpha=0.0$ for the mass function, consistent with recent observations by \citet{Metchev}, 
we predict only 1.1 dwarf in our survey. A more pessimistic $\alpha=0.5$ slope yields 1.5 
stars. Thus, while there is undoubtedly some uncertainty surrounding these predictions, it
does seem unlikely that cool Galactic stars represent a significant contaminant at such
faint limits.  

\subsection{Magnification and Multiple Images}
\label{mult}

Our final test concerning the high redshift nature of our candidate drop-outs concerns their location
within the image plane of the lensing cluster. A key question is whether any might be expected
to be multiply-imaged as was the case, for example, for the object studied by \citet{Kneib}.
Most of the area covered by our NICMOS observations lies within the region of high magnification 
($\mu>2.5$ mags) where multiple images may occur.

Using the mass models for each cluster (updated using Lenstool from those referenced in 
Table \ref{cluster}), we estimate the magnification of each drop-out based on its
photometric redshift and location and examine the likelihood of any counter images as well
as their predicted location and relative fluxes. The Bayesian optimization method incorporated 
in Lenstool also provides the uncertainties in these quantities.  

In the majority of the cases (7 out of the 10 $z$ drop-outs), the model predicts a pair of 
images with similar fluxes (within 0.2-0.4 magnitudes) straddling the critical line 
(Figure \ref{sample}). Single images are expected in two other cases (CL1358z3 and A2667z1) 
and in the final case (CL1358z1), the objects sits on the predicted critical line, but 
is not expected to be viewed as a distinct pair at the finite angular resolution of NICMOS. 
These three cases are consistent with our observations. 
Considering the two J drop-outs, A2219j1 is predicted to be another example of close 
merging unresolved by HST, and A2667j1 is predicted to have much fainter counter-images 
(by 0.8 to 1.5 mags),  below our detection limits.

Our attention is thus focused on the 7 cases where second images are expected. The typical
positional errors are around 1-3 arcsecs. Unfortunately, most are located outside the area
surveyed by NICMOS or in regions close to the edge of the detector or under the central BCG
where the noise level is high. Only in two cases, A2218z1 and A2667z2, does the mass model 
predicts a detectable counter-image in a relatively clean region of the NICMOS detector. Unfortunately,
no significant flux, within the range expected, was seen at either position in the J-band of H-band image. 

%This test provides the only null result in our exhaustive quest to confirm the high redshift nature
%of our candidates. Although it applies to only two of our 10 $z$-band drop-outs, there is no reason 
%if these two are at high redshift that we should not see the counter-image. Nor is there any
%reason to consider these two drop-outs as inferior to the remainder. However, an identical search 
%for multiple images was performed under the assumption that each candidate is at a lower 
%redshift, $z\sim2$. In that case, we would expect counter-images at a different position, closer to 
%the center of the cluster, symmetric to the lower redshift critical line. We detect no flux for any of 
%the candidates either. However, we did recover a counter-image for one of our more marginal first 
%cut $J$-drop candidate in Abell 2390 with $J-H=1.5\pm0.3$ (Fig. \ref{mult_lowz}). We subsequently
%eliminated this candidate and can now verify, from the location of the pair around the critical
%line, that this is a $z\sim2$ object. This discovery clearly illustrates how predictions from the 
%lensing model can discriminate low-redshift contaminants.

Unfortunately, this test, valuable in principle, provides an inconclusive outcome in our exhaustive
quest to confirm the high redshift nature of our candidates.  Although the high redshift test is 
only applicable to two of our 10 $z$-band drop-outs, we are unable to see either of
the counter images. However, an identical search for multiple images was performed under the 
assumption that each candidate is at a lower redshift, $z\sim$2. We would then
expect counter-images at different locations, closer to the cluster center. 
However, we did not detect any of these predicted counter images either, for 4 clear cases. 
Only in a more marginal $J$-drop candidate, 
which we dropped at an early stage, did this test succeed in demonstrating a $z\simeq$2 
solution (Fig. 12). Given the strong likelihood that most of the candidates lie
either at $z\sim$2 or $z\sim$7-8, the test has a confused outcome. On the one hand, for those
high $z$ candidates where we could expect to see a counter-image, none is seen. On the other hand
for a larger sample of candidates, assumed to be at $z\simeq$2, none is seen either. Accordingly, we
deduce the test is not effective as a redshift discriminant.

\subsection{Summary}

We now summarize the possible success rate of our survey in generating high redshift sources,
concentrating on the $z$-band drop-outs. Out of our ten sources, we find that at most one source
is spurious and $\simeq$one is a Galactic star. Thus we conclude 8 are likely to 
be extragalactic sources. Admittedly, about half of our candidates are
unresolved, but we believe this is to be expected given the intrinsically faint limits we are
probing with our lensing method. Discounting the inconclusive test based on counter-images (Section 3.5),
and noting the 35\% contamination from $z\simeq$2 sources, we conclude that, statistically, we can expect 5 of
our 10 sources to be $z>$6.8 star-forming galaxies.

The foregoing analysis, whilst exhaustive, is however, statistical in nature. Our approach
has been to treat all candidates as equally possible and to determine the level of foreground
and spurious contamination as a fraction, without commenting on the nature of each
individual source. Important information is contained in the similarity or otherwise of
the candidate's morphology in the various NICMOS bands and the confidence with which
we see no optical detection at the location of each candidate. For the $z$-band drop-outs,
a marginal detection is permitted in the red wings of the F850LP filter, but any hint of
a signal at the location of the candidate in shorter wavebands would give cause for concern.
It is then a matter of judgement whether to rely primarily on the photometric redshift
solution (\S3.4.1) or to override such information and reject a candidate after visual
inspection.

Concentrating again on the $z$-band drop-outs, two sources, A2219-z1 and A2390-z1, are
resolved and satisfy the more rigorous color cut in Figure \ref{cz}. One might therefore
imagine these are particularly robust candidates.  
A2219-z1 has a detection in the WFPC2 F702W
filter very close to the location of the NICMOS image. Although
reasonably significant (27.1 or 4.5$\sigma$) it does not influence
the photometric solution (Figure 6). Nonetheless, it does raise
doubt about this candidate.

In the case of A2390-z1, there is no detectable optical signal down to F555W$=27.6$ (2.0$\sigma$) F814W$=27.0$ (2.5$\sigma$) and F850LP$=27.8$ (2.0$\sigma$) but
the candidate's morphology differs somewhat between the NIC3 F110W and F160W filters.
This may reflect the presence of two sources, one or both of which is at high redshift,
or a genuine structural difference in the bands, for example as a result of line emission.

The only other source worth commenting on is A2667-z2 for which there is a marginal
WFPC2 F606W detection at the position of the NIC3 source (28.25, or 2.0$\sigma$).
This is reflected in the fact that the photometric redshift solution is fairly ambiguous for this
source (Figure 6).

Finally, we note as in Section \ref{dropouts} that the
$J-H$ color of most of our $z$ drop-outs are significantly bluer than predicted for a
normal SED at $z\simeq$7. Shifting the sources to lower redshift would not
significantly resolve this interesting observation.

Concerning the $J$-band drop-outs, neither are particularly compelling. A2667-J1
has a similar morphological difference between the F160W and faint F110W image
and A2219-J1 has a marginal detection (27.0, 3.5$\sigma$) in the ACS
F850LP filter.

\section{Spectroscopic Follow-up}

Although deep imaging with HST has delivered candidates whose photometric 
redshifts lie beyond $z\simeq$7, to date there has been not a single spectroscopic confirmation despite
heroic efforts. A case in point is the $z\simeq$6.8 lensed system in Abell 2218 \citep{Kneib} which was 
the subject of 9.2ks exposure with the LRIS optical spectrograph and a marathon 33ks exposure 
with the NIRSPEC infrared spectrograph. A marginal continuum was seen but no emission
lines were detected. This contrasts with the successful detection of Lyman $\alpha$ in IOK-1
\citep{Iye} in 31ks. The latter source has an implied star formation rate of $\simeq10\,M_{\odot}$
yr$^{-1}$ whereas the source in Abell 2218, when magnified, was expected to have an observed 
line flux equivalent to an unlensed system with a star formation rate of 
2.6$\times25 \simeq 50\,M_{\odot}$ yr$^{-1}$. A tantalizing explanation for the non-detection of
Lyman $\alpha$ in the object in Abell 2218 is preferential damping by neutral hydrogen in lower 
luminosity sources.

The presence of the Ly$\alpha$ line provides a critical feature for
confirming the nature of candidate high-redshift galaxies.  However,
the line is relatively easily attenuated and therefore may well be
obscured in actively star forming galaxies. Therefore, the absence of
the line does not provide evidence that high-redshift candidates
are false. 
Nevertheless, we conducted an ambitious spectroscopic 
campaign at Keck for some of our candidates. We naturally hoped that we might also secure
the first spectroscopic verification of a $z>$7 source.

\subsection{Observations}

We used the Near InfraRed SPECtrograph (NIRSPEC, \citet{nirspec}) on the 10 m Keck II Telescope 
to follow-up the majority of our candidates in the window $0.964$--$1.120$ $\mu $m,  
corresponding to the redshift range $6.9$--$8.2$ for the Lyman-$\alpha$ line (1216 \AA\ ).
Observations where conducted in several runs between January and September 2007 and
we secured good data for 7 $z$-band drop-outs in total (see Table  \ref{table_spectro}). 

We used a 42\arcsec\ long and 0.76\arcsec\ wide slit, offering a resolving power of $R\sim 1500$ 
and used dithered exposures of 10 minutes each. We adjusted the dithering distance (in the 
range 3-8\arcsec) in each case (third column of Table  \ref{table_spectro}), to prevent 
overlap between a candidate and another bright source. Occasionally it was possible to observe two 
candidates simultaneously. For two candidates where we expect multiple images, the location of 
the expected counter-image (see Sect. \ref{mult}) was used to optimize the slit positions (Figure 
\ref{sample}). 

The NIRSPEC spectra were reduced following the flat-fielding, sky-subtraction, distortion 
corrections and flux calibration procedures described by \citet{Stark07b}. These reduction techniques 
ensure an improved removal of the sky background by subtraction prior to resampling. 
We observed standard stars each night and used these to flux-calibrate the final spectra and
determine the associated variance and hence the 5$\sigma$ limiting line flux. Each position 
was observed for about 3.5 to 4 hours in total, yield a limiting line flux of 
$\sim3\times10^{-18}$ erg\ cm$^{-2}$\ s$^{-1}$ in regions of minimum OH contamination 
(Figure \ref{flim}), assuming a line width $\sigma_\alpha\sim 300$ km\ s$^{-1}$, as 
measured in the well-studied lensed system by \citet{Ellis}. For a different line 
width value, this sensitivity would vary as $\sqrt{\sigma_\alpha}$.

We inspected each reduced 2-D spectrum for faint emission lines at the position of the 
candidate and, where relevant, that of the counter-image. No significant signal was detected 
for any of the candidates.

\subsection{Implications}

While the outcome of our spectroscopic campaign is certainly disappointing, the presence of strong 
OH lines in $z$-band means that our limiting Ly-$\alpha$ flux ($\sim3\times10^{-18}$ erg\ cm$^{-2}$\ 
s$^{-1}$) applies only across 50\% of the observed wavelength range. Thus we would only expect 
partial success even if all of our sources had intense emission lines. Nonetheless it is informative
to consider what the absence of any emission might mean given the star formation implied by our
continuum detections.  We can convert our flux limit into a rest-frame equivalent width using our 
HST photometry. For our candidates, we find a typical upper limit of $W_{\lambda}\sim5-20$ \AA\ . 

\citet{Stanway07} have recently studied a sample of faint Lyman-$\alpha$ emitters at $z\sim6.0$ 
selected from their photometry to be $i$-band drop-outs. They found a tail of high values for 
the $W_{\lambda}$ distribution compared to similar studies undertaken at $z\sim3.0$ \citep{Shapley}. 
They attribute this evolution to a tendency for stronger line emission in intrinsically faint sources. 
58\% of their sample has $W_\lambda>25$ \AA\ rest-frame. Including the lower and upper limit measurements 
of $W_{\lambda}$ from this sample, which most likely contains lower redshift contaminants, 
the fraction is lowered to 34\%.

Assuming no evolution in this distribution from $z\sim6$ to $z\simeq$7-8, we would expect in the optimistic 
case (58\% value) $\sim4$ Lyman-$\alpha$ emitters in our spectroscopic campaign, prior to considerations of 
the OH spectrum. The probability that all 4 objects lie in a region of the spectrum contaminated 
by OH emission is thus $(0.5)^4\sim0.06$, which is low. Even if only 4 of the 7 candidates 
we examined were at high redshift (based on our statistical estimates given in \S3), we should 
expect to detect the emission for $\simeq3$ cases. Here there would be only a 12\% probability of 
each one being occulted by OH emission. In the most pessimistic case from \citealt{Stanway07} 
(34\%), we would expect to detect only one source, with 50\% probability of OH contamination. 

Thus, as in the case of the $z\simeq$6.8 source in Abell 2218, the absence of emission in 7
candidates is somewhat surprising. Assuming a significant fraction are at high redshift as
discussed in \S3, this may be an important indication of the evolution in the intergalactic medium
above $z\simeq$6. Regardless of the cause, it adds to the challenge of making progress in
verifying high redshift candidates.

Reconciling the above with the abundance of intrinsically faint Lyman $\alpha$ emitters claimed
by \citet{Stark07b} is admittedly difficult. Should the bulk of the drop-out population
at $z>$7 continue to reveal no emission, this would suggest a moderate neutral fraction
that would challenge the transparency of the IGM at $z\simeq$8-10 implied by the presence
of feeble Lyman $\alpha$ emitters. The enigma simply reinforces the importance of
continuing to attempt the detection and verification of line emission in very faint sources.

\section{Discussion}

In the foregoing we have described a concerted effort to quantify the abundance of low
luminosity star forming galaxies conducted in parallel to a similar spectroscopic campaign
which has examined the abundance of $z\simeq$8-10 Lyman $\alpha$ emitters \citep{Stark07b}.
That study claimed that if even a small fraction of the candidates is truly at high redshift,
a significant contribution to reionization is provided by low luminosity galaxies. In a similar
manner, recognizing the limitations of our small samples, we now examine the luminosity
function at $z\simeq$7.5 and the possible contribution that our lensed drop-out sources may 
make to cosmic reionization. 

\subsection{Number Densities and the Source Luminosity Function}
\label{ndensity}

The intrinsic area of sky (i.e. that in the source plane) covered by the NICMOS images 
is strongly dependent on the geometry of the critical lines (or caustics), which varies from 
cluster to cluster. Furthermore, multiple images occur in the central regions, duplicating the 
corresponding source plane area. The result of both effects is a smaller survey, reduced by 
a typical factor of $\simeq$10 in the source plane, with an increased depth whose value 
varies across the field of view. 

In order to derive the source density of our $z$-band drop-outs, and to compare our results 
with those conducted in blank fields,  we used the lensing models for each cluster to compute 
the sky area effectively observed in the source plane, down to a given intrinsic AB magnitude. 
We assumed our survey covers the image plane down to the measured 5$\sigma$ depth 
$J_{110}\sim26.2$ in the central NICMOS region, and scaled according to the relative exposure 
time near the edges. We also removed $\sim10\%$ of the NICMOS area affected by bright 
galaxies. We supposed an average redshift of $z=7.5$ to compute the magnification factors.  

Errors in the magnification factors estimated from the lensing models will affect 
the source plane areas and unlensed magnitudes in an opposite way. For an individual cluster, 
the typical uncertainty is about 10\% in the resulting area. This error is even smaller for the entire 
sample of six clusters.

Our total surveyed area is a factor $\sim5$ smaller than the UDF in the same magnitude 
range ($AB<27.7$, Fig. \ref{magsource}). However, the increased depth enables to reach 
AB$\sim28-30$ in this area. A very similar result is found in the case of $J$-band drop-outs, 
assuming $z$=9.

We used the estimated color selection contamination factor, $f_{cont}$, and the selection completeness 
factor, $f_{comp}$ (Sect. \ref{dropouts} and Table \ref{magtable}), to correct each $z$ drop-out 
individually to derive intrinsic number densities. Because of the strong variations in the magnification 
factor across the NICMOS field of view, we corrected for observed completeness by computing the 
completeness factors in intrinsic (unlensed) $<J+H>$ magnitude. This combines both the observed 
completeness factor, $S_{comp}$, given in Table \ref{clusterimages} and the surface reduction in 
the source plane.  Error bars in the number densities were computed using Poisson noise estimates. 
We present the cumulative UV luminosity function of the $z$-band drop-outs in the magnitude range 
$27.0<AB<30.0$ in Fig. \ref{ncounts}. 

As discussed, it is likely that 5 out of our 10 sources are at high redshift. Accordingly, in Fig. \ref{ncounts},
we randomly selected 100 $\times$ 5 sources from our sample to take into account object-to-object variations
in the magnification factor and used this to estimate more realistic error bars. 
Errors in the individual magnification factors vary between 0.05 and 0.2 mag. When compared
with the luminosity bins we used (0.5 mags), this effect has very low significance on the results, and the 
errors are mainly recovered when randomly choosing 5 candidates from the sample.

For comparison, we overplot in Fig. \ref{ncounts} the luminosity function found by 
\citet{Bouwens06} in the UDF, including redshift evolution between $z=6.0$ and $z=7.5$ 
 assuming the observed size scaling as $(1+z)^{-1}$ for fixed luminosity \citep{Ferguson}. 
Likewise, we overplot the best Schechter function fits recently claimed  by \citet{Bouwens08} from an analysis of 
their sample of $z$-band drop-outs. Not surprisingly, there is no overlap between these blank field 
measures and our, much deeper, lensed survey. All that can be said is that our results, which probe 
more than $\simeq$2.5 magnitudes fainter are marginally consistent down to AB$\sim28.5$, and 
higher by $\sim0.3-0.6$ dex at fainter luminosities. 

\subsection{Contribution to Cosmic Reionization}
\label{reiontext}

We finally investigate whether the likely abundance of low luminosity sources found in our survey 
could make a significant contribution to cosmic reionization. The approach we use is somewhat 
similar to the one described by \citet{Stark07b}, who estimated the comoving number density of 
sources necessary to keep the intergalactic medium (IGM) reionized under reasonable assumptions, 
and compare those to the abundances derived from candidate Lyman-$\alpha$ emitters at high redshift. 

In our case, we can estimate the star formation rate density, measured in individual objects 
from their UV rest-frame luminosity, after applying the same completeness corrections described in 
Sect. \ref{ndensity}. We converted the intrinsic (unlensed) $J+H$ magnitudes of our candidates into a 
UV luminosity, $L_{1500}$, and infer the related star formation rate (SFR) by adopting the 
\citet{Kennicutt} calibration. All $z$-band drop-outs span the range SFR$\sim$0.1-1.0 
$M_\odot\ \rm{yr}^{-1}$, thus the overall star-formation rate observed yields the contribution of 
low star-formation rate sources to the entire star-formation rate density $\rho_{\rm{SFR}}$. 

We used the \citet{Madau99} formalism to estimate the amount of star-formation necessary to 
keep the IGM reionized at a given redshift. One important factor in this calculation, that would 
modify the efficiency of star-forming sources to reionize the IGM, is the $H_{II}$ clumping factor 
$C$, defined as $C=<n^{2}_{HII}>/<n_{HII}>^2$ with $N_{HII}$ being the density of ionized hydrogen. 
This factor measures the inhomogeneity of ionized hydrogen in the IGM which will likely increase
between $z=10$ and $z=6$ due to the growth of structure. Assuming an IMF with a Salpeter slope 
with stellar masses ranging from 1 M$_\odot$ to 100 M$_\odot$, and a solar metallicity $Z$=0.02  
, the photon budget from star-forming sources necessary to reionize the IGM can be 
written as:

\begin{eqnarray}
%\rm n_{gal} =  2~\Big(\frac{B}{10}\Big) \left({\rm~n_H \over 10^{-7} \rm~cm^{-3}}\right) 
%\Big(\frac{f_{esc}}{0.05}\Big)^{-1} \Big(\frac{\dot{M_\star}}{0.1~M_\odot~yr^{-1}}\Big)^{-1} \rm \Big(\frac{\Delta t}{575~Myr}\Big)^{-1} ~Mpc^{-3}
\dot{\rho}_{SFR}\simeq (0.031 {\rm M}_\odot\ {\rm yr}^{-1}\ {\rm Mpc}^{-3})\Big(\frac{f_{esc,rel}}{0.5}\Big)^{-1}\ \Big(\frac{C}{10}\Big) \Big(\frac{1+z}{8.5}\Big)^{3} 
\label{madau}
\end{eqnarray}

where f$_{esc,rel}$ is the escape fraction of ionizing photons. We assumed an escape fraction 
f$_{esc,rel}$=0.5 in our calculations. However, values as low as f$_{esc,rel}$=0.02 have been measured 
in $z\sim3$ galaxies by \citet{Shapley06}. Lower escape fractions would increase the amount of 
star formation necessary to reionize the Universe, so adopting  f$_{esc,rel}$=0.5 gives us a lower limit 
on $\rho_{SFR}$. On the other hand, top-heavy IMFs and differences in metallicities would make 
galaxies produce more ionizing photons per star-formation rate, but this effect is less significant 
than variations in f$_{esc,rel}$ and $C$. \citet{Bolton07} have critically reviewed possible values for 
the clumping factor $C$. Many authors \citep{Bunker,Bouwens06} have assumed $C=30$, but 
much lower values are predicted from radiative transfer simulations (\citealt{Iliev} find $C<2$ at $z>11$). 

Fig.\ \ref{reion} illustrates the star formation rate densities obtained by integrating down to a given 
SFR for the two extreme luminosity functions derived by \citet{Bouwens08} when fitting their number 
densities at higher luminosities (equivalent to SFR$>1.0$ M$_\odot$\ yr$^{-1}$). We overplot on 
this figure the contributions derived from Eq. \ref{madau} with clumping factor varying between 
$C=2$ and $C=30$.

As suggested before, luminous galaxies do not produce enough star formation to reionize the IGM
at these redshifts, even when a low $C=2$ clumping factor is assumed. Our sample enables us 
to test whether lower luminosity galaxies help to solve this discrepancy.
Combining the source density from our NICMOS survey with
those claimed at SFR$>1.0$ M$_\odot$ estimated by integrating the luminosity function from 
\citet{Bouwens08}, we can compare the photon budget down to SFR$\sim0.1$ M$_\odot$\ yr$^{-1}$ 
with the amount of star-formation rate necessary to reionize the IGM. The results, shown in the case of 
5 objects randomly chosen from our sample of 10 $z$-dropouts, suggest a contribution compatible with that 
necessary for reionization for clumping factors in the range $2<C<10$.

\section{Conclusions}

The overall goal of this project has been to constrain the abundance of low-luminosity 
star forming galaxies at $z\sim7-10$, selected as $z$- and $J$-band drop-outs in the fields of 6 
lensing clusters observed with ACS and NICMOS onboard the Hubble Space Telescope, and the 
IRAC camera onboard the Spitzer Space Telescope. We summarize our results as follows:

\begin{enumerate}
\item{We have identified 12 high redshift candidates (10 $z$-band drop-outs and 2 $J$-band 
drop-outs) according to carefully-determined photometric selection criteria. These are located in 5 of 
the 6 clusters and span the observed magnitude range $J_{110}\simeq$25-26. Each is typically magnified 
by 1.5 to 4 magnitudes.}

\item{Based on a comprehensive set of tests, we estimate the fraction of sources that might represent
spurious detections and the extent to which low mass stars and low redshift interlopers may contaminate
our sample. Collectively, these tests suggest that around 5 of of our 10 $z$ drop-outs are possible high redshift $z>$7 objects. }

\item{By stacking the available ACS, NICMOS, ground-based $K$-band and IRAC images, we 
investigated further the averaged properties of our lensed $z$ drop-outs. We find a UV spectral 
slope $\beta\sim2.8^{+0.05}_{-0.2}$ similar to that of higher luminosity candidates from the 
UDF. Such a slope suggests a very young stellar population with little reddening and strengthens
our case that the bulk of our candidates are high redshift sources.}

\item{We searched for possible counter-images for our candidates based on the most recent lensing 
models for each cluster. Unfortunately, our results are inconclusive. Many of the counter-images either 
lie outside our NICMOS field or are close to foreground sources. We fail to detect a counter-image in 
two apparently clean cases but a further two sources may be potentially merging on the critical line.}

\item{We undertook follow-up spectroscopy with NIRSPEC for 7 of our 10 $z$-band drop-out candidates 
in the hope of seeing confirmatory Lyman $\alpha$ emission. No emission was found in any candidate
(or its counter-image location) to a flux limit corresponding to $3\times10^{-18}$ ergs\ s$^{-1}$\ cm$^{-2}$ 
in the clean part of the OH spectrum. One explanation is possible evolution in the Lyman-$\alpha$ 
rest-frame equivalent width distribution, compared to previous results by \citet{Stanway07} at $z\sim6$, 
such as might be expected if the neutral fraction rises with redshift. Such a deduction would be difficult
to reconcile with the presence of intrinsically-faint lensed Lyman $\alpha$ emitters at $z\simeq$10 
\citealt{Stark07b}}

\item{Our inferred luminosity function at $z\sim7.5$, after correcting for contamination and incompleteness,
is marginally consistent with an extrapolation of available constraints at brighter luminosities, with a slightly 
higher normalization by 0.3-0.6 dex. If even a modest fraction of our sources are at high redshift,
our results strengthen the suggestion that sources with star formation rates $\sim0.1-1.0$ 
M$_\odot\rm{yr}^{-1}$ contribute significantly to cosmic reionization.}

\end{enumerate}

As we approach the era of JWST and the ELTs, the outcome of our project in lensed fields has been 
to provide new constraints on the faint part of the luminosity function at $z>7$, which confirmed 
the trends seen at higher luminosities. Despite being restricted to a small field after demagnification in the 
source plane, we expect that these results will be readily confirmed by the upcoming Wide Field 
Camera (WFC3) on HST, combining extremely sensitive infrared channels with a field of view much 
larger than NICMOS.

\acknowledgments

We thank the anonymous referee for a report that has improved the content of this paper, and Rychard Bouwens 
for his very helpful comments on an earlier version of the manuscript. We
also acknowledge discussions with Rodger Thompson, Elizabeth Stanway, Roser Pell\'o, Daniel Schaerer, 
Kelle Cruz and Adam Kraus. Andrew Blain and Naveen Reddy kindly provided the Keck NIRC observations of 
Abell 2218. We are thankful to Ichi Tanaka for his support in the reduction of MOIRCS imaging data. We 
acknowledge funding from NASA grant HST-GO-10504.01-A and Spitzer program GO-20439. The authors recognize 
and acknowledge the very significant cultural role and reverance that the  summit of Mauna Kea has always 
had within the indigenous Hawaiian community.  We are most fortunate to have the opportunity to conduct 
observations from this mountain. This program is based on observations made with the NASA/ESA Hubble Space 
Telescope, which is operated by the Association of Universities for Research in Astronomy, Inc., under NASA 
contract NAS 5-26555, the Subaru Telescope, which is operated by the National Astronomical Observatory of Japan, 
the Spitzer Space Telescope, which is operated by the Jet Propulsion Laboratory, California Institute of Technology under a contract with NASA, 
and the Canada-France-Hawaii Telescope (CFHT),  which is operated by the National 
Research Council of Canada, the Institut National des Sciences de l'Univers of the Centre National de 
la Recherche Scientifique of France, and the University of Hawaii.

\bibliography{references}

%---------------------  TABLES  -------------------------------------------------------------

\newpage

\begin{table*}
\begin{tabular}{llllll}
Cluster & RA & Dec & $z$ & Mass model & $N_{mult} (N_z)$ \\
\tableline
\tableline
Abell 2218  & 248.95625 & 66.21444 & $0.176$& \citet{ardis}   & 37 (26)\\
Abell 2219  & 250.08541 & 46.70833 & $0.226$& \citet{Smith05} & 14 (6)\\
Abell 2390  & 328.40086 & 17.69603 & $0.228$& \citet{Swinbank}& 11 (5)\\
Abell 2667  & 357.91387 &-26.08541 & $0.233$& \citet{Covone}  & 10 (5)\\
Cl0024+16   &   6.65122 & 17.16060 & $0.390$& \citet{Kneib03} & 9  (5)\\
Cl1358+62   & 209.96069 & 62.51808 & $0.328$& \citet{Franx}   & 5  (3)\\
\tableline
\end{tabular}
\caption{\label{cluster} Lensing Cluster Sample: $N_{mult}$: number of multiple images used 
in the lens model ($N_z$ number with spectroscopic redshifts)
}
\end{table*}

\newpage
\begin{deluxetable}{llllrll}
\tabletypesize{\scriptsize}
\tablecaption{
Imaging Data:. For a given cluster each entry presents 
the instrument, filter, HST/Spitzer program ID, date of observation, exposure time, and 
final image quality (depth and completeness).
 }

\tablehead{
\colhead{Cluster} &
\colhead{Filter} &
\colhead{Program} &
\colhead{Date} &
\colhead{Exp. time} &
\colhead{Depth (5$\sigma$)} &
\colhead{Completeness} 
}
\startdata

A2218 & ACS$_{435}$ & 9717 & Aug04 & 7048 & 27.74 & \\
& ACS$_{475}$ & 10325& Aug04 & 5640 & 27.95 & \\
& ACS$_{555}$ & 9717 & Aug04 & 7048 & 27.79 & \\
& ACS$_{625}$ & 9717 & Aug04 & 8386 & 27.93 &\\
& ACS$_{775}$ & 10325& Aug04 & 9285 & 27.73 &\\
& ACS$_{850}$ & 9292/9452/10325& Apr02/Aug02/Aug04& 19630 & 27.32 &\\
& NIC3$_{110}$ & 9452/10504 & Apr03/Dec05 & 8446 & 26.26 & 25.93\\
& NIC3$_{160}$ & 9452/10504 & Apr03/Dec05 & 10559 & 26.76 & 26.07\\
& NIRC$_{K}$ & & Jul02& 7200   & 25.5 &\\
& NIRC$_{K}$ & & Jul06& 13620  & 25.9 &\\
& MOIRCS$_{K}$ & & May07 & 18000 & 26.1 & \\
& IRAC$_{3.6}$ & 83/20439 & Dec03/Oct05/Dec05& 37700 & 25.5 &\\
& IRAC$_{4.5}$ & 83/20439 & Dec03/Oct05/Dec05& 37700 & 25.5 &\\
\tableline
A2219 & WFPC2$_{702}$ & 6488 & Aug99 & 14400 & 27.00\\
& ACS$_{850}$ & 10504 & Apr06 & 8374 & 26.75 & \\
& NIC3$_{110}$ & 10504 & May06/Jun06/Jun07 & 9216 & 26.23 & 25.97\\
& NIC3$_{160}$ & 10504 & May06/Jun06/Jun07 & 11519 & 26.73 & 26.14\\
& NIRC$_{K}$ & & Jul06& 22980 & 26.3 & \\
& MOIRCS$_{K}$ & & Aug06 & 17550 & 26.1 & \\
& IRAC$_{3.6}$ & 83 & Feb04/Mar05 & 2400 & 23.9 & \\
& IRAC$_{4.5}$ & 83 & Feb04/Mar05 & 2400 & 23.9 & \\
\tableline
A2390 & WFPC2$_{555}$ & 5352 & Dec94 & 8400 &26.6\\
& WFPC2$_{814}$ & 5352 & Dec94 & 8400 &26.2\\
& ACS$_{850}$ & 9292/10504 & May02/May06 & 8847 & 26.82 & \\
& NIC3$_{110}$ & 10504 & Jun06/Jul06/Jun07& 9470 & 26.27 & 25.89\\
& NIC3$_{160}$ & 10504 & Jun06/Jul06/Jun07& 11839 & 26.54 & 26.06\\
& MOIRCS$_{K}$ & & May07 & 15900 & 26.0 & \\
& IRAC$_{3.6}$ & 83 & Jun04/Nov04 & 2400 & 23.9 &\\
& IRAC$_{4.5}$ & 83 & Jun04/Nov04 & 2400 & 23.9 &\\
\tableline
A2667 & WFPC2$_{450}$ & 8882 & Oct01 & 9600 & 26.26 \\
& WFPC2$_{606}$ & 8882 & Oct01 & 4000 & 26.94 \\
& WFPC2$_{814}$ & 8882 & Oct01 & 4000 & 26.11 \\
& ACS$_{850}$ & 10504 & Jul06 & 8765 & 26.70\\
& NIC3$_{110}$ & 10504 & Aug06 & 9343 & 26.22 & 25.93\\
& NIC3$_{160}$ & 10504 & Aug06 & 11711& 26.51 & 26.01\\
& IRAC$_{3.6}$ & 83 & Dec03 & 2400 & 23.9 &\\
& IRAC$_{4.5}$ & 83 & Dec03 & 2400 & 23.9 &\\
\tableline
CL0024 & ACS$_{435}$ & 10325 & Nov04 & 6435 & 27.67\\
& ACS$_{475}$ & 10325 & Nov04 & 5072 & 27.81\\
& ACS$_{555}$ & 10325 & Nov04 & 5072 & 27.47\\
& ACS$_{625}$ & 10325 & Nov04 & 8971 & 27.75\\
& ACS$_{775}$ & 10325 & Nov04 & 10144 & 27.67\\
& ACS$_{850}$ & 10325 & Nov04 & 16328 & 27.28\\
& NIC3$_{110}$ & 10504 & Aug06 & 9472 &  26.20 & 25.9 \\
& NIC3$_{160}$ & 10504 & Aug06 & 11840 & 26.60 & 26.0 \\
& MOIRCS$_K$ & & Aug06 & 21600 & 26.1 & \\
& IRAC$_{3.6}$ & 64 & Dec03 & 2400 & 23.9 & \\
& IRAC$_{4.5}$ & 64 & Dec03 & 3600 & 24.1 & \\
\tableline
CL1358& ACS$_{435}$ & 9717 & Apr04/May04& 5440 & 27.70\\
& ACS$_{475}$ & 9717 & Apr04/May04& 5470 & 27.96\\
& ACS$_{625}$ & 9717 & Apr04/May04& 6800 & 27.77\\
& ACS$_{775}$ & 9717 & Apr04/May04& 10144 & 27.49\\
& ACS$_{850}$ & 9717 & Apr04/May04& 16328 & 27.13\\
& NIC3$_{110}$ & 10504 & Dec05 & 9216 & 26.34 & 25.92\\
& NIC3$_{160}$ & 10504 & Dec05 & 11519& 26.60 & 26.10\\
& IRAC$_{3.6}$ & 83 & Jan04/Jun05& 2400 & 23.9 & \\
& IRAC$_{4.5}$ & 83 & Jan04/Jun05& 2400 & 23.9 & \\
\tableline
\enddata
\label{clusterimages}
\end{deluxetable}

%\begin{table*}
%\begin{tabular}{lccc}
%Cluster & Filter & Exposure Time & Seeing \\
%\tableline
%\tableline
%A2219 & B & 5400 & 1\arcsec.0 \\
% & R & 6300 & 0\arcsec.8 \\
%& I & 3000 & 0\arcsec.8 \\
%\tableline
%A2390 & B & 2799 & 1\arcsec.1\\
%& R & 5700 & 0\arcsec.7\\
%& I & 3600 & 0\arcsec.9 \\
%\tableline
%\end{tabular}
%\caption{\label{otherim}Additional CFHT optical data \citep{Bardeau}}
%\end{table*}

\begin{table*}[ht]
\footnotesize 
\begin{tabular}{lllccccccccc}
Candidate & R.A. & Dec. & $z_{850}$ (AB) & $J_{110W} (AB)$ & $H_{160W} (AB)$ & K (AB) & $\mu (mags)$ & $S_{comp}$ & $f_{cont}$ & $f_{comp}$ \\
\tableline
\smallskip
A2218-z1  & 248.9713 & +66.2071 & $>27.32$      & 26.1$\pm$0.13 & 25.9$\pm$0.11 & $>25.7$ & 1.9 &  0.59 & 0.14 & 0.66 \\
A2218-z2  & 248.9507 & +66.2150 & $>27.32$      & 26.2$\pm$0.18 & 26.0$\pm$0.11 & $>25.7$ & 2.7 &  0.50 & 0.19 & 0.72\\
A2219-z1  & 250.0803 & +46.7071 & $26.3\pm0.15$ & 24.7$\pm$0.05 & 25.3$\pm$0.06 & $>25.7$ & 3.6 &  0.84 & 0.0  & 0.95\\
A2390-z1  & 328.4130 & +17.6905 & $>26.82$      & 25.5$\pm$0.12 & 26.1$\pm$0.12 & $>25.6$ & 3.5 &  0.68 & 0.17 & 0.79\\
A2390-z2  & 328.4001 & +17.6962 & $>26.82$      & 25.8$\pm$0.15 & 25.8$\pm$0.10 & $>25.6$ & 1.8 &  0.65 & 0.27 & 0.51\\
A2667-z1  & 357.9119 & -26.0949 & $26.7\pm0.36$ & 25.9$\pm$0.15 & 26.1$\pm$0.18 & N/A & 1.6     &  0.59 & 0.34 & 0.51\\
A2667-z2  & 357.9153 & -26.0826 & $26.7\pm0.42$ & 25.7$\pm$0.12 & 25.6$\pm$0.11 & N/A & 2.0     &  0.75 & 0.20 & 0.68\\
CL1358-z1 & 209.9714 & +62.5128 & $>27.33$      & 26.3$\pm$0.17 & 26.1$\pm$0.12 & N/A & 1.9     &  0.43 & 0.15 & 0.63\\
CL1358-z2 & 209.9521 & +62.5108 & $>27.33$      & 26.2$\pm$0.13 & 26.7$\pm$0.28 & N/A & 4.0     &  0.43 & 0.15 & 0.72\\
CL1358-z3 & 209.9549 & +62.5187 & $>27.33$      & 26.3$\pm$0.17 & 26.6$\pm$0.19 & N/A & 4.0     &  0.35 & 0.14 & 0.66\\
\tableline
A2219-j1  & 250.0900 & +46.7040 & $>26.7$       & $>26.3$       & 25.0$\pm$0.05 & $>25.7$ & 4.0 \\
A2667-j1  & 357.9136 & -26.0869 & $>26.7$       & $>26.5$       & 25.1$\pm$0.08 & N/A & 3.6\\
\tableline
\end{tabular}
\caption{\label{magtable}High Redshift Candidate Photometry: total magnitudes and corresponding 
magnification assuming $z=7.5$ for $z$-band drop-outs and $z=9.0$ for $J$-band drop-outs. Each 
$z$ drop-out entry is followed by its observed completeness, selection contamination factor and selection
completeness. Upper limits correspond to 5 $\sigma$.}
\end{table*}

\begin{table*}
\begin{tabular}{ccccccc}
\hline\hline
Dropouts & z$_{850LP}$ & J$_{110W}$ & H$_{160W}$ & K & IRAC$_{3.6\mu \rm m}$ & IRAC$_{4.5\mu \rm m}$ \\
\hline
(all 10) & 28.59$\pm0.21$ & 25.72$\pm 0.14$ & 25.70$\pm 0.14$ & \\
(all 4 with K-band) & 28.99$\pm0.32$  & 25.71$\pm0.14$ & 25.81$\pm 0.12$ & $>26.2$ & \\
(all 6 with IRAC)   & 29.10 $\pm0.23$ & 26.11$\pm0.16$ & 26.29$\pm 0.16$ &       & $>25.0$ & $>24.8$\\
\end{tabular}
\caption{\label{stackphot}Stacked Photometry of the $z$-band drop-outs. Upper limits correspond to 5 $\sigma$.}
\end{table*}

\begin{table*}
\begin{tabular}{lccccc}
Run & Candidate(s)  & Dither & Exposure Time & Seeing & Notes\\
\tableline
\tableline
Jan 2007 & CL1358z1 and z2 & 3\arcsec &  9 $\times$ 600s& 0.5 & \\
\tableline
May 2007 & CL1358z1 and z2 & 3\arcsec & 15 $\times$ 600s& 0.5-0.6\arcsec & \\
         & A2219z1         & 5\arcsec & 13 $\times$ 600s& 0.5-0.8\arcsec & \\
         &                 &          & 12 $\times$ 600s& 0.5\arcsec     & \\
         & A2218z1         & 5\arcsec & 18 $\times$ 600s& 0.9\arcsec & candidate + counter-image\\
         &                 &          & 13 $\times$ 600s& 0.5\arcsec     & \\
         & A2390z1         & 5\arcsec &  4 $\times$ 600s& 0.8\arcsec & \\
         &                 &          & 12 $\times$ 600s& 0.5\arcsec     & \\
\tableline
Sep 2007 & A2390z2         & 8\arcsec & 21 $\times$ 600s& 0.5\arcsec & candidate + counter-image\\
         & A2667z1         & 6\arcsec & 17 $\times$ 600s& 0.5\arcsec & \\ 
\tableline
\hline\\
\end{tabular}
\caption{\label{table_spectro}
Log of the spectroscopic observations performed on 7 of the $z$ band drop-outs. From left to right: epoch of observation, candidate name, spatial dithering between individual exposures, expoure time, seeing conditions. For two $z$ drop-outs, we managed to observe the predicted location of the counter-image at the same time as the candidate.
}
\end{table*}

\begin{table*}[ht]
\begin{tabular}{ccccc}
Ref. & $z$ & $\Phi^*$ (Mpc$^{-3}$) & $M^*$ (mag) & $\alpha$ \\
\hline
\citet{Bouwens06} & $6$ & 1.4e-3 & -20.25 & -1.73 \\
\hline
\citet{Bouwens08} & $7.4$ & 1.1e-3 & -19.80 & -1.74 \\
                      &       & 1.7.8e-3& -19.60 & -1.4  \\
                      &       & 8e-4 & -19.90 & -2.0  \\
\hline
\end{tabular}
\caption{\label{lfparam} Best-fit Schechter parameters of the high redshift UV luminosity function, from earlier results found in the UDF, and overplotted in Fig. \ref{ncounts}. For each work we give the normalization $\Phi^*$, the absolute magnitude $M^*$ at the exponential cutoff and the faint-end slope $\alpha$. \citet{Bouwens08} explore three possible evolutions of the $L^*$ and $\Phi^*$ parameters for different fixed slopes $\alpha$. 
}
\end{table*}
\cleardoublepage

%---------------------  FIGURES  -------------------------------------------------------------

\clearpage

\clearpage

\begin{figure}[ht]
\begin{minipage}{11.cm}
\includegraphics[width=\textwidth]{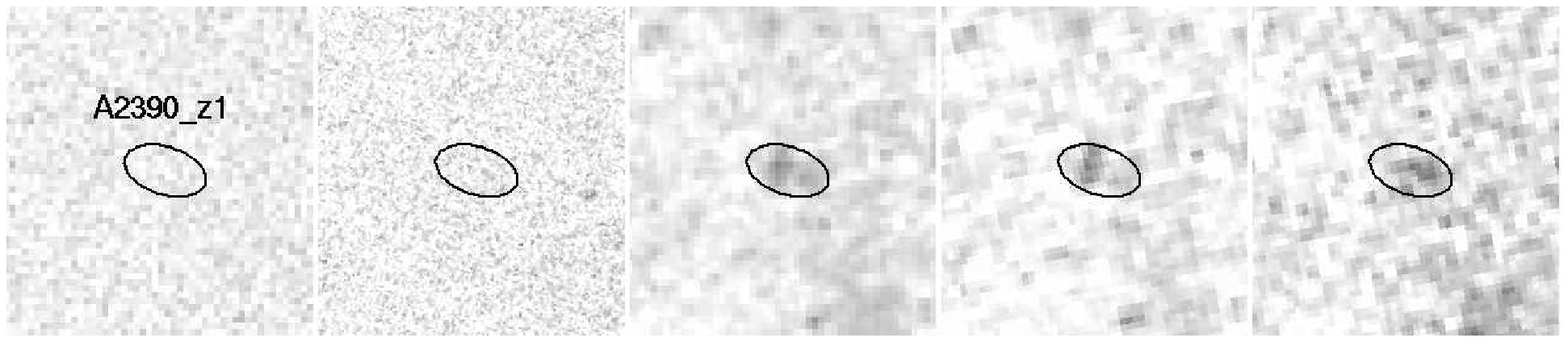}
\end{minipage}
\begin{minipage}{5.cm}
\includegraphics[width=\textwidth]{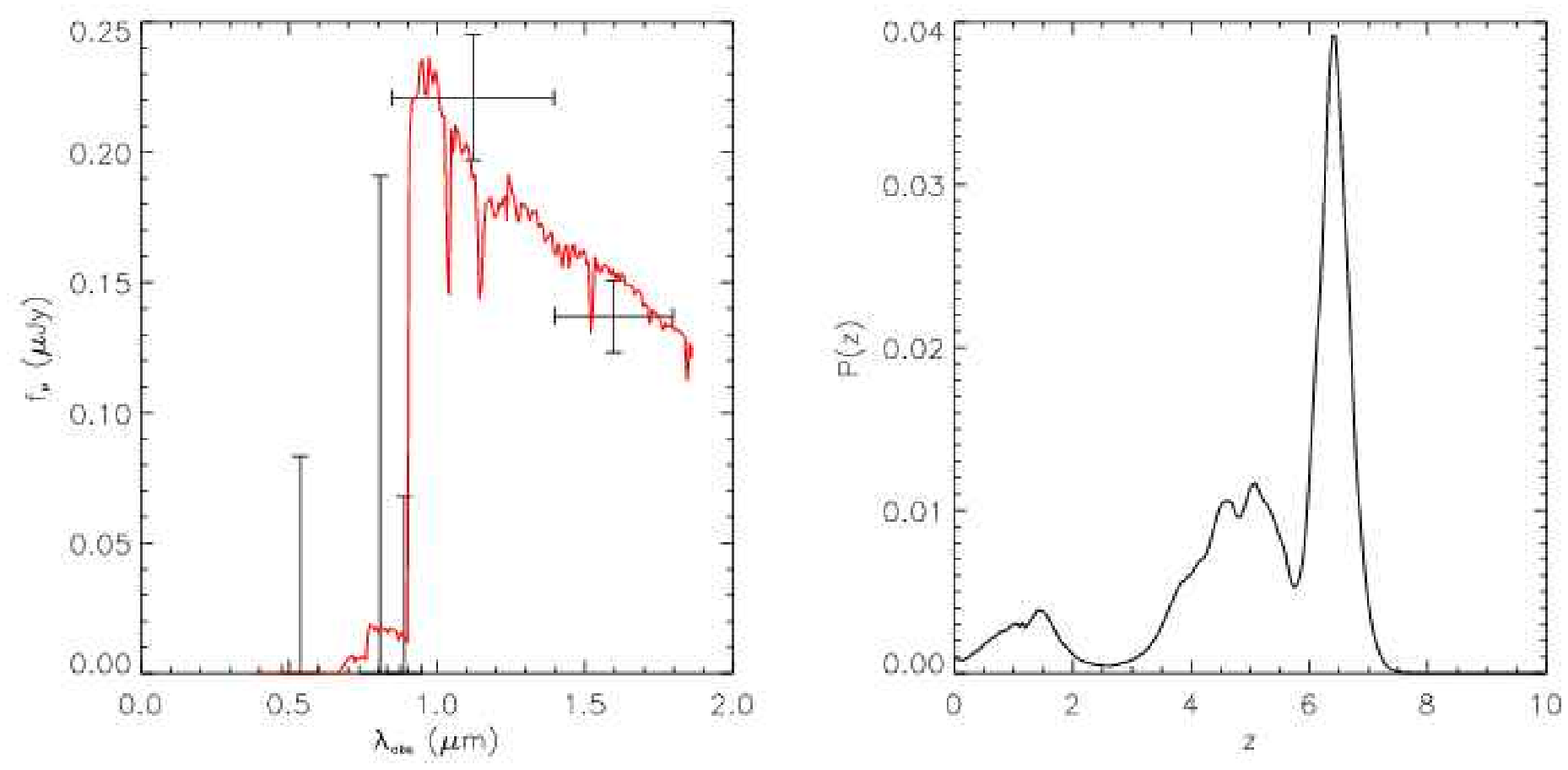}
\end{minipage}
\begin{minipage}{11.cm}
\includegraphics[width=\textwidth]{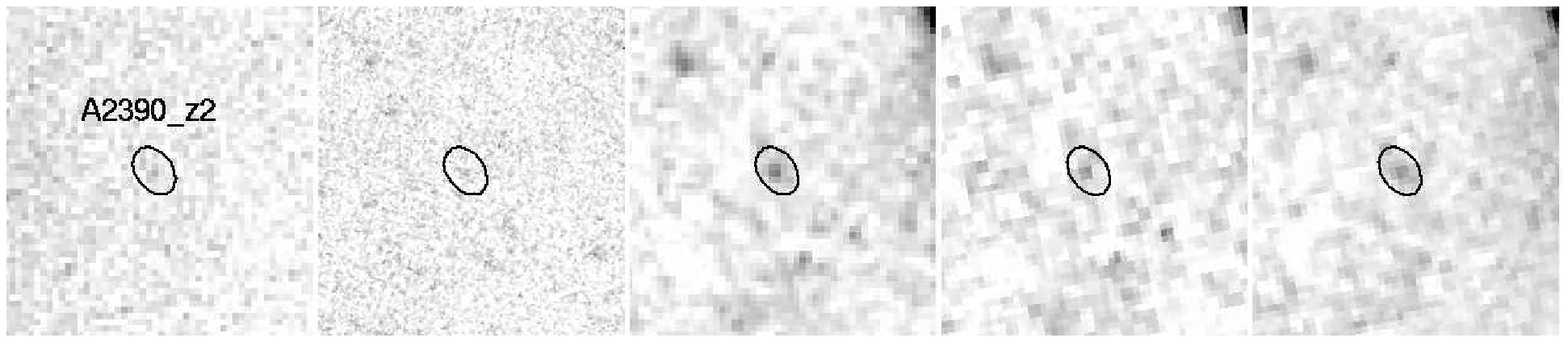}
\end{minipage}
\begin{minipage}{5.cm}
\includegraphics[width=\textwidth]{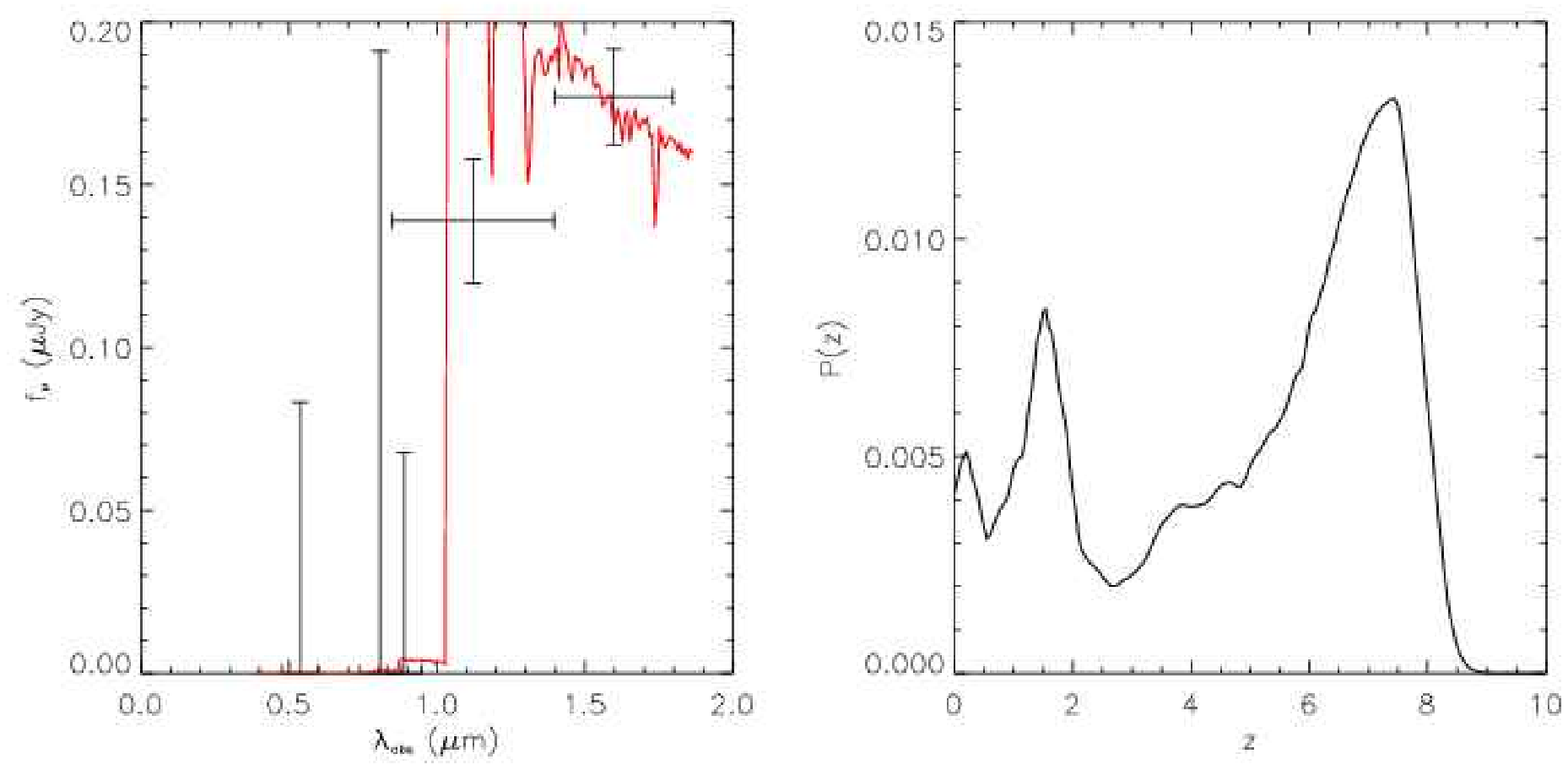}
\end{minipage}
\begin{minipage}{11.cm}
\includegraphics[width=\textwidth]{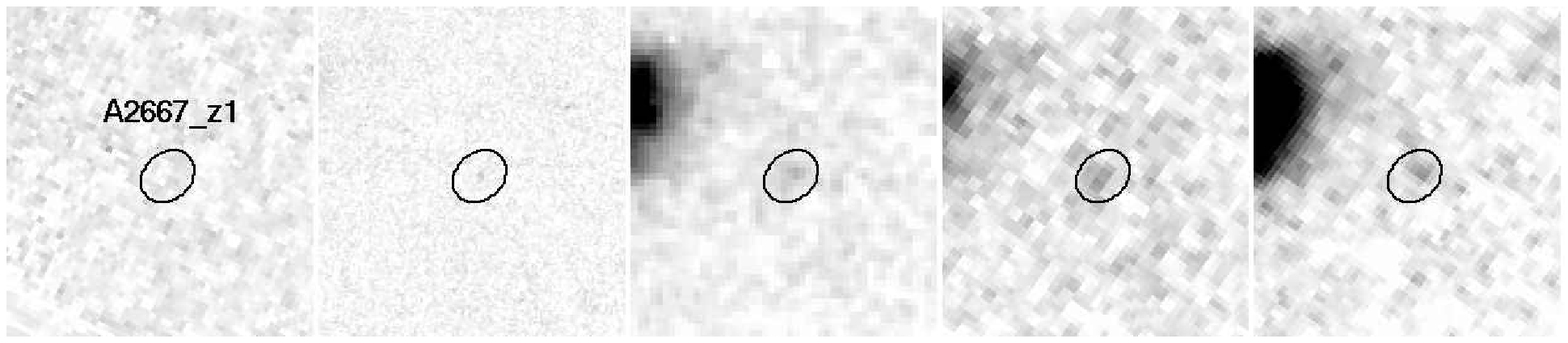}
\end{minipage}
\begin{minipage}{5.cm}
\includegraphics[width=\textwidth]{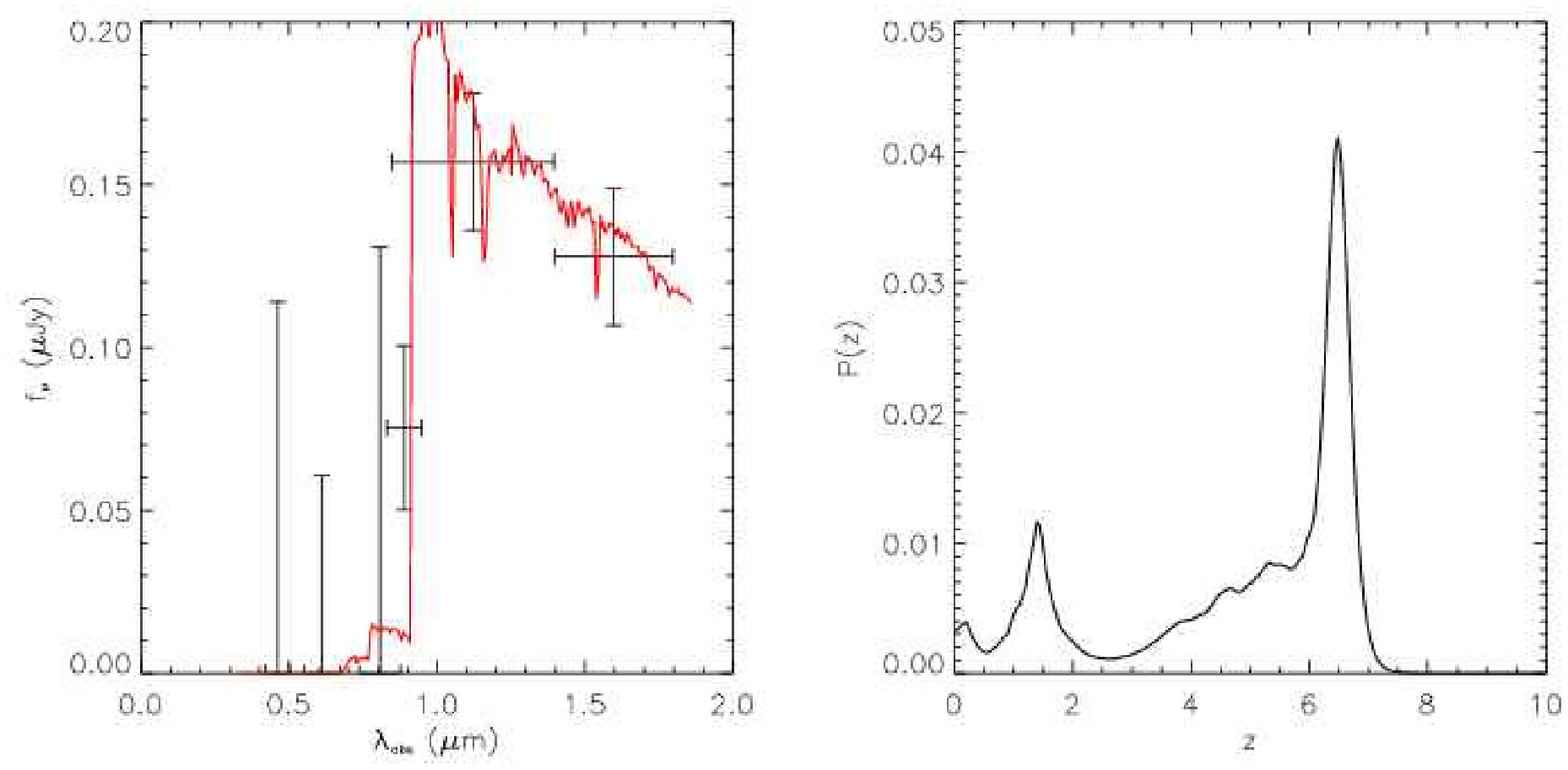}
\end{minipage}
\begin{minipage}{11.4cm}
\includegraphics[width=11cm]{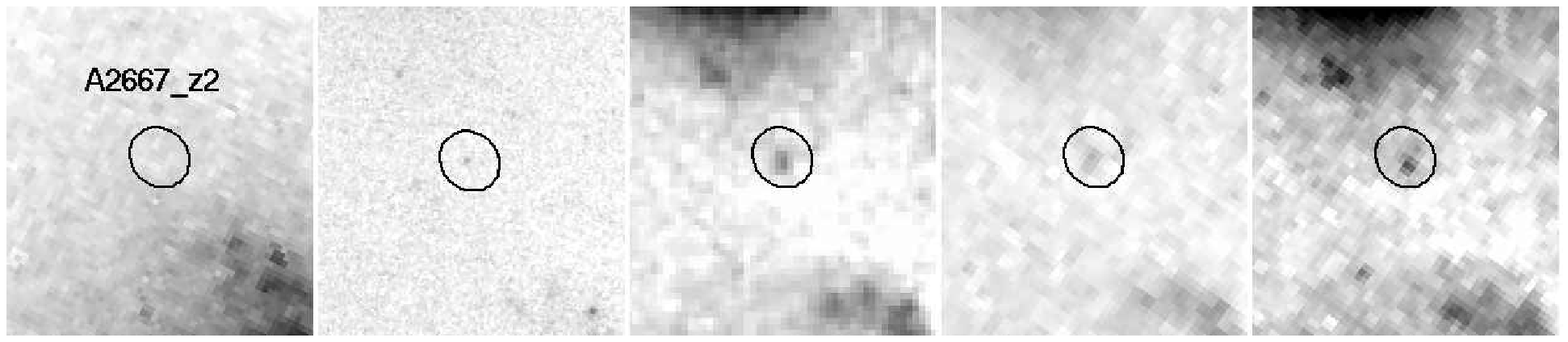}
\end{minipage}
\begin{minipage}{5.cm}
\includegraphics[width=\textwidth]{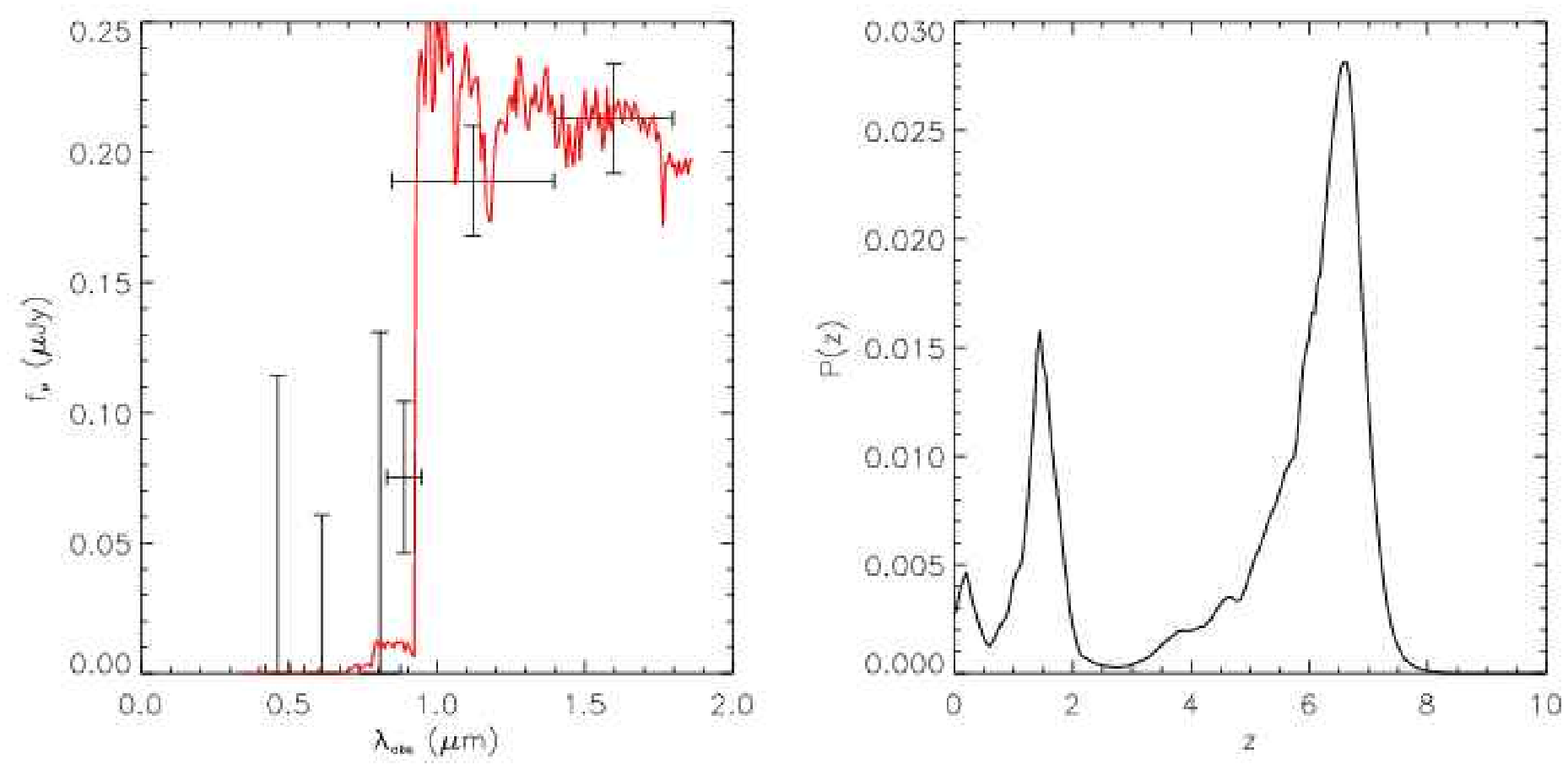}
\end{minipage}
\medskip\par
\begin{minipage}{11.cm}
\includegraphics[width=\textwidth]{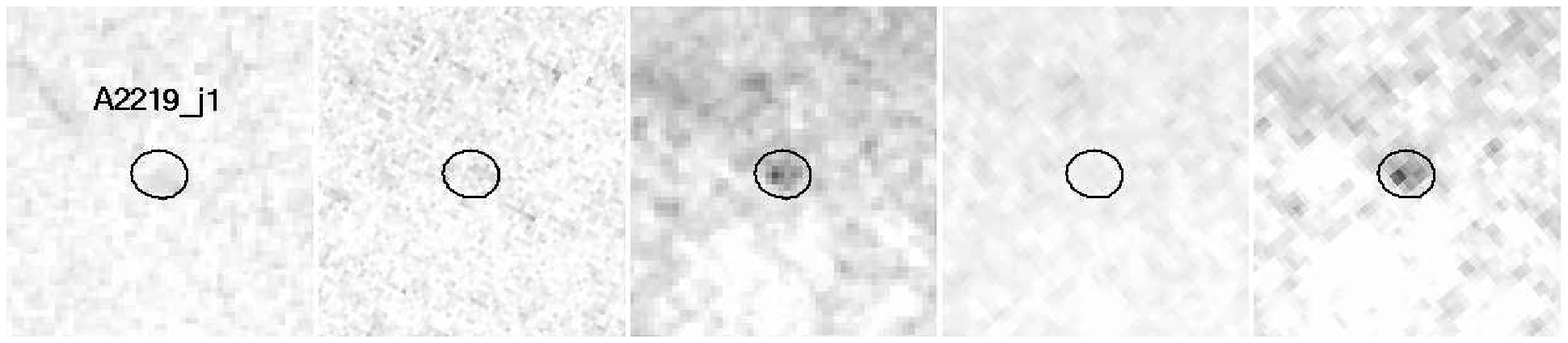}
\includegraphics[width=\textwidth]{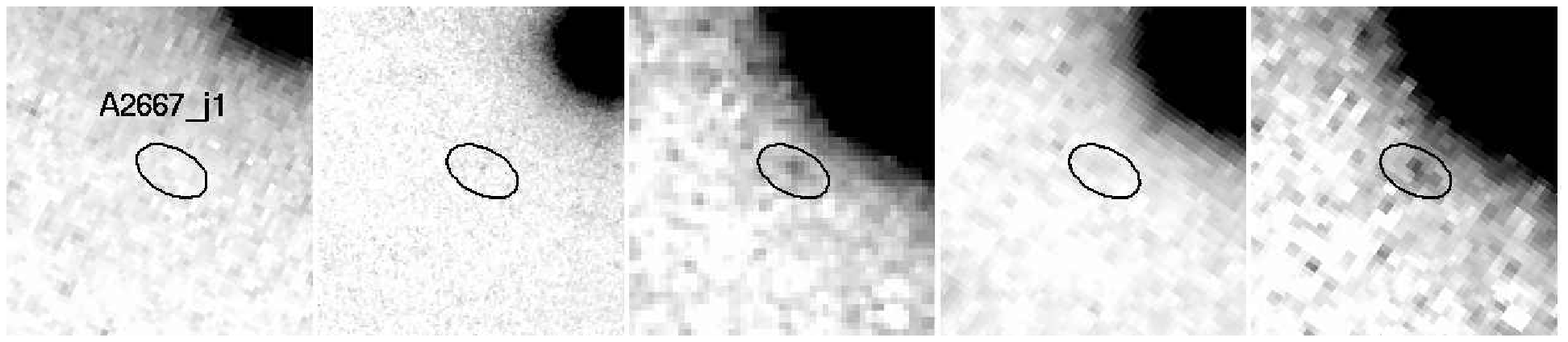}
\end{minipage}

\begin{minipage}{18.cm}
\includegraphics[width=\textwidth]{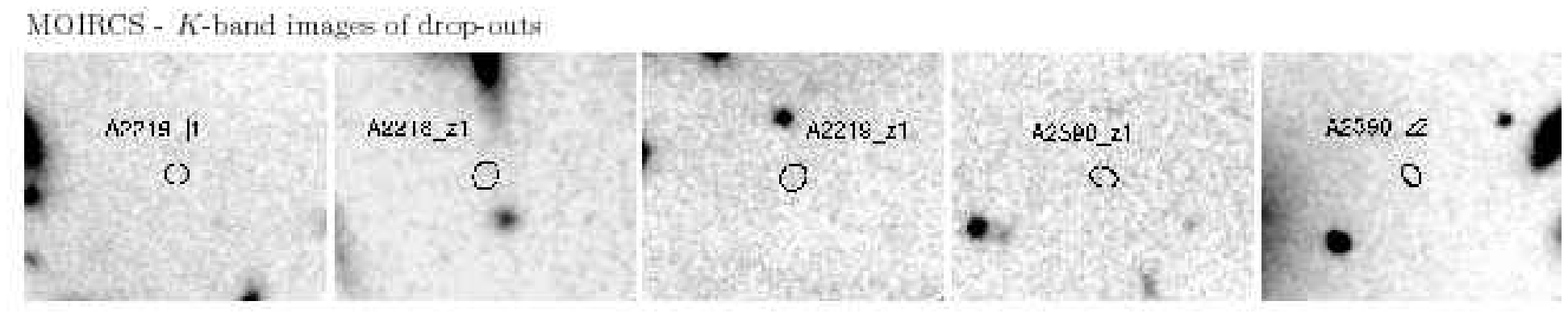}
\end{minipage}
\end{figure}
\cleardoublepage

\begin{figure}
\includegraphics[height=3.4cm]{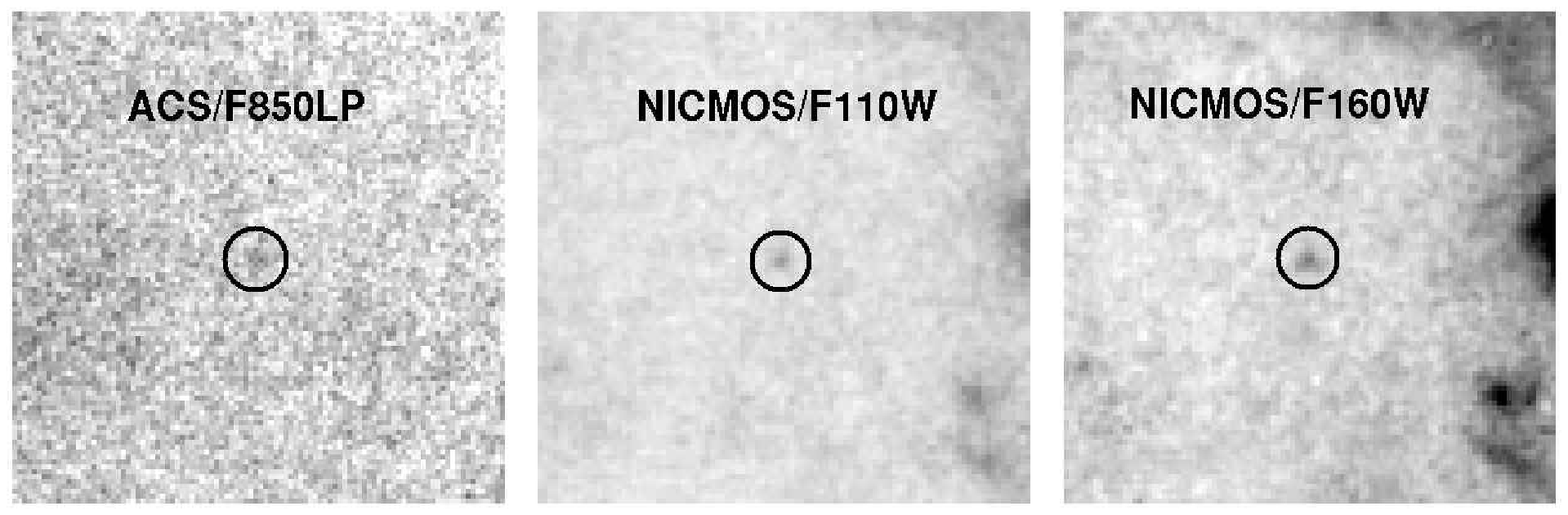}\\
\includegraphics[height=3.2cm]{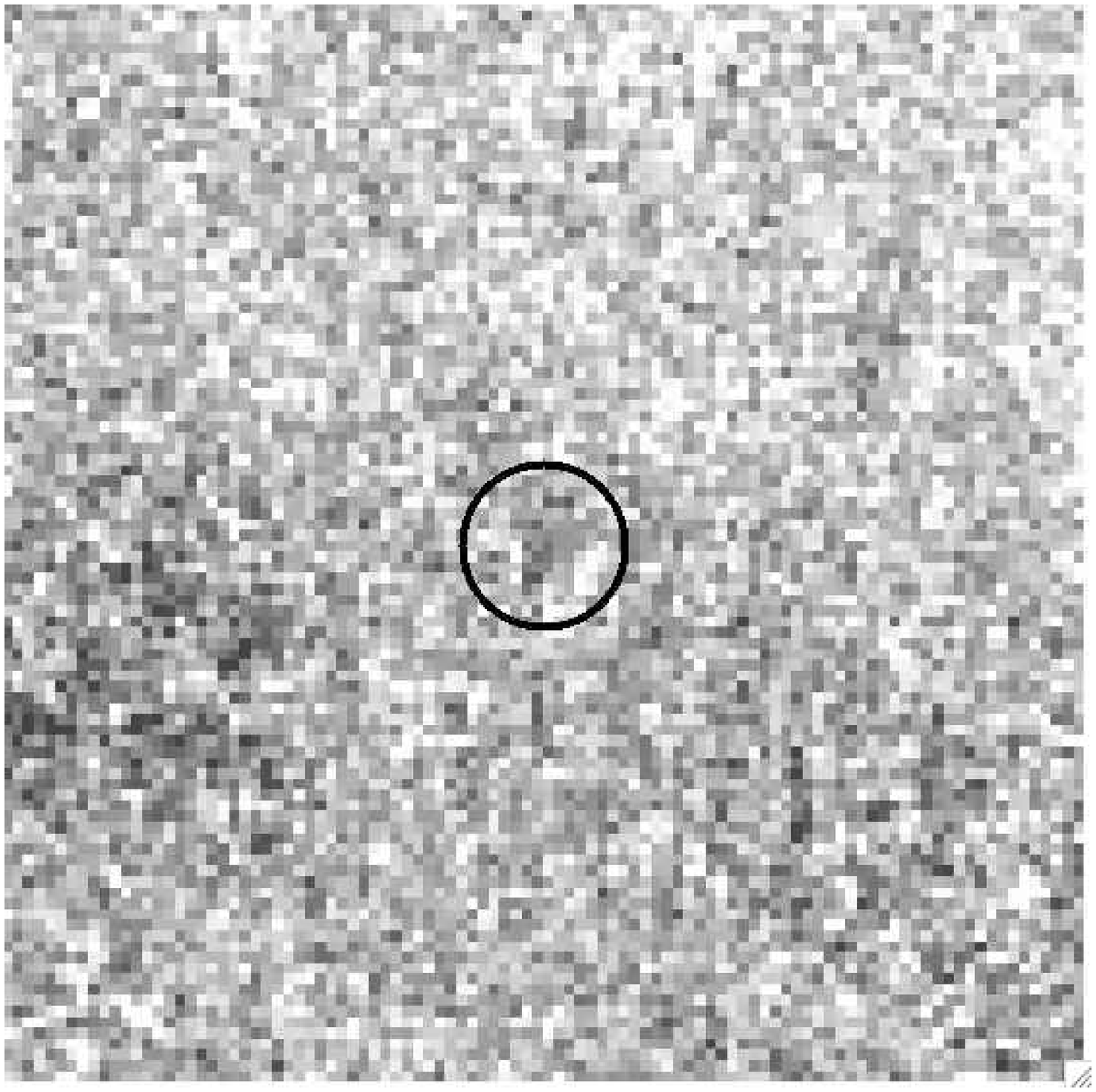}
\includegraphics[height=3.2cm]{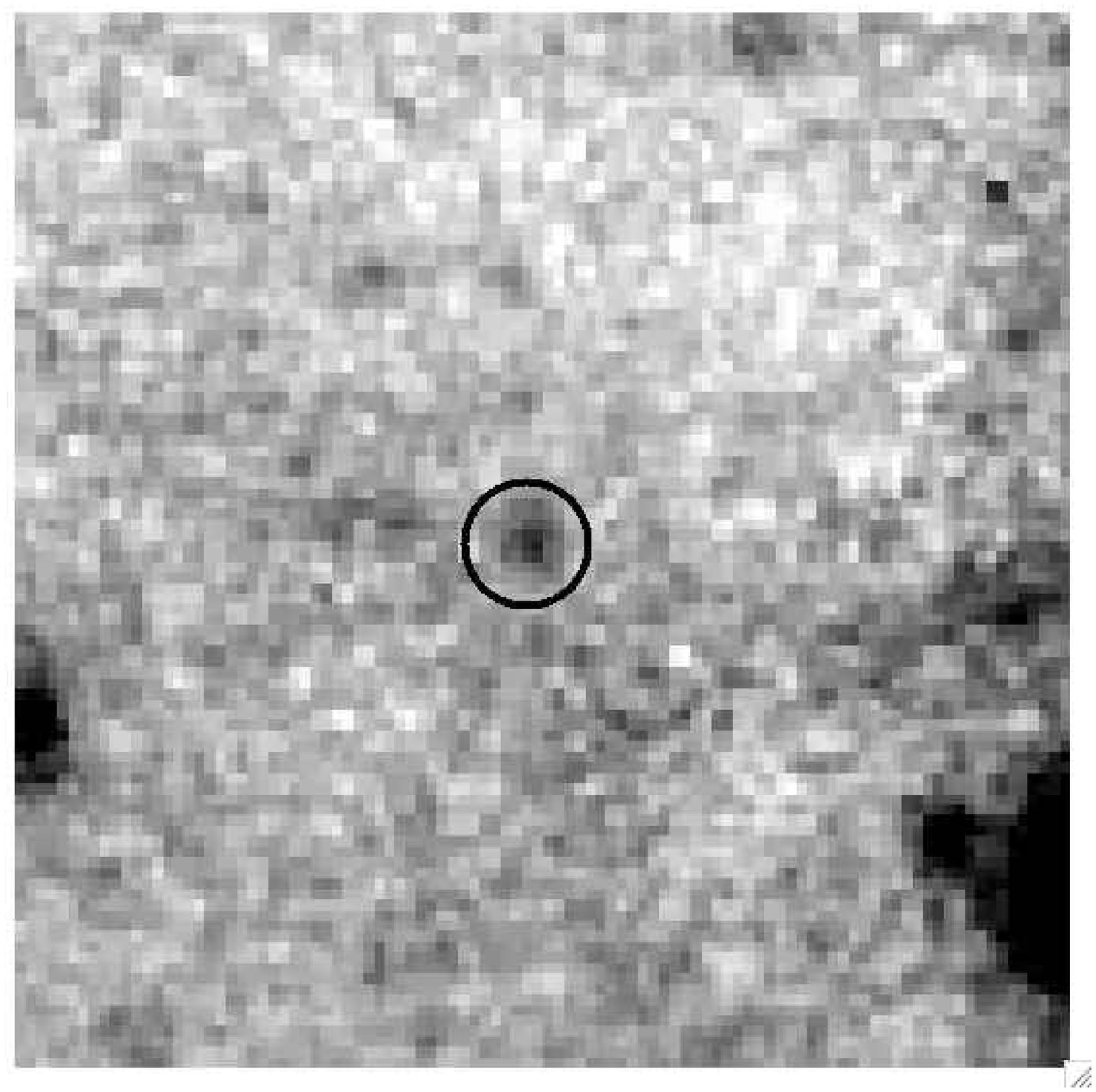}
\includegraphics[height=3.2cm]{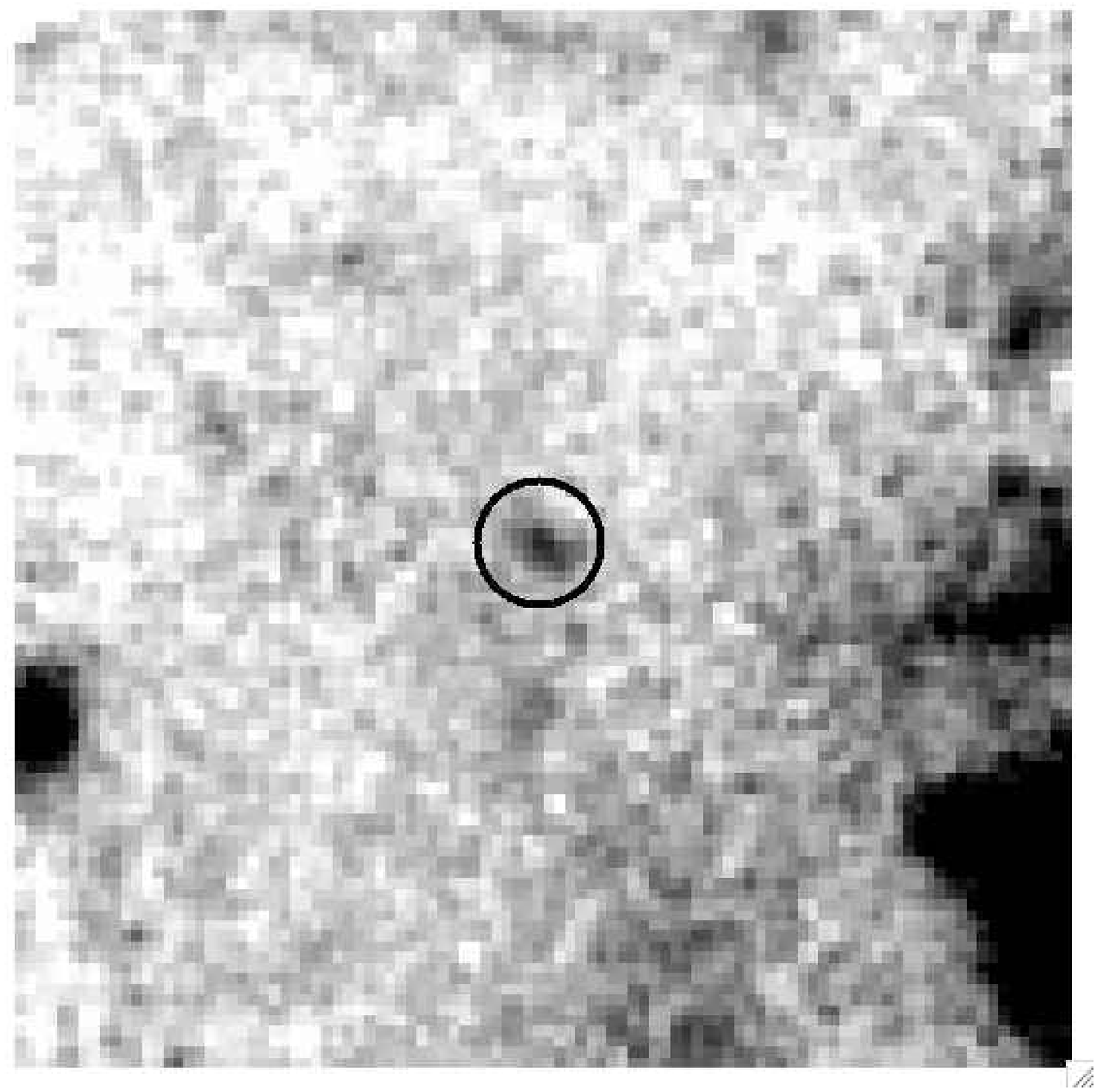}
\includegraphics[height=3.2cm]{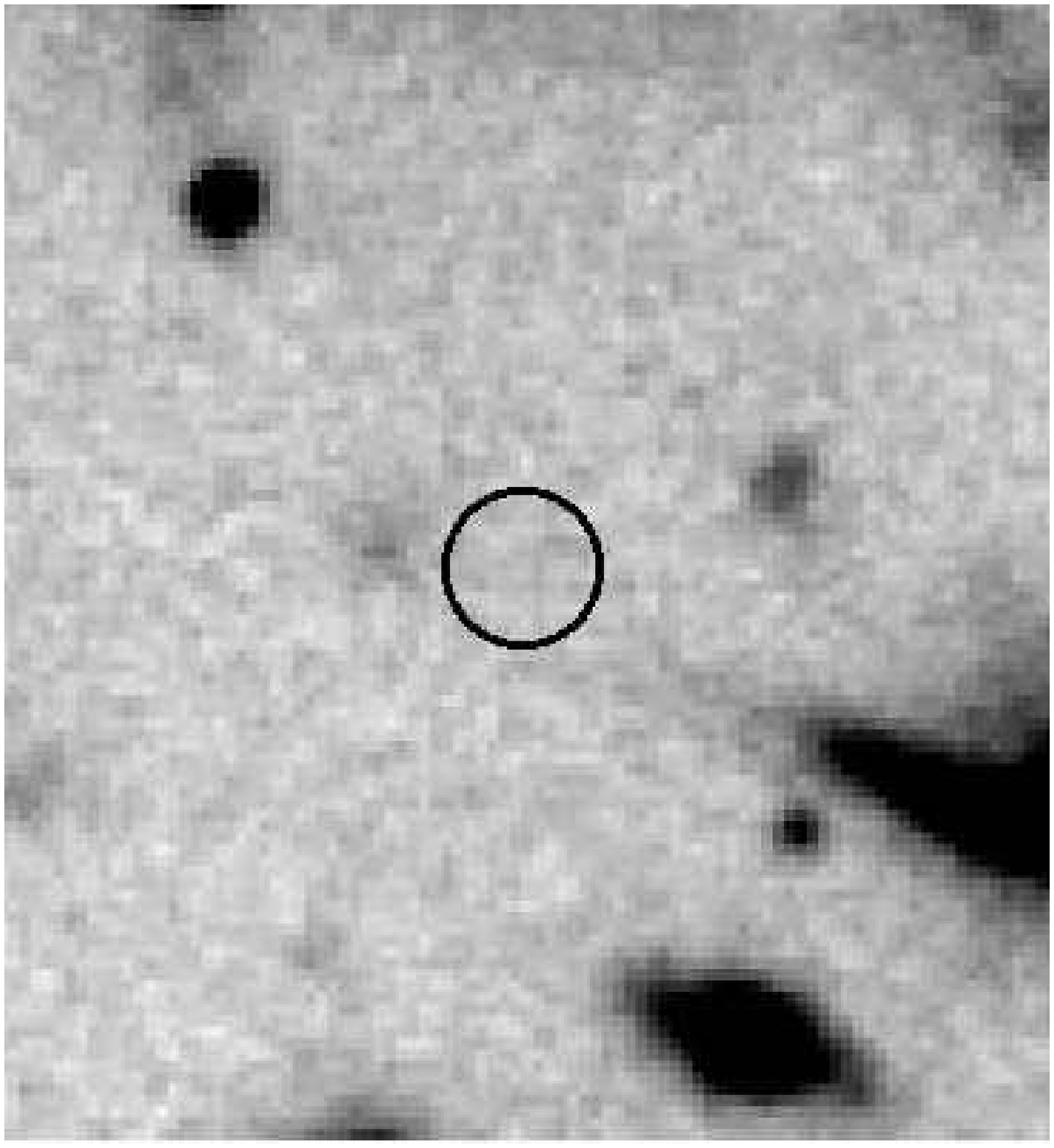}\\
\includegraphics[height=3.2cm]{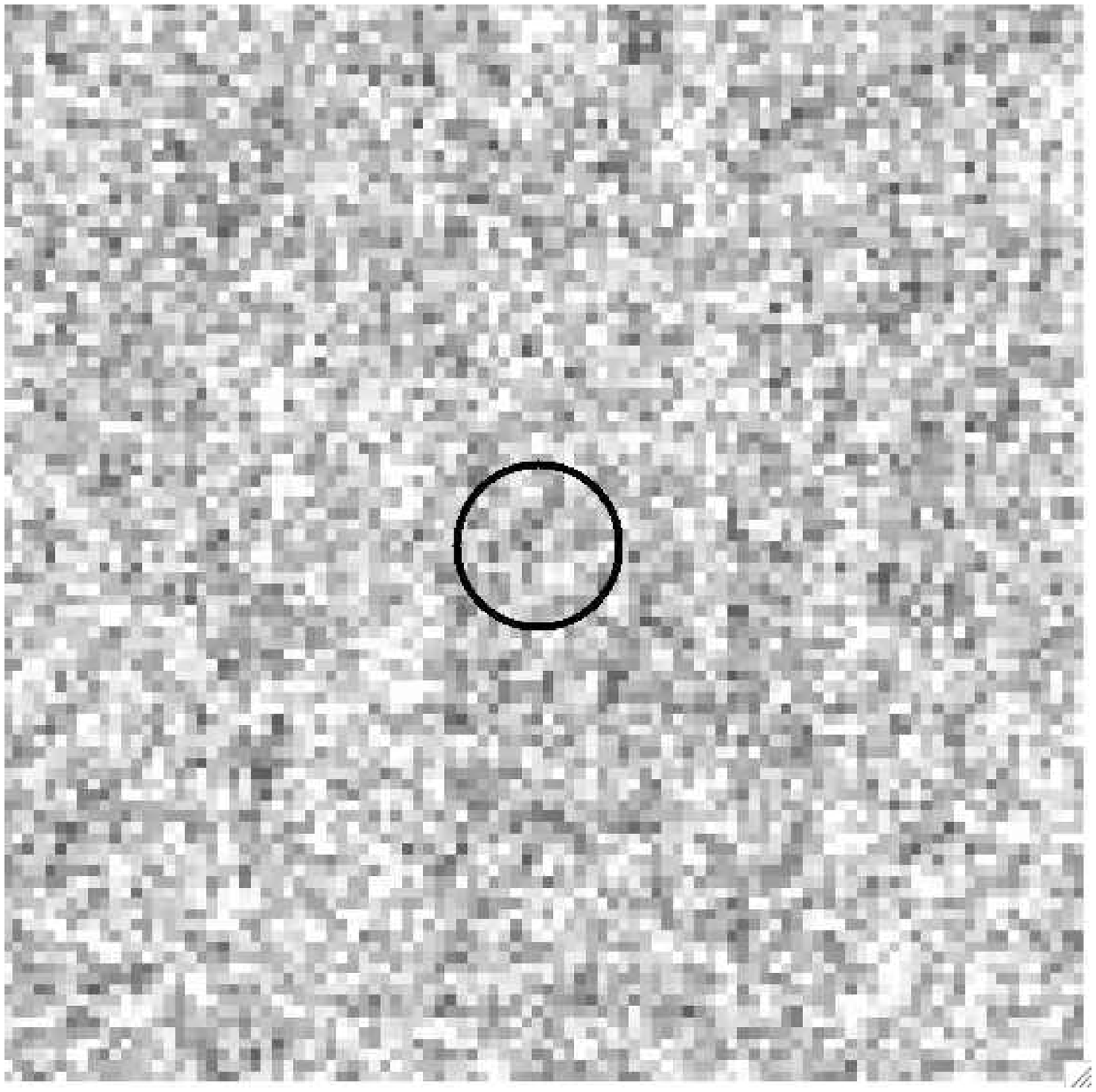}
\includegraphics[height=3.2cm]{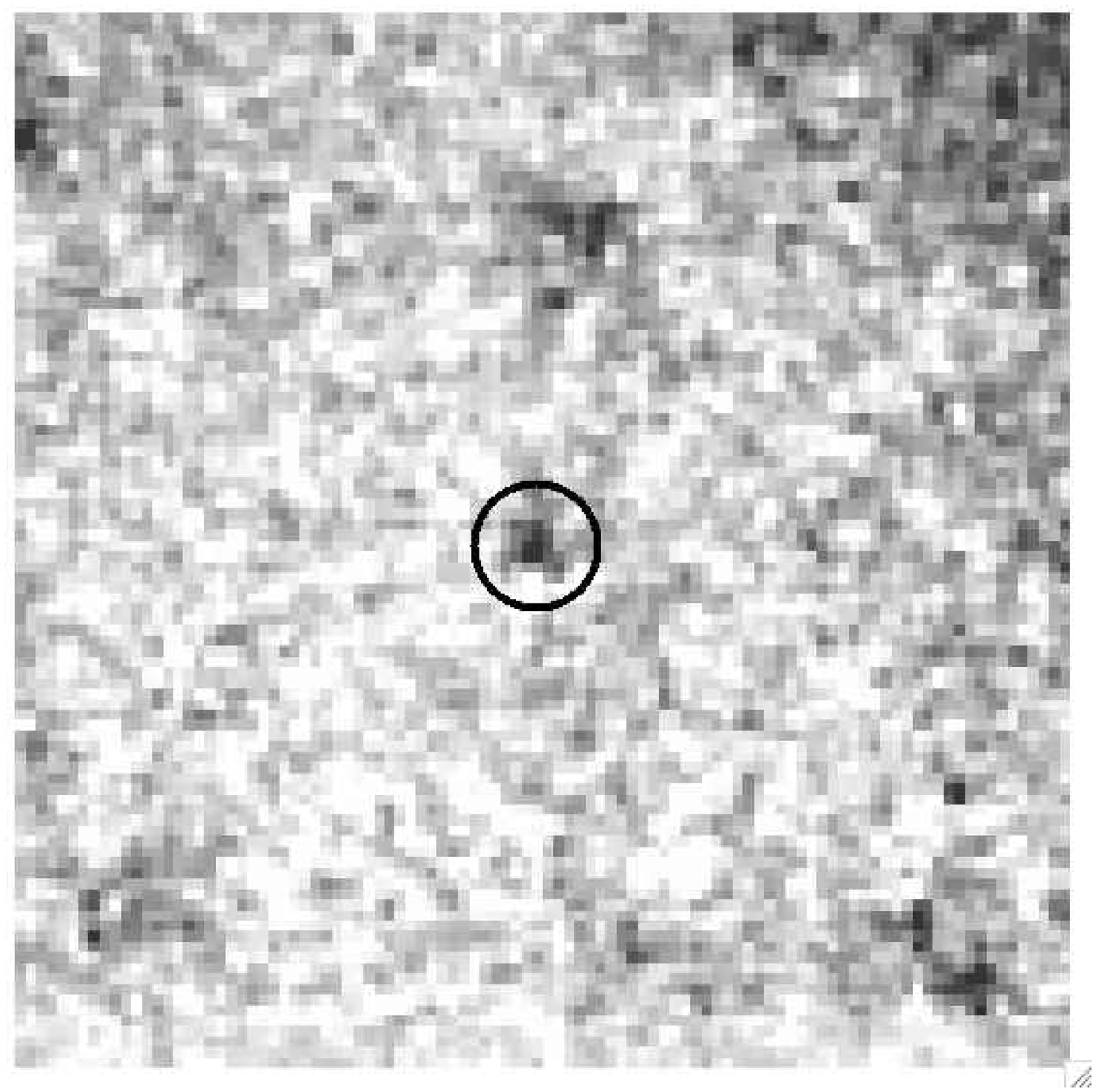}
\includegraphics[height=3.2cm]{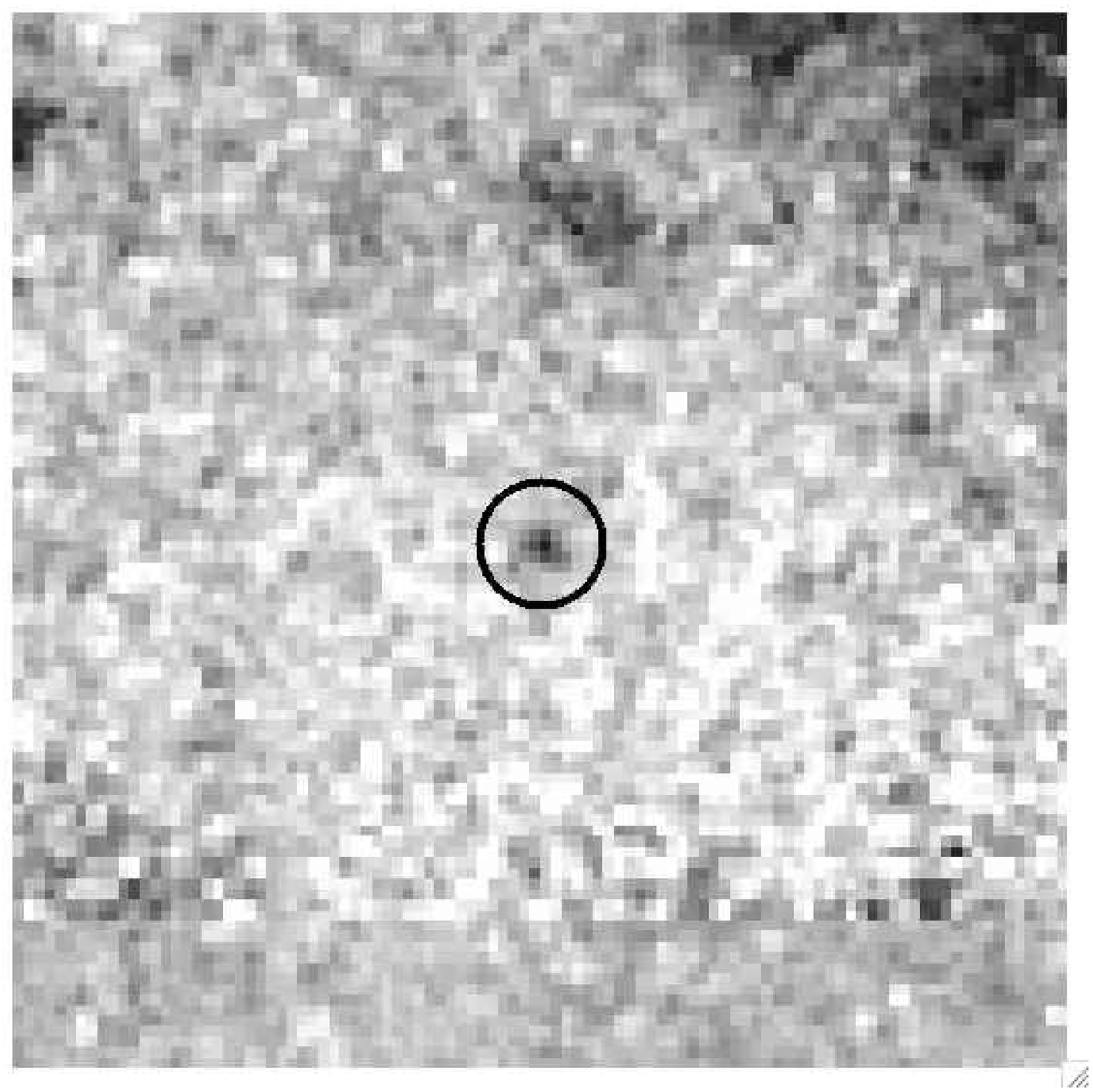}
\includegraphics[height=3.2cm]{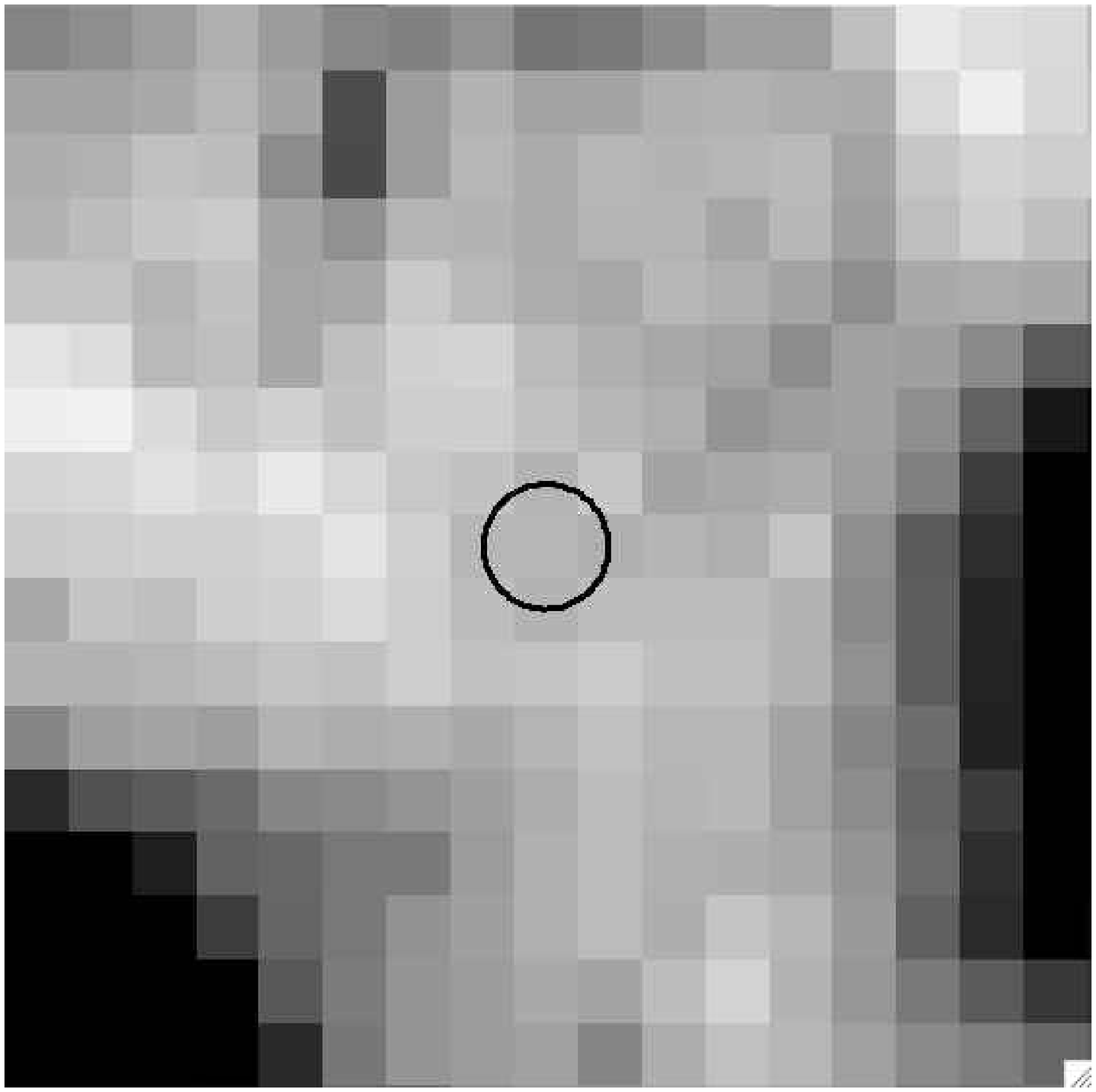}
\includegraphics[height=3.2cm]{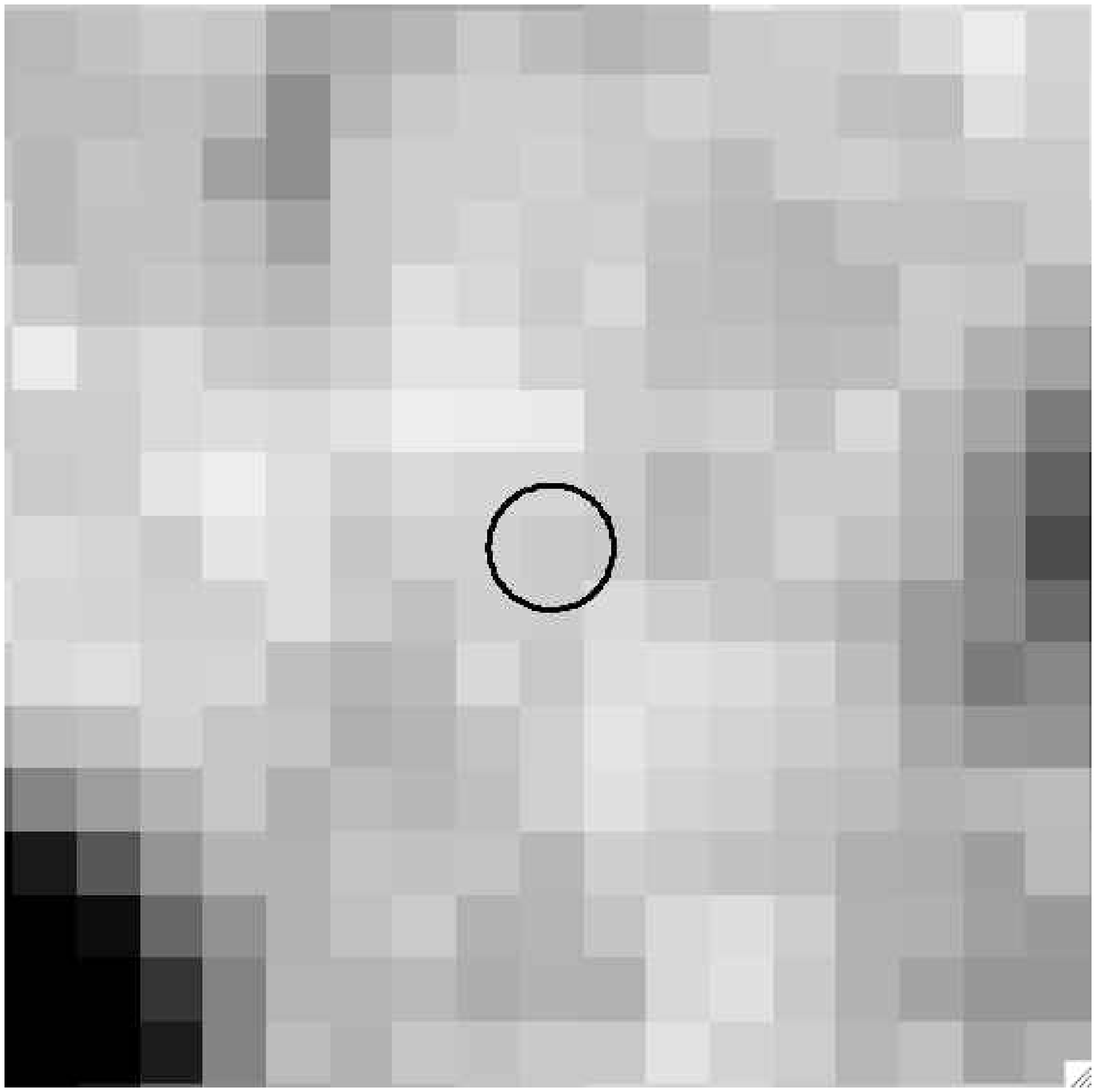}
\caption{\label{stackim}Stacked images of the $z$-band drop-outs. 
First row: ACS and NICMOS images of all 10 sources. Second row: stacked images 
for the 4 drop-outs with $K$-band imaging (rightmost image). Third row: stacked images 
for those 6 sources with unconfused IRAC data.}
\end{figure}

\clearpage

\begin{figure}
\includegraphics[width=5.cm]{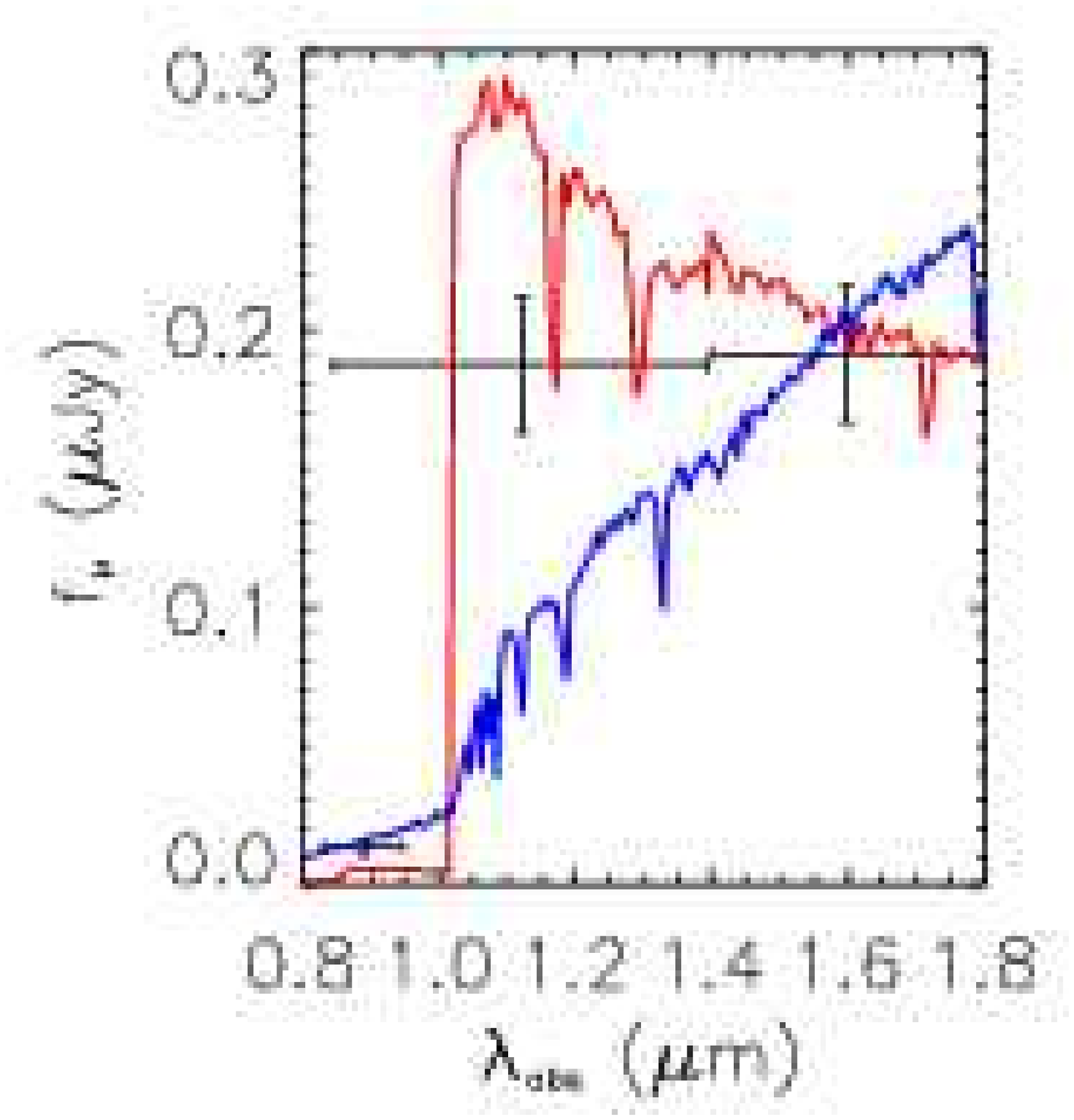}
\includegraphics[width=5.cm]{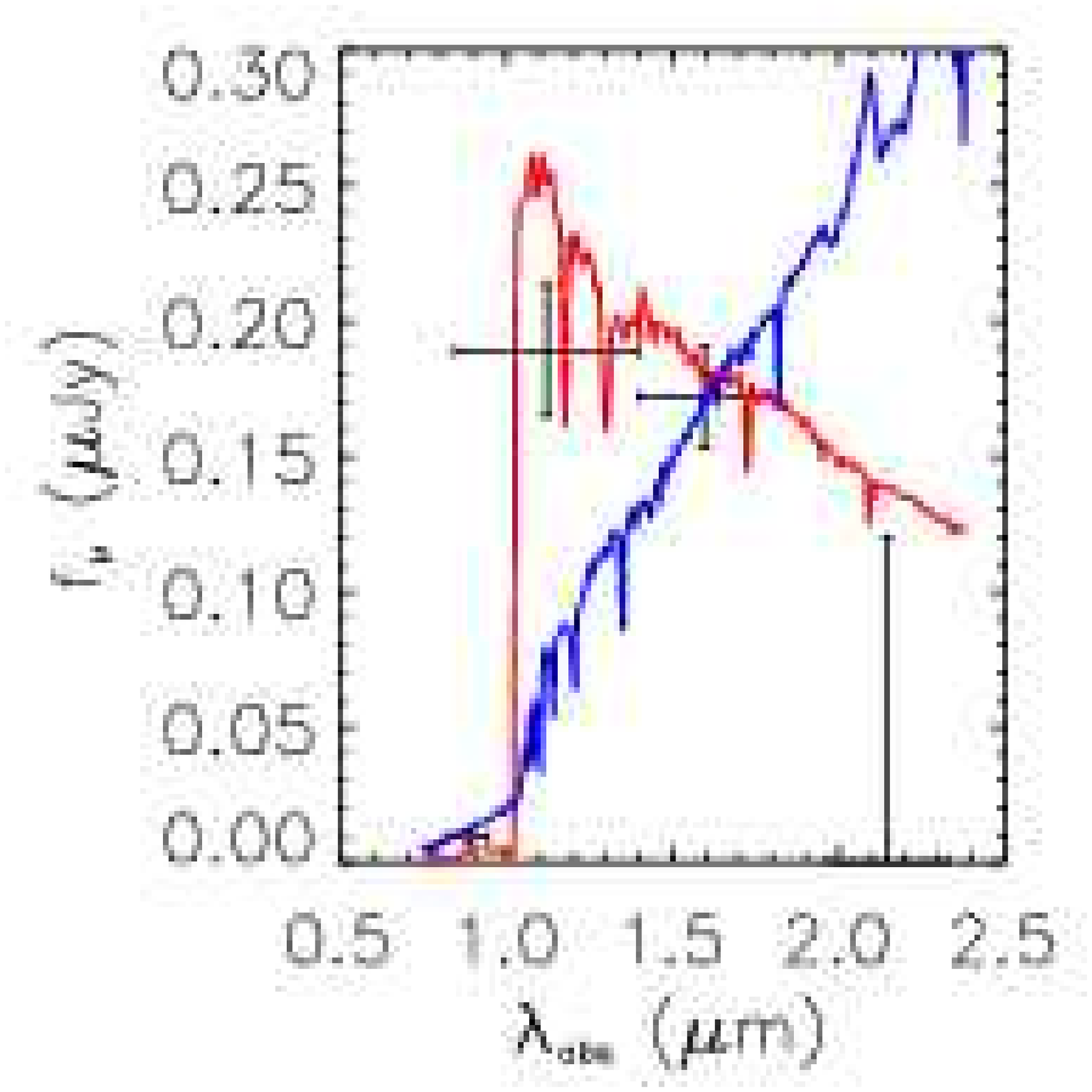}
\includegraphics[width=5.cm]{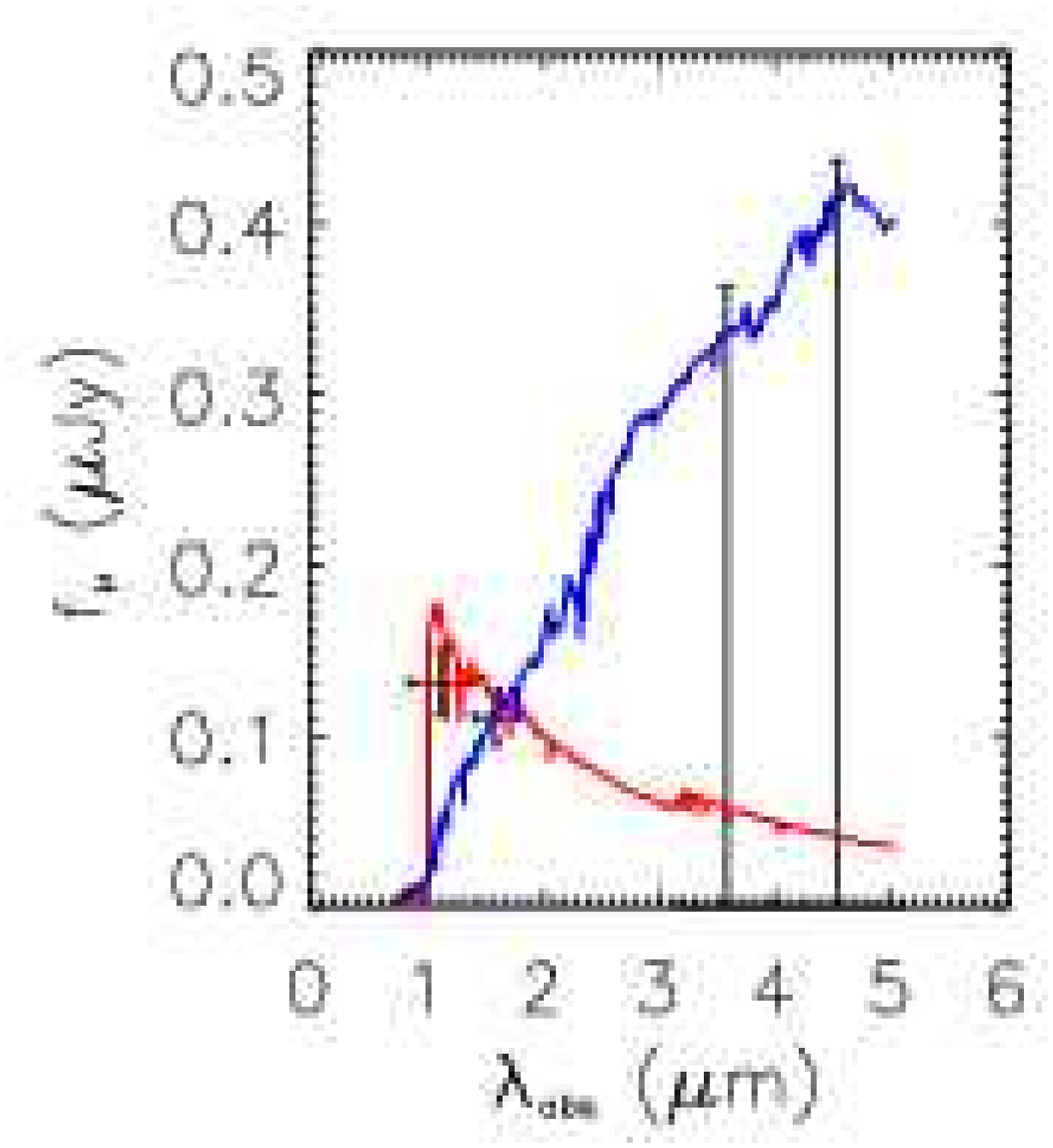}
\caption{\label{stacksed} SED of the $z$-band drop-outs derived from the stacked photometry (see 
Fig. \ref{stackphot}). (Left to right): all 10 drop-outs with ACS/NICMOS data, drop-outs with 
useful ACS/NICMOS/MOIRCS data, drop-outs with useful ACS/NICMOS/IRAC data. In each case, 
the best template found with HyperZ over $0<z<10$ (red curve) or $0<z<3$ (blue curve) is shown.}
\end{figure}

\clearpage

\begin{figure}
\centerline{\mbox{\includegraphics[width=9.5cm]{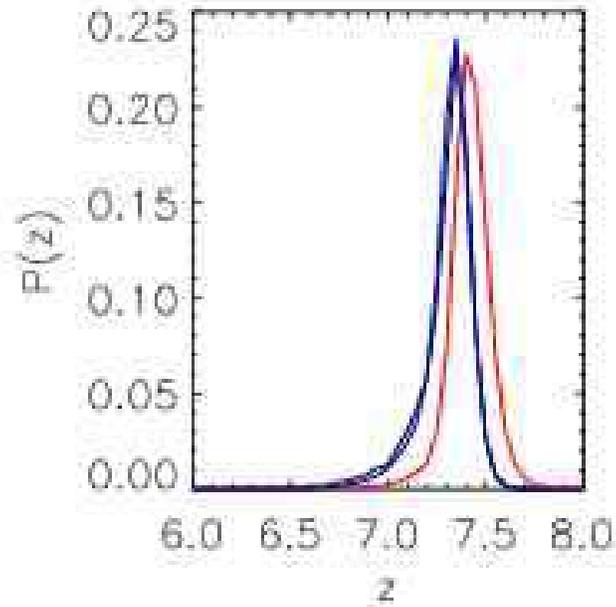}}}
\caption{\label{allpz}
Redshift probability distribution estimated using HyperZ for the stacked SEDs of the $z$-band drop-outs 
shown in Fig. \ref{stacksed}. The black, red and blue curves refer respectively to all 10 dropouts, those 
four with $K$-band imaging, and those 6 sources with useful IRAC upper limits. All three solutions are 
consistent with a mean population redshift of  $z\sim7.35$. 
}
\end{figure}

\clearpage

\begin{figure}
\centerline{\mbox{\includegraphics[width=12cm,angle=270]{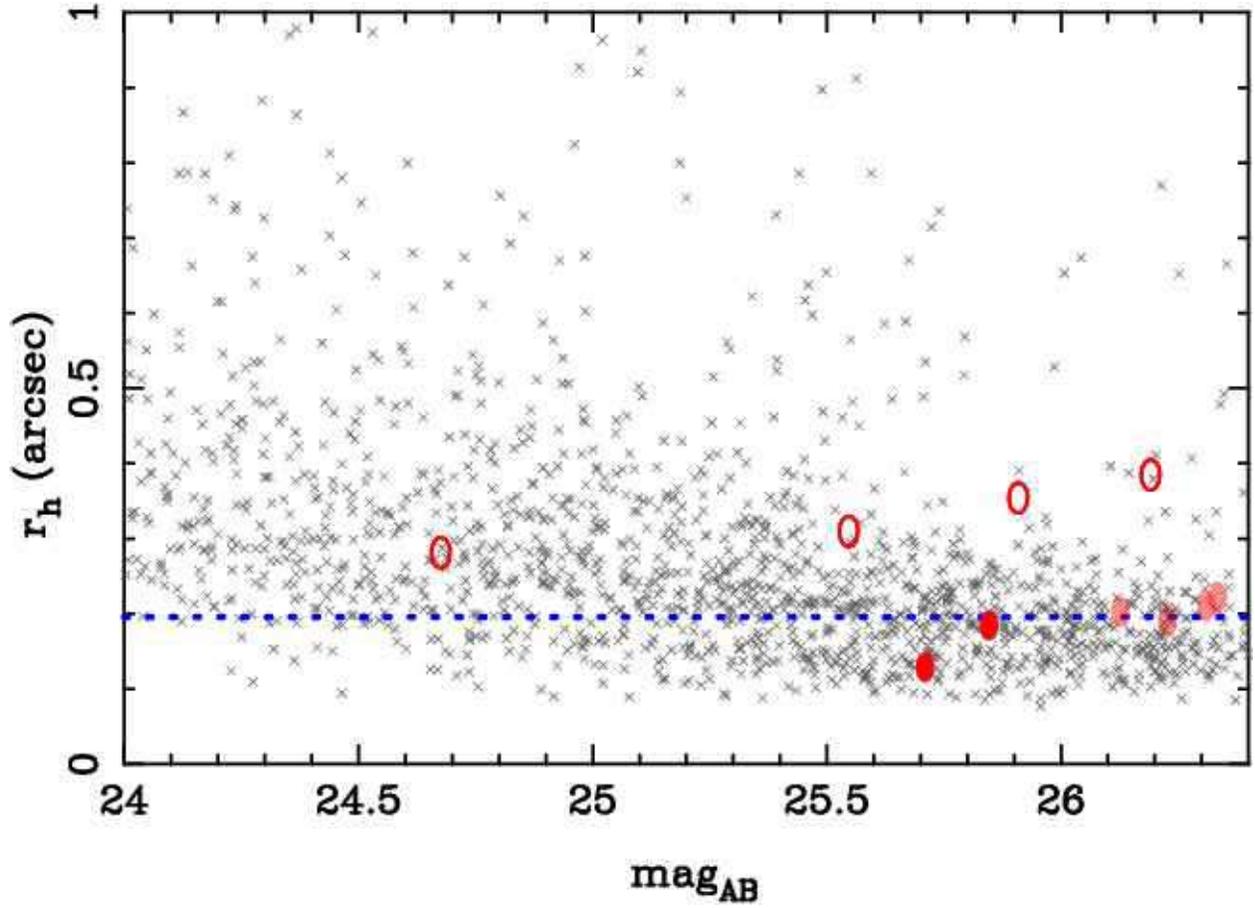}}}
\caption{\label{sizemag}Distribution of half-light radii $r_h$ measured by SExtractor in the photometric 
catalogs, as a function of the total $J$ band magnitude. The blue dashed curve corresponds to the 
measured $r_h$ for bright non-saturating stars. The 10 $z$-band drop-outs are shown as red ellipses.
Two objects are unresolved (filled ellipses) whereas four appear resolved (open ellipses.). The rest
cannot be reliably categorized.}
\end{figure}

\begin{figure}
\centerline{\mbox{\includegraphics[height=6cm]{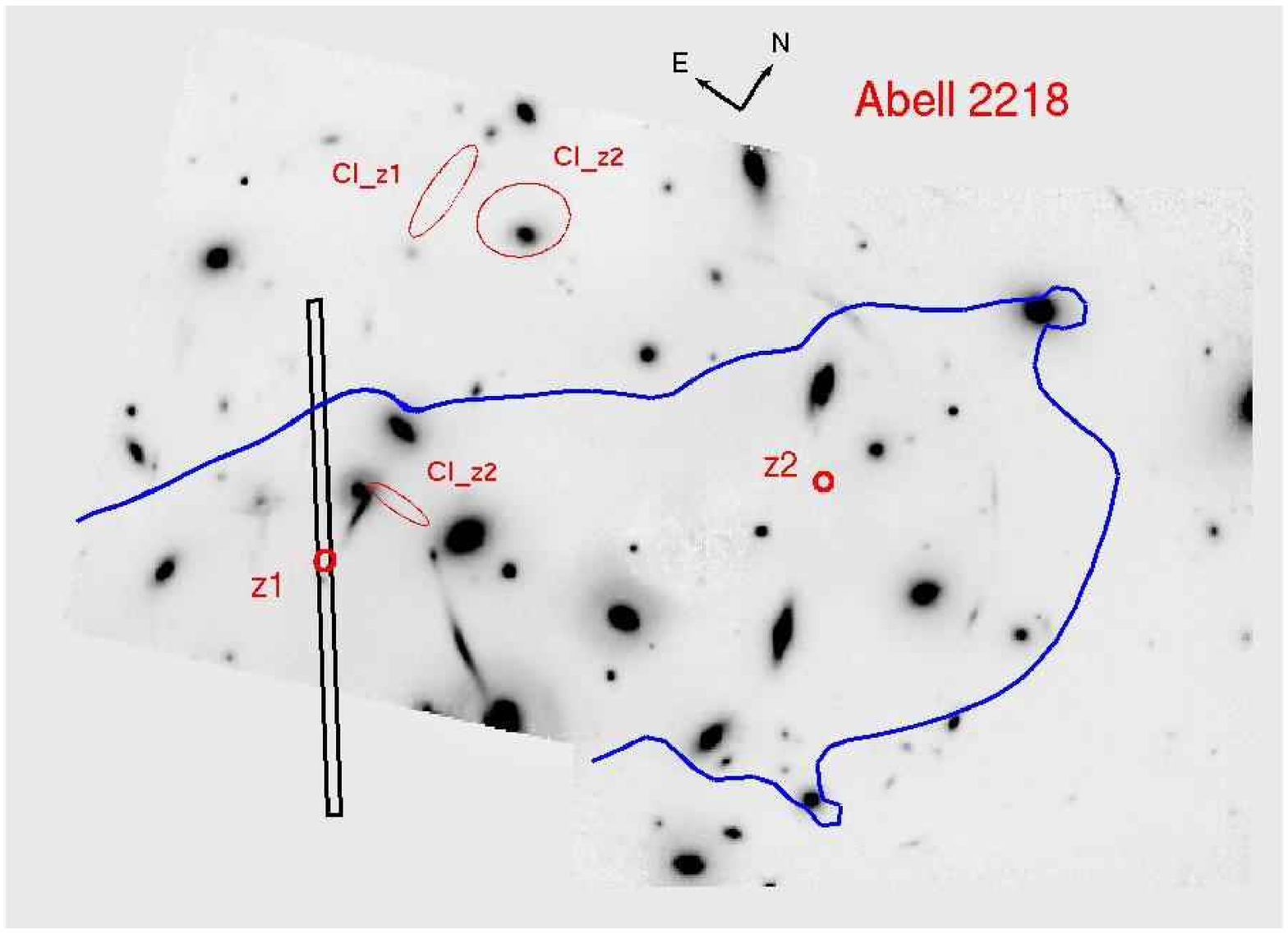}\includegraphics[height=6cm]{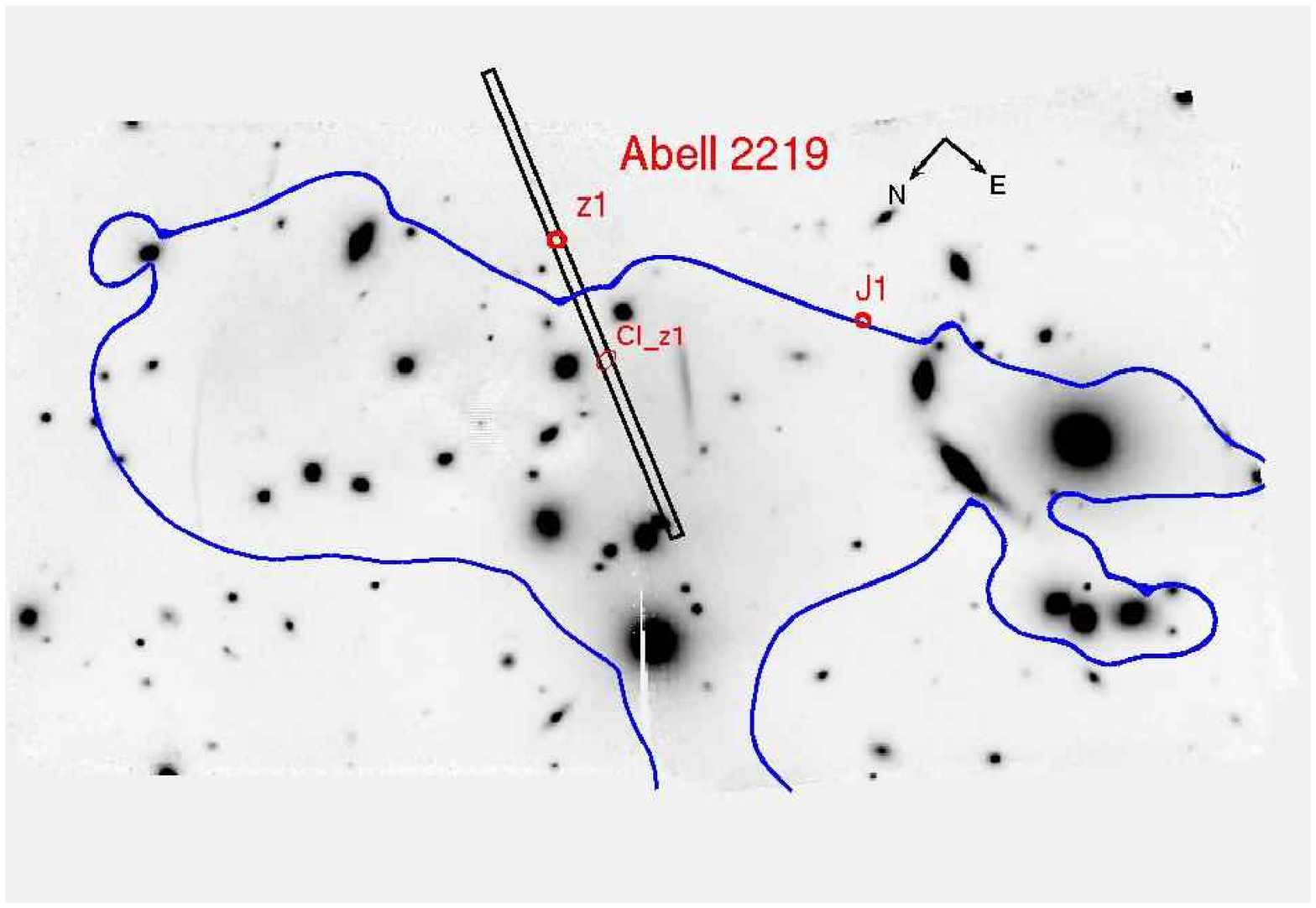}}}
\centerline{\mbox{\includegraphics[height=6cm]{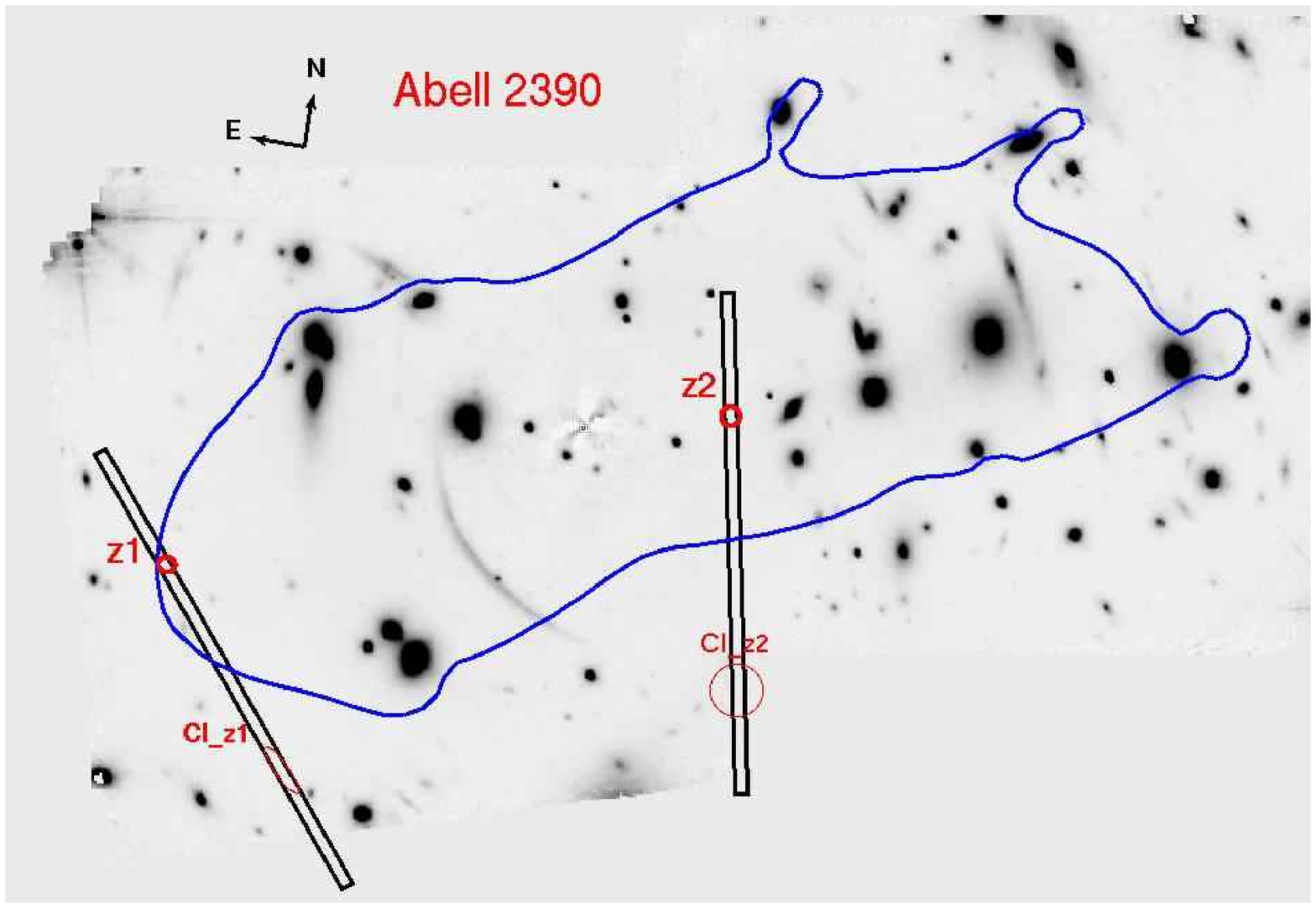}\includegraphics[height=6cm]{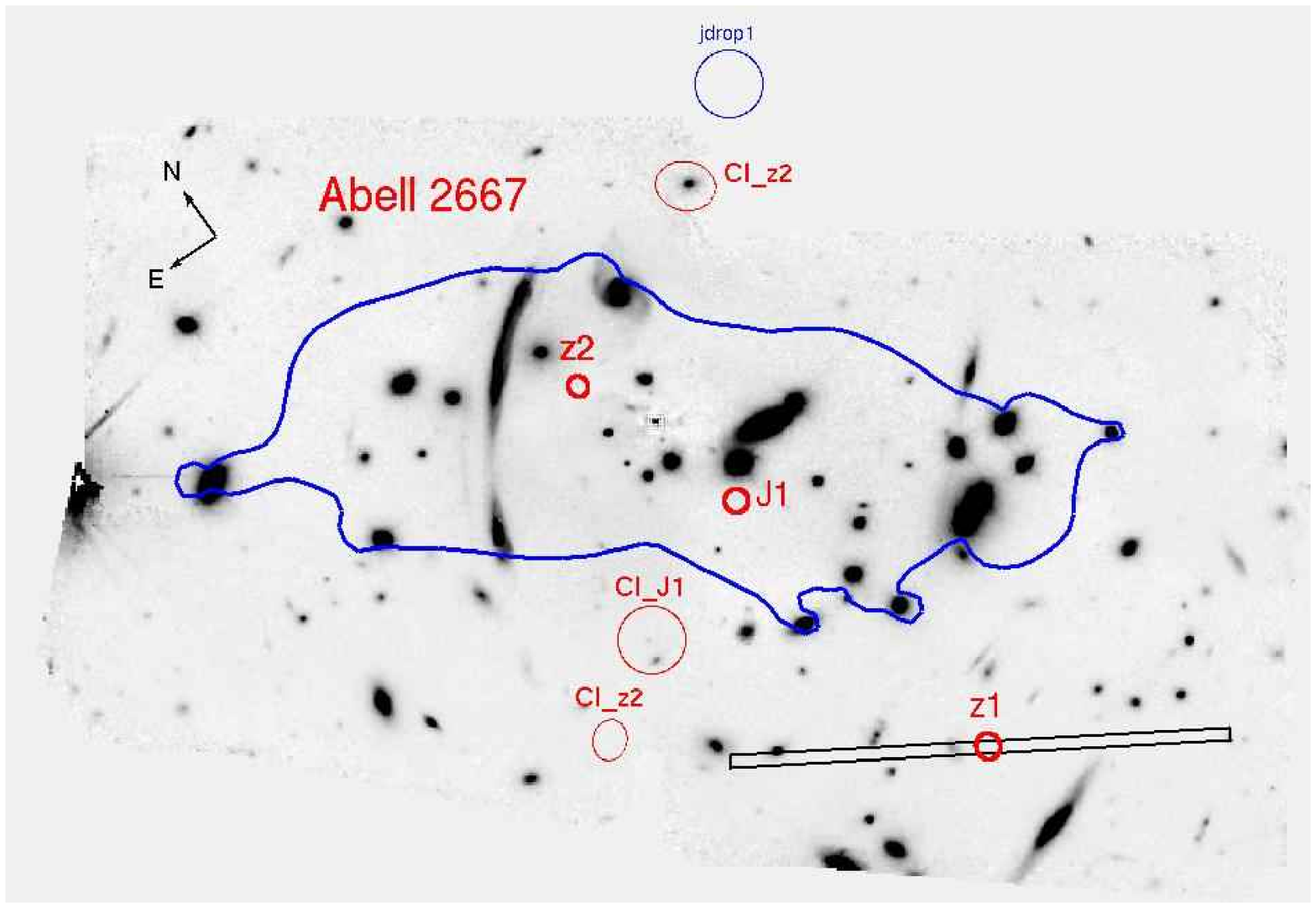}}}
\centerline{\mbox{\includegraphics[height=6cm]{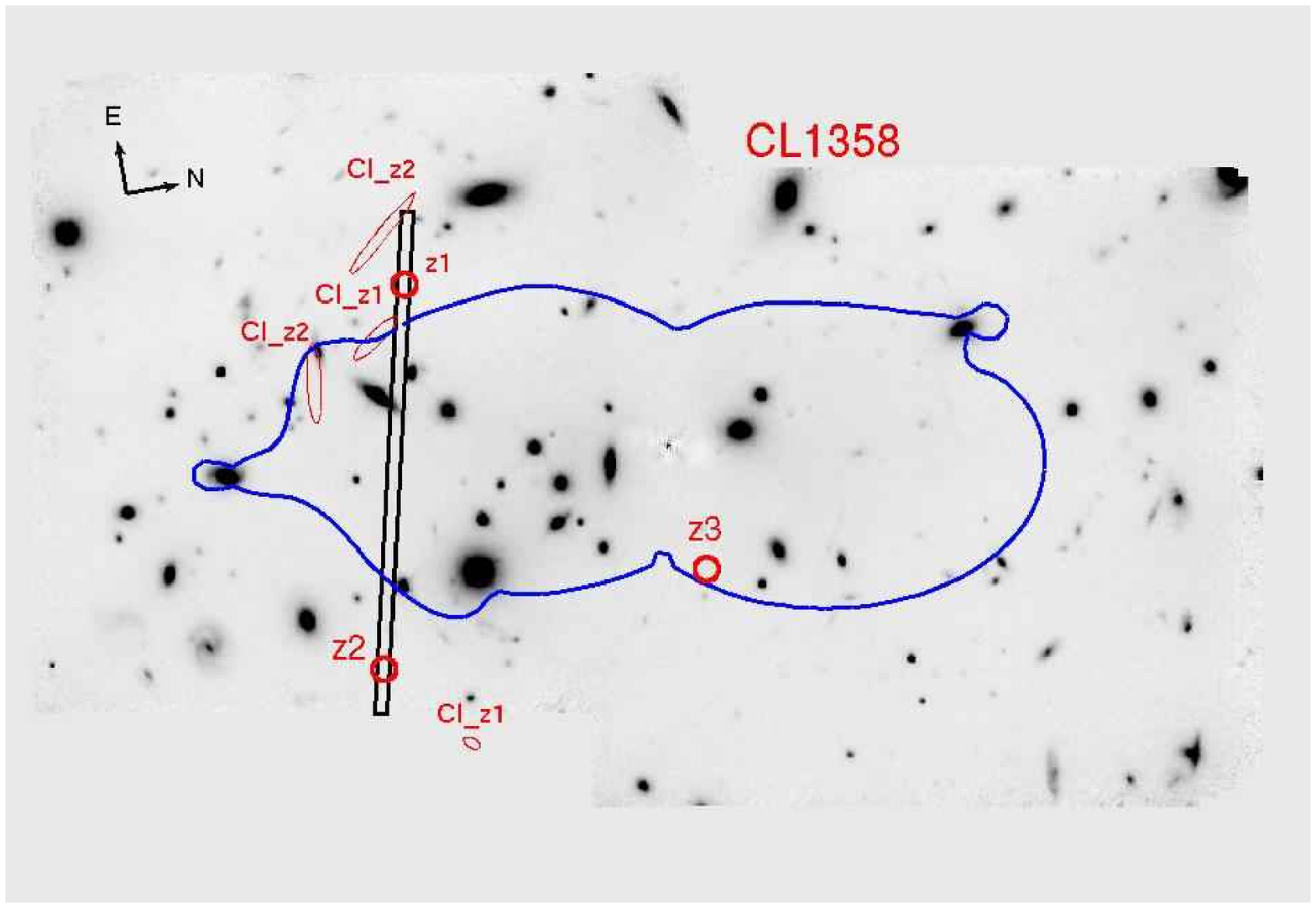}}}
\caption{\label{sample} 
Location of the $z$-band and $J$-band drop-outs with respect to the high redshift ($z=7.5$) critical line for 
each cluster field (blue curve). Ellipses with a ``CI'' label mark the position (and estimated error) of the 
brightest counter-images. The adopted NIRSPEC follow-up slit position angles are overplotted as black 
rectangles.}
\end{figure}

\begin{figure}
\centerline{\mbox{\includegraphics[width=17cm]{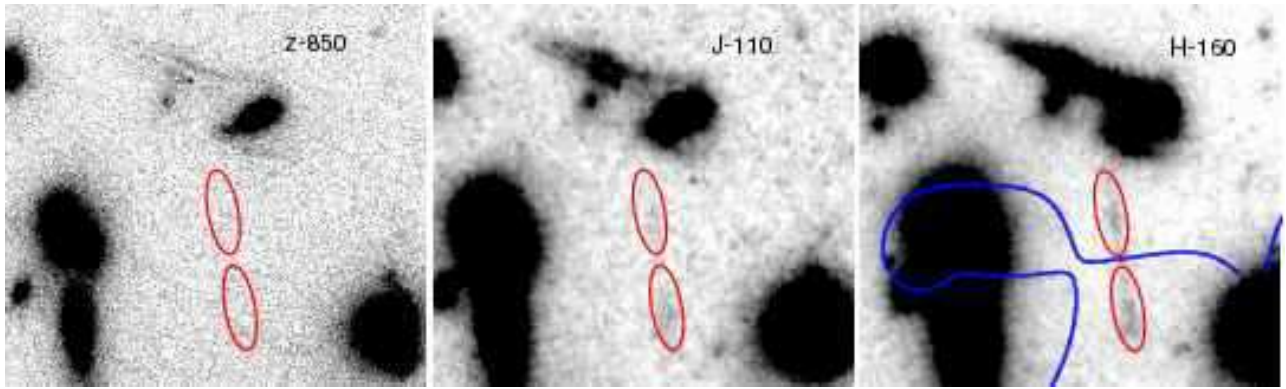}}}
\caption{\label{mult_lowz} A pair of magnified optical drop-outs identified in the NICMOS images. The 
theoretical location of the $z=1.8$ critical line (right panel) confirms this source to be a low-$z$ contaminant.}
\end{figure}

\clearpage

\begin{figure}
\includegraphics[width=15cm]{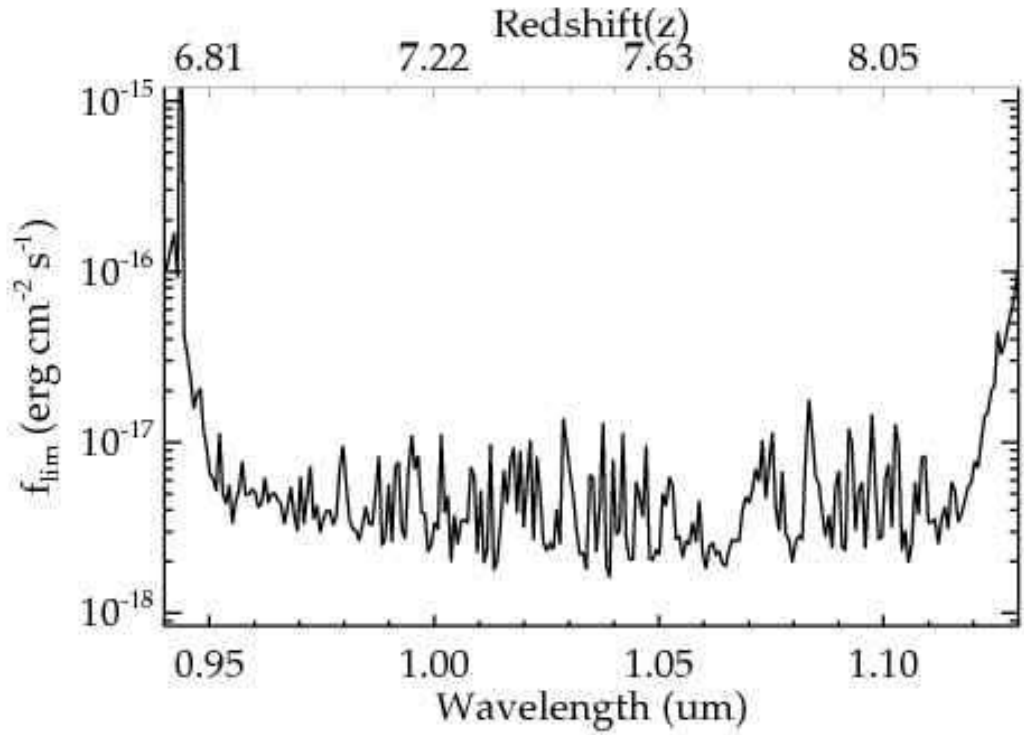}
\caption{\label{flim} Determining the limiting flux from the NIRSPEC follow-up campaign. The plot shows
the 5 $\sigma$ limiting emission line flux versus wavelength (and inferred Lyman $\alpha$ redshift) for 
a typical 3.5 hours integration.}
\end{figure}

\clearpage

\begin{figure}
\includegraphics[width=15cm]{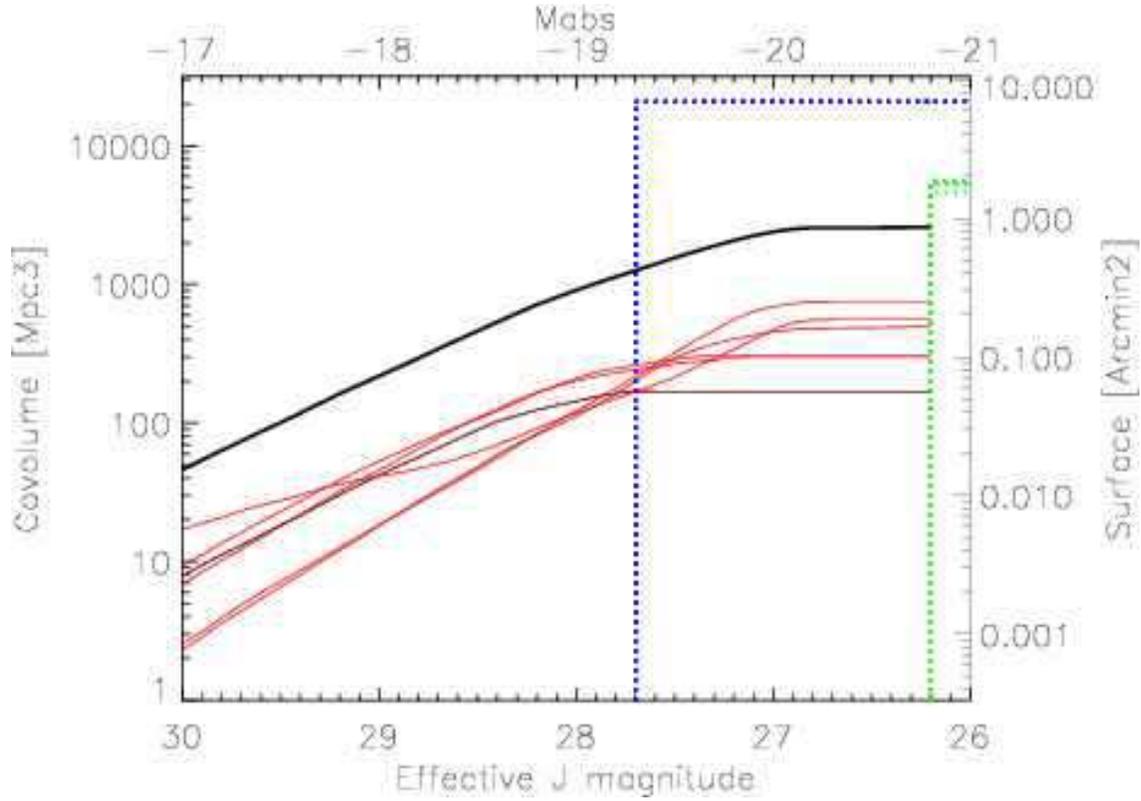}
\caption{\label{magsource} Survey characteristics: the intrinsic (unlensed) surface area sampled in the 
source plane down to a given limiting magnitude for each cluster (thin red lines) and for all six clusters 
(thick black line). The upper scales give the corresponding absolute magnitudes assuming $z\sim7.5$. The green dashed lines illustrate the areas sampled in the absence of lensing. The blue dotted line
shows the equivalent survey parameters for the UDF \citep{Bouwens04,BouwensNature}.
}
\end{figure}

\clearpage

\begin{figure}
\centerline{\mbox{\includegraphics[width=18cm]{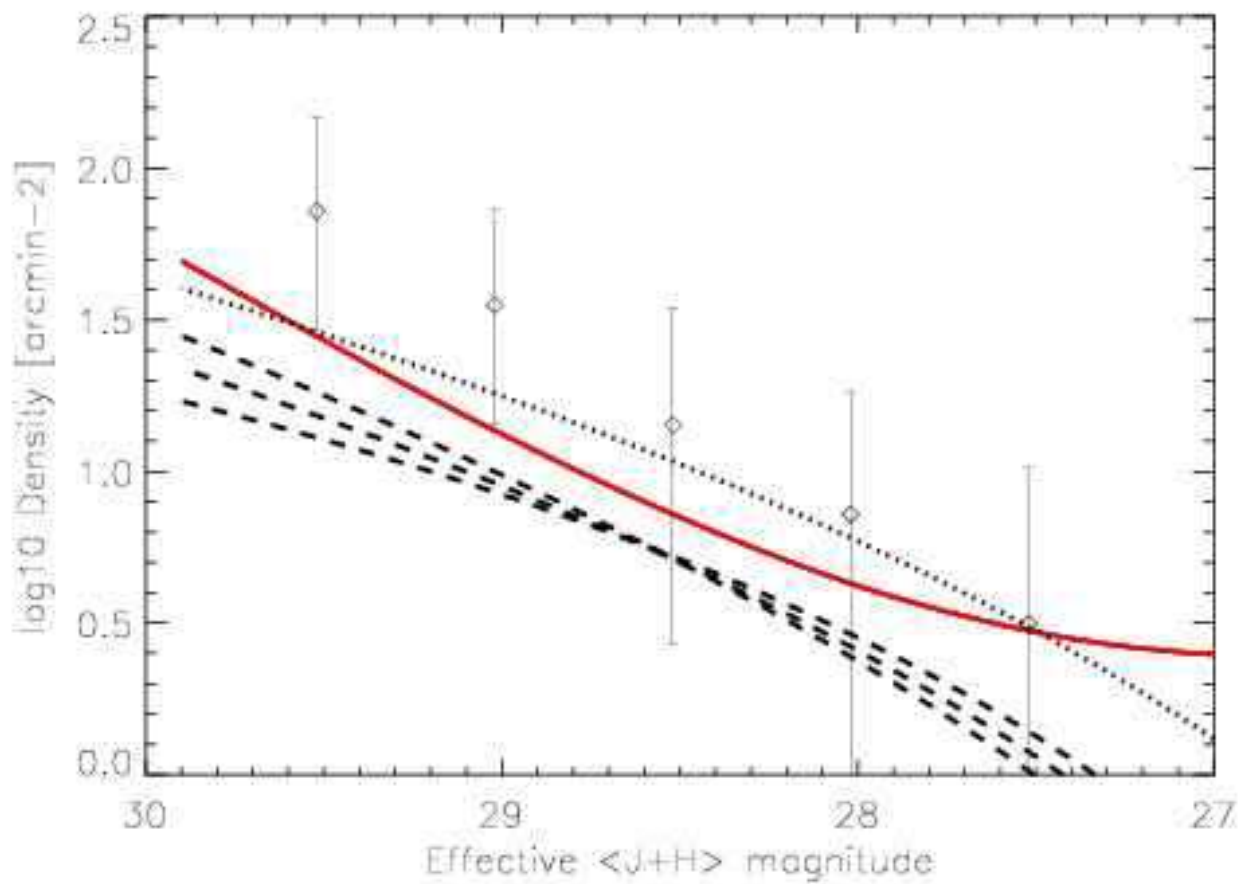}}}
\caption{\label{ncounts}
The cumulative number density of star-forming galaxies at $z\sim7-8$ as a function of the effective 
(unlensed) magnitude. Datapoints and error bars correspond to the range of densities resulting when 
randomly selecting 5 candidates from our sample and adopting Poisson errors (open diamonds, 
offset for clarity). In the most pessimistic case, where no sources are at high redshift, 
the implied upper limit is shown by the thick red curve. We overplot the best fit luminosity functions 
found by \citet{Bouwens06} (light dotted line) and \citet{Bouwens08} (bold dashed lines) in the UDF 
(parameters in Table \ref{lfparam}) 
}
\end{figure}

\begin{figure}
\centerline{\mbox{\includegraphics[width=15cm]{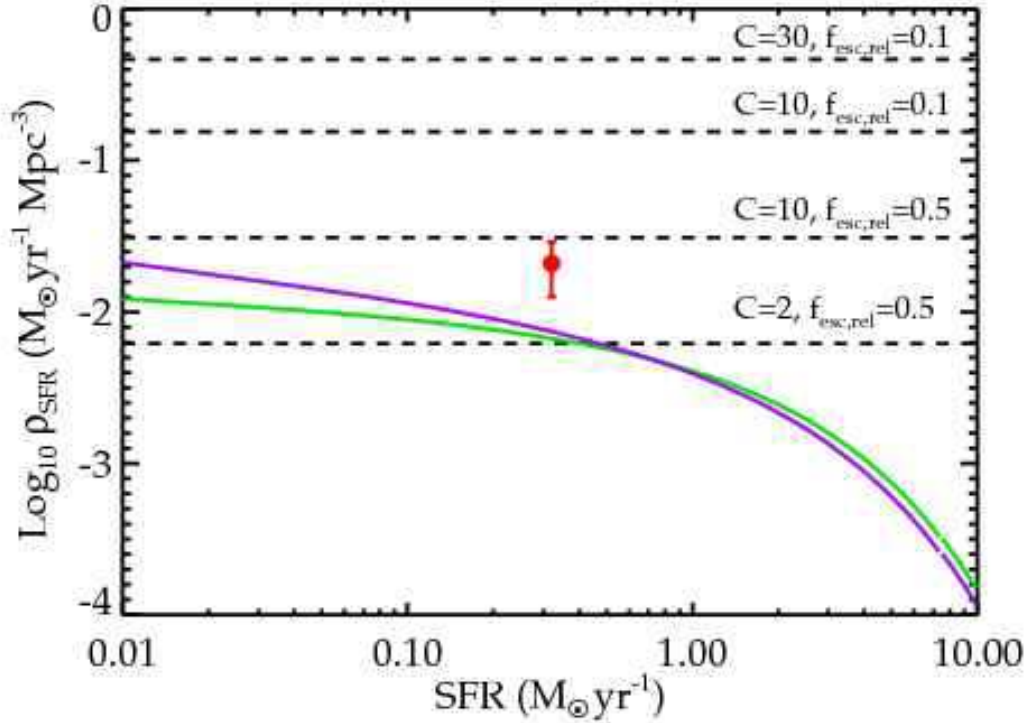}}}
\caption{\label{reion}
The cumulative comoving density of star formation rate at $z\simeq$7.5 derived for the two extreme
(two last entries of Table \ref{lfparam}) luminosity functions from \citet{Bouwens08} with faint end
slopes of $\alpha=2.0$ (purple) or $\alpha=1.4$ (green). The constraints from the present survey are
shown as a red circle, for the average and range of densities resulting when 
randomly selecting 5 candidates from our sample (as in Fig. \ref{ncounts}). The density necessary to keep the IGM reionized at $z=7.5$, calculated from
Eq. \ref{madau} for a range of clumping factors $C$ and escape fraction $f_{escp,rel}$, is shown as the dashed lines.
}
\end{figure}

\end{document}